\documentclass[twocolumn,tighten]{aastex701}

\usepackage{mathtools}
\usepackage{chngcntr}
\usepackage{amssymb}
\usepackage{amsmath}
\usepackage{color}
\usepackage{rotating,graphicx}
\usepackage{times}

\def\ovii{{{\rm O}~{\sc vii}~}}

\def\oviiin{{{\rm O}~{\sc viii}}}
\def\nex{{{\rm Ne}~{\sc x}}}
\def\mgxii{{{\rm Mg}~{\sc xii}}}
\def\sixiv{{{\rm Si}~{\sc xiv}}}
\def\sxvi{{{\rm S}~{\sc xvi}}}

\def\planck{{\it Planck}~}
\def\planckn{{\it Planck}}
\def\act{{\it Atacama Cosmology Telescope}~}
\def\actn{{\it Atacama Cosmology Telescope}}

\def\spt{{\it South Pole Telescope}~}
\def\msun{{M$_{\odot}$}~}
\def\msunn{{M$_{\odot}$}}
\def\mvirial{{M$_{200c}$}~}

\def\rvirial{{R$_{200c}$}~}
\def\rvirialn{{R$_{200c}$}}
\def\Tvirial{{T$_{200c}$}~}
\def\Tvirialn{{T$_{200c}$}}
\def\Tfloor{{T$_{\rm floor}$}~}
\def\Tfloorn{{T$_{\rm floor}$}}
\def\Tavg{{T$_{\rm avg}$}~}
\def\Tavgn{{T$_{\rm avg}$}}

\def\onehalo{{$y_{1h}$}~}
\def\twohalo{{$y_{2h}$}~}
\def\onehalon{{$y_{1h}$}}
\def\twohalon{{$y_{2h}$}}

\defcitealias{Das2023}{D23}
\defcitealias{Das2025}{D25}
\defcitealias{Madhavacheril2020}{M20}
\defcitealias{Zou2019}{Z19}
\defcitealias{Arnaud2010}{A10}
\defcitealias{Kim2022}{K22}
\defcitealias{Amodeo2021}{A21}
\defcitealias{Yang2021}{Y21}

\shorttitle{tSZ effect and richness}
\shortauthors{Das et al.}

\begin{document}

\title{Thermal Sunyaev-Zel'dovich Effect in the circumgalactic medium - III: does richness matter?}

\correspondingauthor{Sanskriti Das}
\email{snskriti@stanford.edu; dassanskriti@gmail.com}

\author[0000-0002-9069-7061]{Sanskriti Das}
\altaffiliation{Hubble Fellow}
\affil{Kavli Institute for Particle Astrophysics and Cosmology, Stanford University, 452 Lomita Mall, Stanford, CA~94305, USA}
\email{snskriti@stanford.edu; dassanskriti@gmail.com}

\author[0000-0001-6320-261X]{Yi-Kuan Chiang}
\affiliation{Academia Sinica Institute of Astronomy and Astrophysics (ASIAA), No. 1, Section 4, Roosevelt Road, Taipei 10617, Taiwan}
\email{ykchiang@asiaa.sinica.edu.tw}

\author[0000-0002-4822-3559]{Smita Mathur}
\affil{Department of Astronomy, The Ohio State University, 140 W. 18th Ave., Columbus, OH 43210, USA}
\affil{Center for Cosmology and AstroParticle Physics, The Ohio State University, 191 W. Woodruff Ave., Columbus, OH 43210, USA}
\email{mathur.17@osu.edu}

\begin{abstract}
We measure the thermal Sunyaev-Zel'dovich Effect in the circumgalactic/intragroup medium (CGM/IGrM) of $\rm 7~million\;M_{200c} = 10^{12-14}$\msun $z\leqslant1$ DESI Legacy Survey galaxies/groups using ACT+\planck observations, and investigate for the first time how the thermal pressure depends on richness. We detect the ``one-halo" term down to 10$^{12.2}$\msun at $5.8\sigma$. While several mass-richness bins are well described by the standard GNFW thermal-pressure profile ($\beta=5.49,r_o=0$), others favor a flatter and/or miscentered profile ($\beta=1.3-4.0,r_o>0.1$\rvirialn), with the mass dependence of this pattern varying with richness. The wide variety of thermal pressure profiles across mass and richness indicates a range of physical regimes, from dominant thermal feedback to efficient cooling with subdominant thermal feedback, and even substantial cooling. In some mass bins, the super-virial ($\approx2-7$~\Tvirialn) temperature floor (the lowest temperature allowed by the data for baryon-sufficient halos) points to significant heating. However, dominant kinetic-mode feedback is not essential anywhere, challenging a universal (ejective)feedback-vs-gravity picture. The slope of the thermal energy-mass relation in IGrM is consistent with self-similarity, but in the CGM it falls below self-similarity by $7.3\sigma$. This makes halos of relatively low mass and richness the dominant contributors to the universe’s thermal energy budget at $z\leqslant1$. We therefore argue that calibrations of thermal energy scaling relations must move beyond the convention of using mass as the sole demarcator and instead incorporate additional intrinsic and environmental properties of galaxies. 
\end{abstract}


\section{Introduction}\label{sec:intro}

A significant fraction of baryons is expected to reside around galactic disks within galactic halos and between galaxies in groups, as the circumgalactic and intragroup medium \citep[CGM/IGrM;][]{Cen1999}, but observational characterization of these diffuse baryons is extremely challenging. Galactic outflows, driven by AGN and/or stellar feedback, can carry substantial energy from galactic disks to halos, but how much of that energy is retained in the halo, recycled to the disk, or lost to intergalactic space remains an open problem. For the baryons, the reference is set by the cosmological baryon fraction, but having such a benchmark for the energy is nontrivial. For halos in hydrostatic equilibrium, the thermal pressure is expected to balance gravity, making the halo mass a natural indicator of thermal energy. In the presence of efficient cooling, the thermal pressure is smaller than the fiducial value, and non-thermal sources can provide the residual pressure. The thermal feedback from galactic winds can provide additional thermal energy without significantly altering the baryon distribution and potentially drive halos out of thermal equilibrium. Kinetic mode feedback primarily affects the baryon distribution and may not significantly contribute to the thermal energy budget. Thus, \textit{missing baryons} may not necessarily track \textit{missing energy} across all galaxy populations. In both scenarios of deficit/excess of thermal pressure, it is essential to test whether mass remains the primary indicator of thermal energy or whether other astrophysical properties, e.g., star formation and nuclear activity (or the lack thereof), accretion and merger history, etc., are equally important determinants.  

For halos of $>10^{12}$\msunn, the most massive and volume-filling phase of the CGM/IGrM is hot and fully ionized. The Sunyaev-Zel'dovich (SZ) effect \citep{Sunyaev1969,Mroczkowski2019}, a distortion in the cosmic microwave background (CMB) spectrum due to the inverse Compton scattering of the CMB photons with the free electrons in the intervening plasma, is a powerful technique to characterize the hot CGM/IGrM because it depends linearly on density and is independent of redshift and metallicity. The thermal SZ (tSZ) effect, the strongest of the different SZ effects, is characterized by the Compton-$y$ parameter: $y(r_\perp) = (\sigma_{\rm T}/m_e c^2) \int {\rm P_e(r)} dr_\parallel$\footnote{$\sigma_{\rm T}, m_e$ and c are Thomson's scattering cross section, the rest mass of the electron, and the speed of light in vacuum, respectively.}. It is a measure of the thermal pressure (${\rm P_e} = n_e \rm k_B T_e$) or thermal energy ($\propto \int{\rm P_e dV} \propto \int y \rm dA$) of the free electrons of the relevant medium integrated along the line-of-sight. 

Thanks to better sensitivity and smaller beam size (FWHM$\approx$$1.\!'4$) of \act (ACT) and \spt (SPT) than \planck (FWHM$\approx$$10'$), we can resolve fainter and smaller halos than was possible before. The ACT{+}\planck and SPT{+}\planck maps have been extensively cross-correlated with catalogs of massive luminous infrared galaxies and galaxy groups out to $z$$\leqslant$1.5 \citep{Schaan2021,Meinke2021,Vavagiakis2021,Pandey2022,Sanchez2023,Liu2025,Pandey2025} and large-scale structure \citep{Chiang2021,Lokken2025,Isopi2025,Raghunathan2026,LaPosta2026}. In \cite{Das2023}, hereafter \citetalias{Das2023}, we extended this study to lower masses using ACT{+}\planck data for 0.6~million $z$$\leqslant$0.3 WISE$\times$SuperCosMos galaxies, and established that the tSZ effect in lower-mass halos (down to M* of $10^{10.8\pm0.2}$\msunn) can be meaningfully constrained. In \cite{Das2025}, hereafter \citetalias{Das2025}, we expanded the analysis to 2.5 million $z$$ \leqslant$1.2 WISE$\times$DESI galaxies of comparable mass to \citetalias{Das2023}, detected tSZ down to M* of $10^{10.6\pm0.2}$\msunn, facilitating the first investigation into any correlation between star-forming activities and thermal pressure in the ionized CGM. The dependence of thermal energy on mass significantly differed in quiescent vs star-forming galaxies, implying that the convention of using mass as the primary demarcator requires a major revision in the low-mass regime. In this paper, we continue to explore the same question in light of the richness of halos. 

The SZ effect studied from a cosmological perspective examines individual halos using a top-down approach, treating them as small-scale structures (i.e., large multipoles in CMB data) primarily driven by gravity. The effects of/on baryons, e.g., ejection and heating, are considered secondary phenomena that cause deviations from a gravity-dominated regime. Massive halos, often targeted for their larger sizes and pronounced SZ effect, align well with this framework, particularly due to AGN feedback commonly present in these halos. 

However, the central galaxies of most halos are typically surrounded by several sub-halos of satellite galaxies. Thus, in addition to feedback, the combined influence of these sub-halos would affect the SZ signal of the central halo in various ways depending on their number, mass, and spatial distribution. The enormous SZ effect of the intracluster medium makes the fractional contribution of sub-halos negligible at the zeroth order. The impact of sub-halos is (partially) accounted for by the ``two-halo" term, where its shape, i.e., the radial profile, is assumed to be agnostic of the mass of the main halo (i.e., the ``one-halo"), and the fiducial amplitude is assumed to increase with the mass of the main halo \citep[e.g.,][]{Vikram2017}. This assumption is valid for massive galaxy clusters, as their richness tends to correlate with mass \citep[e.g.,][]{Murata2018,Orlowski2021}.

\begin{figure}
    \centering
\includegraphics[width=0.99\linewidth]{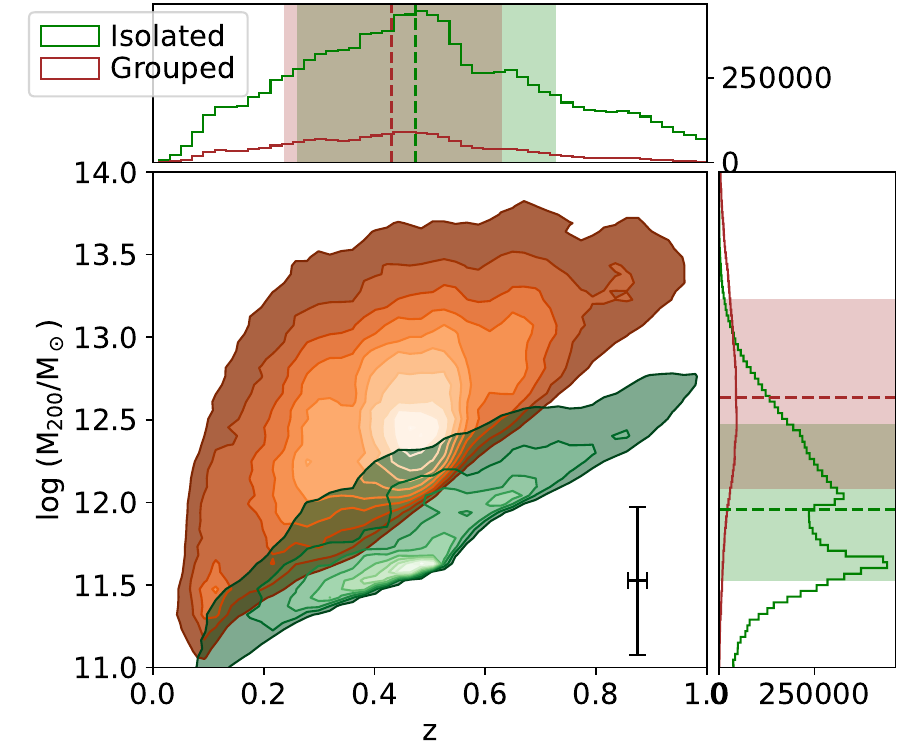}
\includegraphics[width=0.99\linewidth]{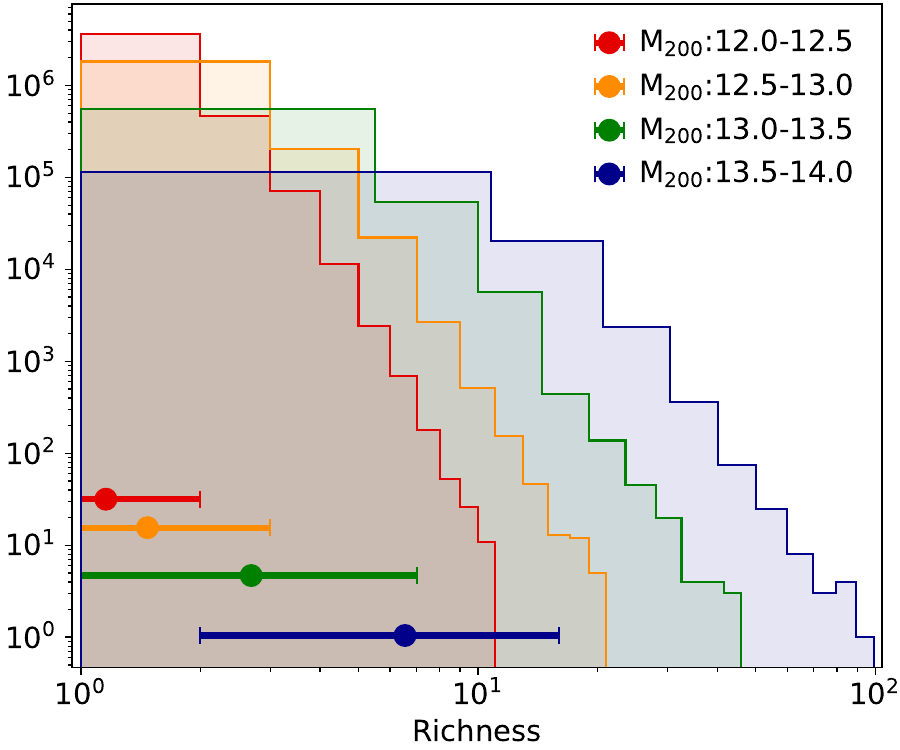}
    \caption{\textbf{Left:} Distribution of halo mass with redshift of our sample from \citetalias{Yang2021}. The green and orange histograms/contours correspond to the isolated galaxies and galaxy groups (across richness), respectively. Average uncertainties in halo mass and redshift are shown in the bottom right corner of the plot; these parameters are generally better constrained at lower redshift and higher mass. \textbf{Right:} Richness distribution in different halo mass ranges. The average richness with 90\% confidence interval is also plotted.}
    \label{fig:Mz}
\end{figure}

In contrast, lower-mass halos in galaxy groups and individual galaxies exhibit large scatter around the richness-mass relation (Figure~\ref{fig:Mz}, bottom panel). Secondly, the mass distribution of member galaxies may vary across halo mass, and a prominent central galaxy may not always be present. In such cases, the barycenter would be offset from the position of the member galaxies (as seen in the Local Group, where M~31 and the Milky~Way have comparable masses). Consequently, the SZ signal stacked at the optical/IR center of galaxies would differ from that stacked at the barycenter. Thirdly, if there is significant interaction between the sub-halos and the central halo, the ``one-halo" term in densely populated environments may exhibit a different shape compared to relatively isolated halos of similar mass. These complexities render a simplistic feedback-vs-gravity model inadequate, underscoring the importance of studying the tSZ effect as a function of richness across a large range of mass with a complementary astrophysical perspective of a bottom-up approach. 

As the first step toward this new paradigm, we expand our analysis to 7~million DESI galaxies and galaxy groups of \mvirial of $10^{12-14}$\msunn, comparable to the mass range of \citetalias{Das2023, Das2025}, covering a similar redshift range of $z\leqslant1$ as \citetalias{Das2025}. It is the first investigation into any correlation between the richness and thermal pressure of the ionized CGM/IGrM. It is the largest sample of halos explored in the tSZ effect to date. 

The paper is organized as follows. In \S\ref{sec:data}, we discuss the details of the data. We present and interpret our results in \S\ref{sec:result} and summarize our results in \S\ref{sec:end}. Throughout the paper, we use the flat $\Lambda$CDM (cold dark matter) cosmology of \cite{Planck2020}: local expansion rate $ H_0 =67.66\rm~ km~s^{-1} Mpc^{-1}$, matter density $\Omega_m=0.30966$, baryon density $\Omega_b=0.04897$, resulting in the dark energy density $\Omega_\Lambda = 1- \Omega_m = 0.69034$ and cosmological baryon fraction $f_{b,cosmo} = \Omega_b/\Omega_m$ of 0.15814. $H(z)$ is the Hubble parameter at redshift $z$, with its redshift evolution $E(z)^2 = H(z)^2/H_o^2 = \Omega_m\times(1+z)^3 + \Omega_\Lambda$. 

\section{Data extraction and analysis}\label{sec:data}

Here we provide the details about the Compton-$y$ map and the galaxy sample to cross-correlate with (\S\ref{sec:galsample}), methods to obtain radial profiles of the thermal pressure from stacked $y$-maps, extraction of the CGM/IGrM signal through parametric component separation of the thermal pressure (\S\ref{sec:model}), and estimation of the thermal energy and \textit{temperature floor} of the CGM/IGrM (\S\ref{sec:physical}). 

\subsection{Compton-$y$ maps and Galaxy/galaxy groups sample}\label{sec:galsample}

We consider the Compton-$y$ maps constructed through ILC-based component separation from \planck and ACTPol~DR4 data\footnote{{The gain in S/N per pixel in the DESI$\times$ACT DR6 data relative to DR4 is marginal. The DR4 sample size is sufficient for our scientific goals: as shown in \citetalias{Das2023,Das2025} and in \S\ref{sec:model}, the uncertainty on the stacked $y$ signal is dominated by halo-to-halo scatter rather than by statistical noise, so stacking additional galaxies would not significantly reduce it. Third, the stacked radial profiles of Compton-$y$ for the same galaxies differ between DR4 and DR6, even when both maps are deprojected with the same cosmic dust properties ($T = 24$ K, $\beta = 1.2$; Figure~\ref{fig:cib}, top right). These differences between the DR4 and DR6 $y$-maps may arise due to the choice of component-separation method, as discussed in Appendix~\ref{sec:cib}. Because the DR4 maps have been extensively validated against multiple galaxy catalogs in the literature, we adopt DR4 for this work.}} in BOSS (Baryon Oscillation Spectroscopic Survey) North ($-117^\circ < \rm RA<150^\circ$, $-2^\circ < \rm DEC < 19^\circ$; area = 1633 deg$^2$) and Deep~56 ($-9^\circ < \rm RA<40^\circ$, $-7^\circ < \rm DEC < 4^\circ$; area = 456 deg$^2$) from \citet[hereafter \citetalias{Madhavacheril2020}]{Madhavacheril2020}. In the following, we use the (cosmic and Galactic) dust-corrected $y$-maps masked at the position of individually detected galaxy clusters and radio sources (see the detailed procedure and the effect of each correction and masking on the stacked profiles in \citetalias{Das2023,Das2025}). 

To cross-correlate with the Compton-$y$ maps, we consider the catalogs of galaxies and galaxy groups based on the DESI Legacy Imaging Survey DR8 \citep{Yang2021} and extract the sky position, redshift, virial mass, and richness {($\lambda$, total number of detected member galaxies within a specific dark matter halo)}, shown in Figure~\ref{fig:Mz}. This catalog contains halos at $0<z\leqslant 1.0$ down to the virial mass of 10$^{11.5}$\msun (with 0.45~dex/0.2~dex uncertainty below/above 10$^{13.5}$\msunn) and richness of 1 with $>$90\% purity. Virial mass was defined as M$_{180m}$, representing an overdensity 180 times the critical matter density of the universe, $\rho_m(z)$. We convert it to \mvirial assuming an NFW profile of the dark matter halo and a mass- and redshift-dependent concentration factor from \cite{Diemer2019}. In the rest of the paper, virial properties represent an overdensity 200 times the critical density of the universe, $\rho_c(z)$. The richness can be over-/underestimated due to the projection effect, miscentering, or orientation bias \citep{Shi2024}, but the average richness in a stacked target may not deviate significantly from the true value. Also, {richness in this case is estimated down to a certain magnitude of the galaxies, implying that members that are bright enough to be detected (hence massive enough to have an observable effect in tSZ) are included in the richness estimate}. Thus, the true richness is not of prime importance -- we are using it as a qualitative measure of the \textit{crowdedness} in the halo. Because the tSZ effect is redshift-independent, galaxy clusters that are not hosting our targets but are along the line of sight of any target can still contaminate our measurement (see Appendix~A of \citetalias{Das2023}). Therefore, we exclude halos in this catalog that intersect the halo of any galaxy cluster individually detected in tSZ. We split halos into three categories based on richness: high richness ($>$5 members; 80436 halos)\footnote{the choice of richness bins is driven by the number of available halos that are large enough to provide a constraint on the tSZ signal}, low richness (2--5 members; 952,550 halos), and isolated (one member; 5,999,385 halos), and explore the dependence of the thermal energy on mass for each category. We also test if, at the same mass, the radial profiles of thermal pressure and integrated thermal energy within the virial radius differ across richness. 

\subsection{Stacking, aperture photometry, and pressure modeling}\label{sec:model}

We developed a novel probabilistic approach to stacking and extracting the radial profiles of Compton-$y$, $y_{\rm stacked}$, by fully accounting for the uncertainties in redshift and halo mass of individual targets in a computationally rigorous but inexpensive way (see Appendix~A of \citetalias{Das2025}). We employ the same method here. The conventional method that stacks the $y$-map at the redshift and halo mass of targets (without uncertainties) at fixed angular radius and then retrieves the radial profile of the stacked $y$-map may result in less accurate and misleadingly precise estimates. Instead, we split the entire mass and redshift range of the halos into a 2-d grid of infinitesimal \textit{cells} of 0.02~dex and average the radial profile of Compton-$y$ at a fixed fraction of the virial radius across halos, weighted by the probability of an individual halo to be in the redshift-mass of that \textit{cell} (equation~\ref{eq:stack}). It provides a more accurate estimate of the stacked $y$-value at the cost of precision, as the measurement uncertainties are fully included. Finally, the unbiased mean and the covariance matrix, \textbf{C$_{\rm stacked}$}, are calculated by bootstrapping the sample in each mass and richness bin over 1000 replicas. \textbf{C$_{\rm stacked}$} includes both statistical and systematic uncertainties and is dominated by systematic uncertainties {caused by large halo-to-halo scatter in the intrinsic Compton-y profile of galaxies/groups}.  

 \begin{widetext}
\begin{equation}\label{eq:stack}
    y_{\rm stacked}(x_\perp) =
    \sum_i w({\rm M}_{200c,i}) \sum_j  w(z_j) y_{avg}(\theta_{ij} |{\rm M}_{200c,i},z_j); \;{\rm where}\; \theta_{ij} = r_{\perp,i}/{\rm D}_{{\rm A},j}; \;\;    x_\perp = r_{\perp,i}/{\rm R}_{200c,i} 
\end{equation}  
\end{widetext}

$y_{\rm stacked}$ in equation~\ref{eq:stack} consists of three components: 1) the signal or the ``1-halo" term (\onehalon), i.e., the projected radial profile of Compton-$y$ of stacked targets, 2) the ``2-halo" term, \twohalon, the tSZ signal of the surrounding halos, and 3) any zero-point offset, $y_{zp}$, the tSZ signal of halos in the foreground/background of stacked targets, global Compton-$y$ from reionization, residual contamination from dust, and any calibration uncertainties. Same as \citetalias{Das2025}, to fit \onehalon, we use the generalized NFW (GNFW) profile of thermal pressure of free electrons \citep[Equation~\ref{eq:P};][]{Nagai2007,Amodeo2021}.  Additionally, we explore the possibility of miscentering in stacked targets by employing a conditionally variable normalized centroid, $x_o=r_o/R_{200c}$; this is the \textit{first time} miscentering has been taken into account in stacking. In equation~\ref{eq:P}, $\frac{\mu}{\mu_e}$ converts the gas pressure to free electron pressure; $\mu$ and $\mu_e$ are the mean molecular weight and the same per free electron. We adopt $\mu$ = 0.59 and $\mu_e$ = 1.14, the values appropriate for a fully ionized medium with primordial abundances. $x_c$ is the normalized core radius $r_c/R_{200c}$ and $\alpha$, $\beta$, and $\gamma$ are the slopes of the pressure profile at {intermediate, large, and small radius, i.e,} $x\approx x_c$, $x\gg x_c$, and $x\ll x_c$, respectively. Most halos of our sample are unresolved by the ACT beam, so the impact of $\alpha$ and $\gamma$ is negligible on the pressure profile. Therefore, to avoid degeneracy among the GNFW parameters, we freeze $x_c$, $\alpha$ and $\gamma$, at their empirically derived best-fitted value of 0.6, 1.015, and $-$0.308 for M$_{500}$$>$ 10$^{14}\rm M_\odot$ galaxy clusters at $z<0.2$, and conditionally vary the fiducial $\beta$ of 5.49 \citep{Arnaud2010}. 

\begin{widetext}
\begin{subequations}  
\begin{equation}\label{eq:P}
P_e(x|M,z) = \Big[200\frac{\mu}{\mu_e}\frac{GM_{200c}}{2R_{200c}}\rho_{c}(z)f_{b,cosmo}\Big] P_o \Big(\frac{|x-x_o|}{x_c}\Big)^\gamma \Big[ 1+ \Big(\frac{|x-x_o|}{x_c}\Big)^\alpha \Big]^{(\beta-\gamma)/\alpha}; \; where\;x = r/R_{200c}
\end{equation}   
\begin{equation}\label{eq:y1h}
y_{1h}(x_\perp|M,z) = \frac{\sigma_T}{m_e c^2}\int_{-\infty}^\infty P_e(\sqrt{x_\perp^2 + x_\parallel^2}|M,z) dx_\parallel \circledast Beam(x_\perp) 
\end{equation}     
\end{subequations}
\end{widetext}

We convolve the projected pressure profile with a Gaussian Beam of FWHM $2.4'$ relevant for cosmic-dust-deprojected Compton-$y$ maps \citepalias{Madhavacheril2020} to construct $y_{1h}$ (equation~\ref{eq:y1h}). We use $y_{2h}$ from \cite{Vikram2017}, scaled to the average mass and redshift of our sample, with variable amplitude $A_{2h}$. Thus, the free parameters are $\rm P_o$ (the amplitude of \onehalon), $A_{2h}$, and $y_{zp}$. We set uniform priors of $0\leqslant\rm P_o\leqslant100$, and $0\leqslant\rm A_{2h}\leqslant 100$. 

Compared to previous studies, we have substantially modified and introduced a novel technique for modeling the pressure. We undertake an Occam's razor approach to rigorously test whether the data \textit{statistically require} a deviation from the fiducial shape and centroid in terms of $\beta$ and $r_o$, and in case either of these is \textit{required}, we undertake a fit-by-part strategy. Below, we describe the technique and the philosophy behind this in detail.  

To the fitting routine, \onehalo and \twohalo are essentially two components with different \textit{shapes} without their physical information. {When \onehalo dominates the total $y$-value at the fiducial $\beta$, the data can distinguish between flatter (smaller than fiducial) vs steeper (closer to fiducial) profiles and unambiguously prefer a value near the true value of $\beta$.} But if \twohalo dominates the total $y$-value at the fiducial $\beta$ of \onehalon, {the modeling with freed $\beta$ becomes indecisive.} As \onehalo becomes flatter at reduced $\beta$, it starts to take on a shape that the fitting routine cannot separate from \twohalon. Beyond a point, a weaker \twohalo with a flat \onehalo of comparable strength can be \textit{statistically indistinguishable} from a stronger \twohalo and a subdominant steep \onehalon. Thus, if $\beta$ is allowed to vary simultaneously with other parameters, results may be degenerate with weaker marginalized constraints on and larger covariance between all parameters, without providing additional physical knowledge. Moreover, the dependence on $\beta$ could be non-monotonic, with local minima far from the global minimum. It can happen for any free parameter, but as the model depends linearly on the other three parameters, it is easier to track. As the projected pressure is affected both in shape and amplitude by $\beta$, the tracking of local minima would be nontrivial. The same issues apply for $r_o$ as well. 

In previous studies \citepalias{Das2023,Das2025}, we varied $\beta$ simultaneously with the other three free parameters, but it did not converge, and the final probability distribution of $\beta$ was practically uniform across the entire range. That implied the data could not determine if the true  $\beta$ was different from the fiducial $\beta$. Therefore, we went a step back, kept $\beta$ frozen at the fiducial value, and fitted the data with the three free parameters. Because our current data are more precise due to a larger sample size of 7 million (compared to 2.5/0.6 million in \citetalias{Das2025}/\citetalias{Das2023}), and every sample should be tested independently, we look for the optimum shape of the pressure profile from scratch.

To avoid the degeneracy as discussed in the previous paragraphs, we do not simultaneously vary $\beta$ with the other free parameters. Instead, we fit the data for the three free parameters by freezing $\beta$ at different values and obtain the fit statistics in the form of the Bayesian information criterion \citep[BIC;][]{Neath2012} for the best fit at each frozen $\beta$. Effectively, it is a fit-by-part with four (three+one) parameters. At large impact parameter $(r\gg r_c)$, the pressure is proportional to $(r/r_c)^{-\beta}, {\rm therefore}, y_{1h} \propto \int P dr \propto \frac{r^{1-\beta}}{1-\beta}$. Thus, the observed tSZ effect won't converge if $\beta\leqslant1$. Therefore, we consider $\beta$ only above 1. Following the standard convention of BIC, the range of $\beta$ with $\rm \Delta BIC = BIC - BIC_{min}\leqslant2$ is \textit{statistically indistinguishable} from the $\beta$ with minimum BIC, i.e., maximum likelihood. The data will have \textit{moderate evidence} and \textit{strong evidence} against the $\beta$ with $2<\rm \Delta BIC\leqslant6$ and $6<\rm \Delta BIC\leqslant10$, respectively, while $\beta$ with $\rm \Delta BIC>10$ will be \textit{ruled out}. 

We split possible results broadly into two scenarios: 1) We choose the $\beta$ with maximum likelihood, $\beta_{best}$, only if the data have at least \textit{moderate evidence} against the fiducial $\beta$. We consider the range of $\beta$ that is \textit{statistically indistinguishable} from $\beta_{best}$ to estimate the systematic uncertainty, which is propagated to the statistical uncertainties of the three free parameters, including the covariance between $P_o$ and $\beta$. 2) Same as \citetalias{Das2023,Das2025}, if the fiducial $\beta$ is \textit{statistically indistinguishable} from $\beta_{best}$, we choose the fiducial $\beta$.

We have stacked the galaxies/groups at their optically identified dynamical center, which may not coincide with the true centroid of the thermal pressure distribution. A flatter-than-fiducial pressure profile could have observational signatures similar to those of a mis-centered pressure profile at fiducial slope. However, the physical interpretation of these two cases is entirely different. The first case is related to thermodynamic properties, e.g., thermal/kinetic feedback, cooling, etc. But the second case is related to hydrodynamical properties, e.g., the lack of dynamical relaxation, stripped baryons from sub-halos, and merger history. Therefore, if the data \textit{statistically require} a deviation from the fiducial $\beta$, we additionally test whether the deviation is true or a manifestation of a miscentering (i.e., $r_o$ differing from zero). Note that $r_c$ and $r_o$ are strongly degenerate, but as most of the halos are unresolved by the ACT beam, any physically meaningful values of $r_c$ and $r_o$ will be below the angular resolution. Therefore, to avoid overfitting, we keep the $r_c$ frozen as before and fit the data for the original three free parameters while freezing $r_o$ and $\beta$ in a 2-d mesh of physically feasible values ($\beta = 1-7$, $r_o=0-0.35 R_{200c}$). Thus, it is effectively a fit-by-part with five (three+two) parameters. We choose the best combination of $r_o$ and $\beta$ based on $\rm \Delta BIC$. A nonzero $r_o$ and/or different-from-fiducial $\beta$ are chosen only if it is \textit{statistically required}, and the systematic uncertainties are derived from the range of $r_o$ and $\beta$ that are \textit{statistically indistinguishable} from the values with minimum BIC. 

\begin{figure*}
\includegraphics[trim=15 0 45 40, clip,width=0.33\linewidth]{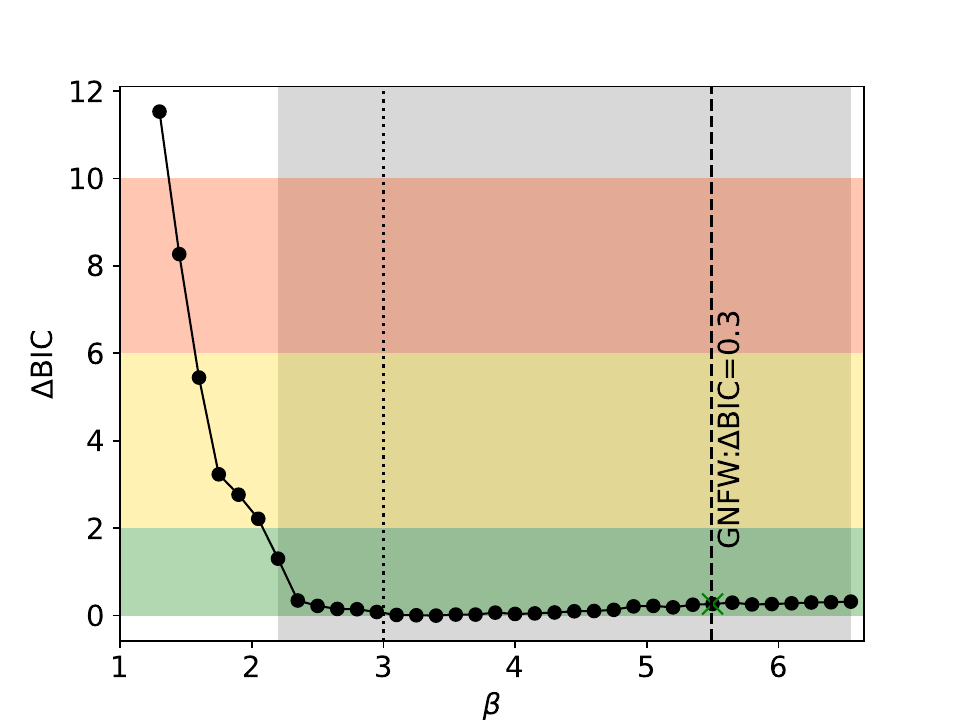}
    \includegraphics[width=0.33\linewidth]{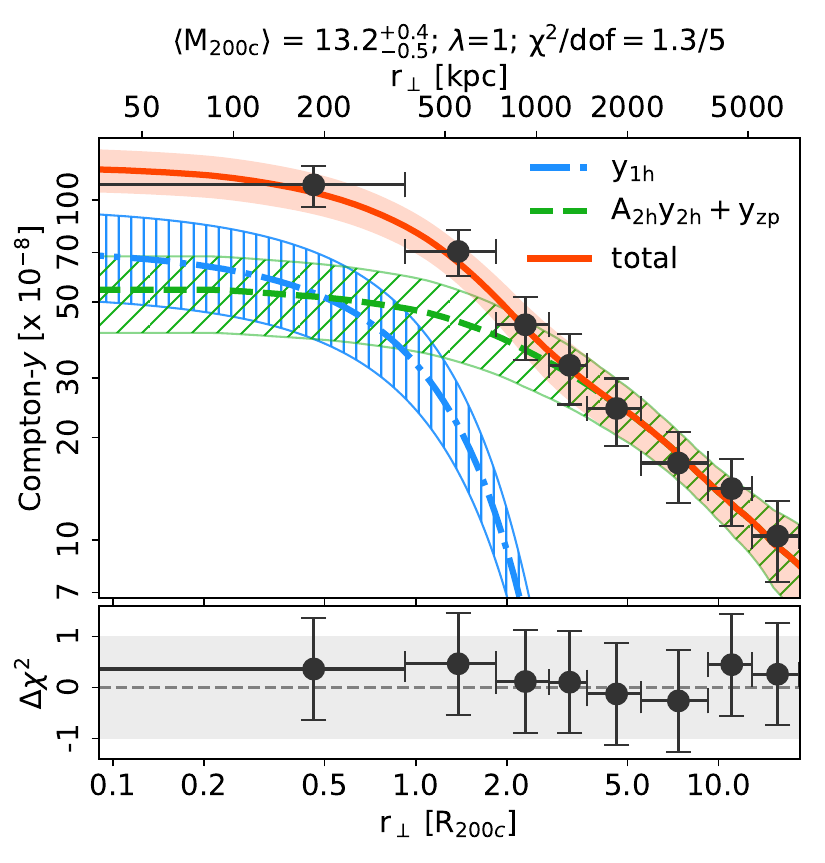}
    \includegraphics[width=0.33\linewidth]{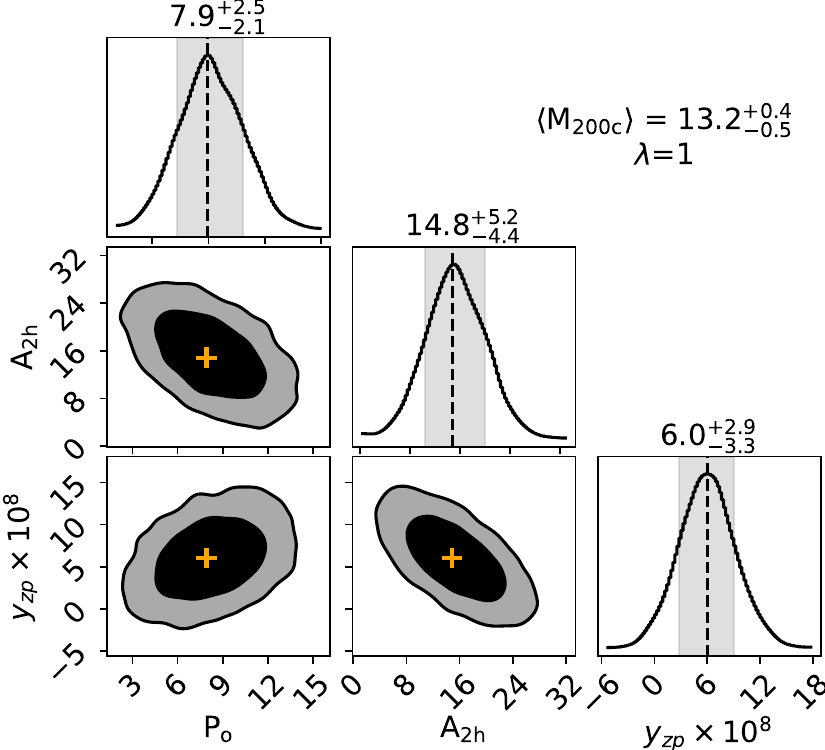}
    \caption{Results of stacking, aperture photometry, and pressure modeling for one of the mass bins of isolated galaxies, $\lambda=1$. Results for all other mass and richness bins are shown in Figure~\ref{fig:others}. \textbf{Left:} Fit statistic at different values of $\beta$. The green, yellow, and red horizontal patches denote regions of \textit{negligible evidence}, \textit{moderate evidence}, and \textit{strong evidence} of deviation from the maximum likelihood, respectively. The vertical dotted line marks the limit below which the thermal energy would not converge to a finite value at a large impact parameter. The vertical gray patch corresponds to the range of $\beta$ that is \textit{statistically indistinguishable} from the $\beta$ with maximum likelihood. The fiducial $\beta$ of GNFW is labeled with a vertical dashed line. In this case, a deviation from GNFW is \textit{not required} ($\Delta \rm BIC<$2); therefore, the plots in the right two panels are shown for the fiducial $\beta$. \textbf{Middle:} Projected radial profile of Compton-$y$, with the top and bottom x-axes showing the projected physical radius (kpc) and virial radius (\rvirialn). The width of each annular aperture is shown with the error bars along the x-axis. The error bars along the y-axis are derived from the diagonal terms of the covariance matrix \textbf{C$_{\rm \bf stacked}$}. These error bars are shown only for illustration; the full \textbf{C$_{\rm \bf stacked}$} is considered in the analyses. The dash-dotted blue curve and the dashed green curve are the best-fit ``1-halo" term, $y_{1h}$, and tSZ background (${\rm A_{2h}}y_{2h} + y_{zp}$), respectively, with the hatched area around them corresponding to the 1$\sigma$ uncertainty. The solid red curve and the shaded area around it are the best-fit models with 1$\sigma$ uncertainty. \textbf{Right:} Posterior probability distributions of the amplitude of the thermal pressure profile: $\rm P_o$, normalization of the 2-halo term: $\rm A_{2h}$, and zero-point offset: $y_{zp}$. The contour plots, drawn from \textbf{C$_{\rm \bf emcee}$}, correspond to 68\% and 95\% confidence intervals, with the most likely values marked with `+'. The vertical dashed lines and the shaded regions in the diagonal plots correspond to the most likely value and 68\% confidence interval of the marginalized distribution of the free parameters; the values are mentioned in the respective titles.}\label{fig:main}
\end{figure*}

Assuming the likelihood (e$^{-\chi^2/2}$; Equation~\ref{eq:chisq}) to be Gaussian, we estimate the posterior probability distribution functions of the free parameters using the ensemble sampler in \texttt{emcee} \citep{Foreman-Mackey2013}. We run multiple \texttt{emcee} ensembles with 32 walkers and 5000 independent sets of chains. This time, we change the default Affine-Invariant ensemble method of ``stretch move" to a mix of 25\% DE (differential evolution) move and 75\% DESnooker move. This significantly improves the sampling of the highly
\begin{equation}\label{eq:chisq}
\begin{split}
    \rm  \chi^2 =  [\overrightarrow{\rm Data} - \overrightarrow{\rm Model}(\mathbf{p})]^T   \mathbf{C_{\rm \bf stacked}}^{-1}[\overrightarrow{\rm Data} - \overrightarrow{\rm Model}(\mathbf{p})] \\
        \overrightarrow{\rm Model} = (y_{1h}+ A_{2h}y_{2h}+y_{zp}), 
        {\rm and}\; \mathbf{p} = [P_o, A_{2h}, y_{zp}]
\end{split}
\end{equation}
correlated parameter space and converges faster by reducing the burn-in phase and increasing the acceptance fraction of walkers\footnote{For a walker $i$ and a complementary set of walkers $j$, a new coordinate of the walker is $X_{i+1} = X_i + \sqrt{1/f}(X_j-X_i)$ in the stretch move, where the scale factor $f$ is a random variable drawn from a uniform distribution of $\mathcal{U}$[1/2,2]. DE move \citep{Nelson2014} determines the step size parameter ($\epsilon$), step size factor ($\gamma_o$), and direction of walkers by taking the difference between two random chains in the complementary set: $X_{i+1} = X_i + \gamma_o(X_{j1}-X_{j2}) + \epsilon$. By default, $\gamma_o$ is $2.38/\sqrt{2d}$ and $\epsilon$ is drawn from a $d$-dimensional Gaussian distribution of $\mathcal{N}_d(0,\sigma)$, where $d$ is the number of free parameters, and $\sigma$ is 10$^{-5}$. DESnooker move \citep{ter2008} is a more advanced variant of DE move: $X_{i+1} = X_i + \gamma(X'_{j1}-X'_{j2}) + \epsilon$, designed for better exploration by ``snookering" towards the current best-known position, with $\gamma$ being drawn from a uniform distribution of $\mathcal{U}$[1.2,2.2]. Here, three random chains ($X_{j1}, X_{j2}, X_{j3}$) are drawn from the complementary set, and the orthogonal projection of the first two chains ($X_{j1}, X_{j2}$) onto the difference between the third chain and $X_i$, i,e., ($X_{j3}-X_i$), yields $X'_{j1}$ and $X'_{j2}$}. We ensure that the total number of chains is larger than 50 times the integrated autocorrelation time, $\tau$, of each free parameter, and the Gelman-Rubin convergence parameter is within 0.9--1.1. We remove the first 2$\tau_{max}$ steps from the sampler to remove the burn-in phase. We consider the most likely values of the free parameters, i.e., values corresponding to the minimum $\chi^2$ (equation~\ref{eq:chisq}), and the covariance matrices of their posterior probability distributions, \textbf{C$_{\rm \bf emcee}$}, for the following calculations. 

\subsection{Thermal energy and temperature floor}\label{sec:physical}

For each mass and richness bin, we calculate $\rm \tilde Y^{sph}_{R200}$, the redshift-normalized thermal energy within a spherical volume of radius R$_{200c}$ from the best-fit pressure profiles (equation~\ref{eq:Y}). 

\begin{equation}\label{eq:Y}
\begin{split}
        {\rm \tilde Y^{sph}_{R200} =}(\sigma_{\rm T}/m_e c^2)E(z)^{-2/3} \\ \int_{0}^{\rm R_{200c}} \Big(\frac{3\mu_e}{2\mu}\Big){\rm P_e}(r) ~4\pi r^2 dr /(500~\rm Mpc)^2 
\end{split}
\end{equation}     

Expecting \mvirial and $\rm \tilde Y^{sph}_{R200}$ to follow a power-law relation (Equation~\ref{eq:YM}), we employ a hierarchical Bayesian model \citep[\href{https://github.com/jmeyers314/linmix}{linmix};][]{Kelly2007} for fitting a straight line to ln($\rm M_{200c}$) and ln($\rm \tilde Y^{sph}_{R200}$). 

\begin{equation}\label{eq:YM}
\begin{split}
        {\rm \tilde Y^{sph}_{R200} = \tilde Y_{0}(M_{200c}/M_{0})^\Gamma}, \\
        {\rm i.e., ln(\tilde Y^{sph}_{R200})} = a_0 + a_1 \rm ln(M_{200c})
\\
{\rm where}\; a_0 = {\rm ln(\tilde Y_{0})-\Gamma ln(M_{0})};\;{\rm and}~ a_1 = \Gamma
\end{split}
\end{equation}

We assume that $\rm M_{200c}$ and $\rm \tilde Y^{sph}_{R200}$ are drawn from a 2-d log-normal distribution $\mathcal{N}_2(\mu,\mathbf C_{YM})$. Here, the mean $\mu = (\mu_M,\mu_Y)$ represent the unobserved true  values of ln($\rm M_{200c}$) and ln($\rm \tilde Y^{sph}_{R200}$), and the covariance matrix $\mathbf C_{YM}$ contains the measured errors of ln($\rm M_{200c}$) and ln($\rm \tilde Y^{sph}_{R200}$). $\mu_M$ and $\mu_Y$ are connected through ${\rm \mathcal{P}(\mu_Y|\mu_M)} = \mathcal{N}(a_0 + a_1 \mu_M, \sigma_Y^2)$, where the regression parameters $a_0, a_1$, and $\rm \sigma_Y^2$ denote the intercept, slope, and Gaussian intrinsic variance of $\mu_Y$ around the regression line. We set uniform priors on $a_0, a_1$, and $\rm \sigma_Y^2$. 

\begin{figure*}
\gridline{\fig{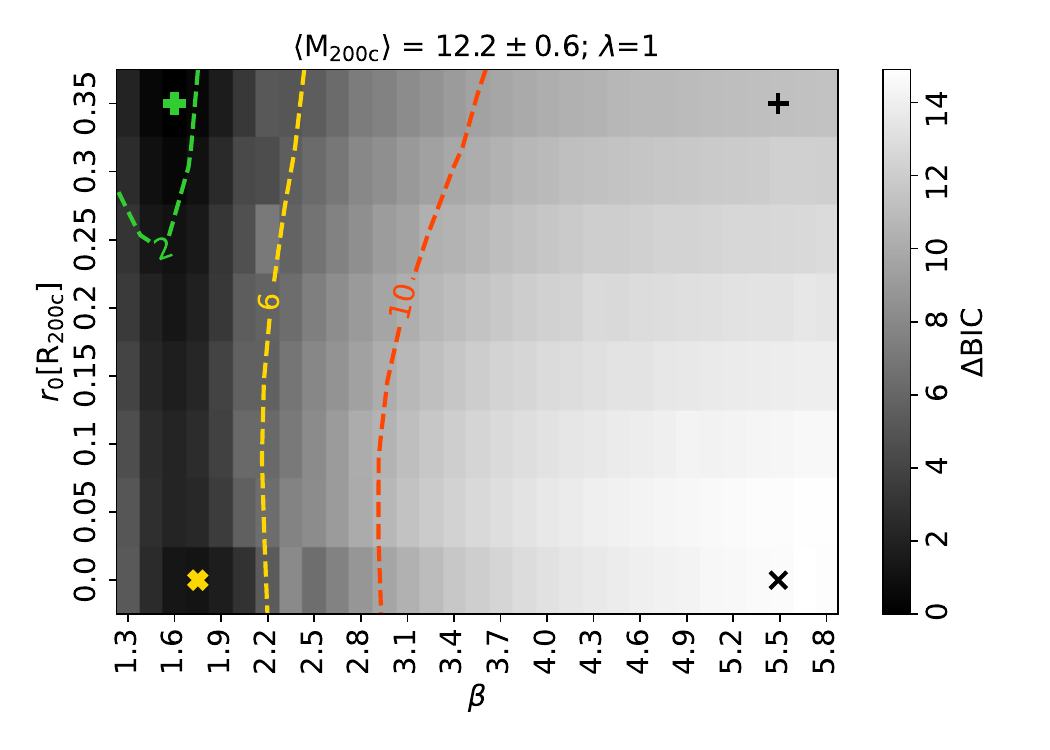}{0.25\linewidth}{(a)}
          \fig{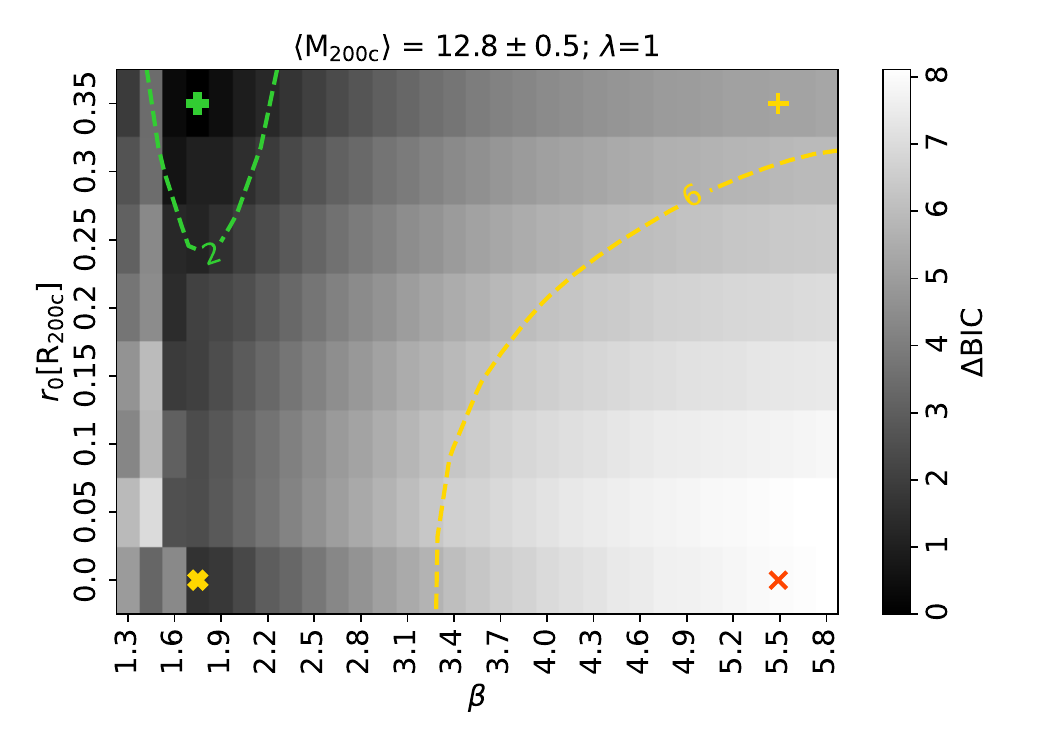}{0.25\linewidth}{(b)}
          \fig{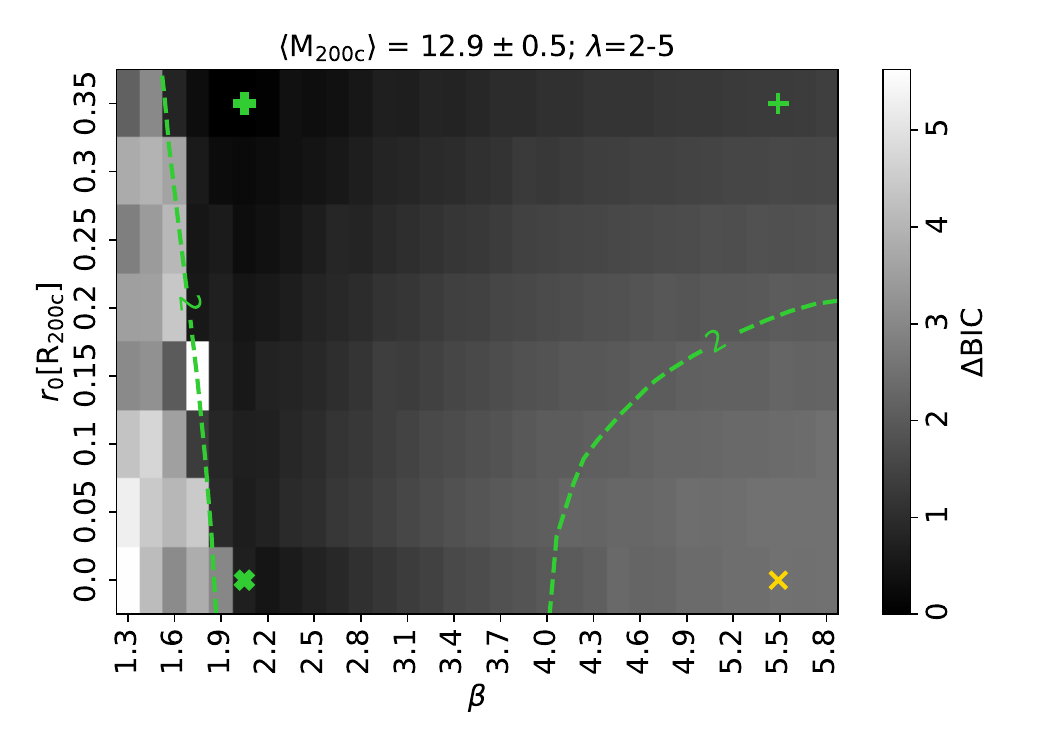}{0.25\linewidth}{(c)}
          \fig{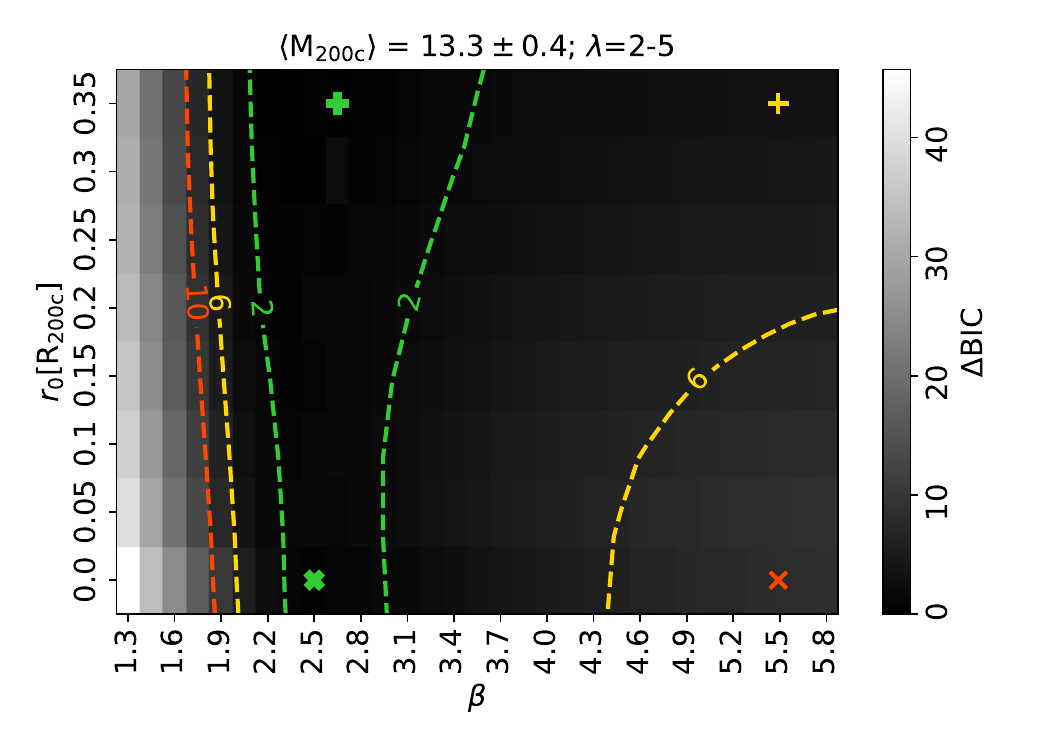}{0.25\linewidth}{(d)}}
\gridline{\fig{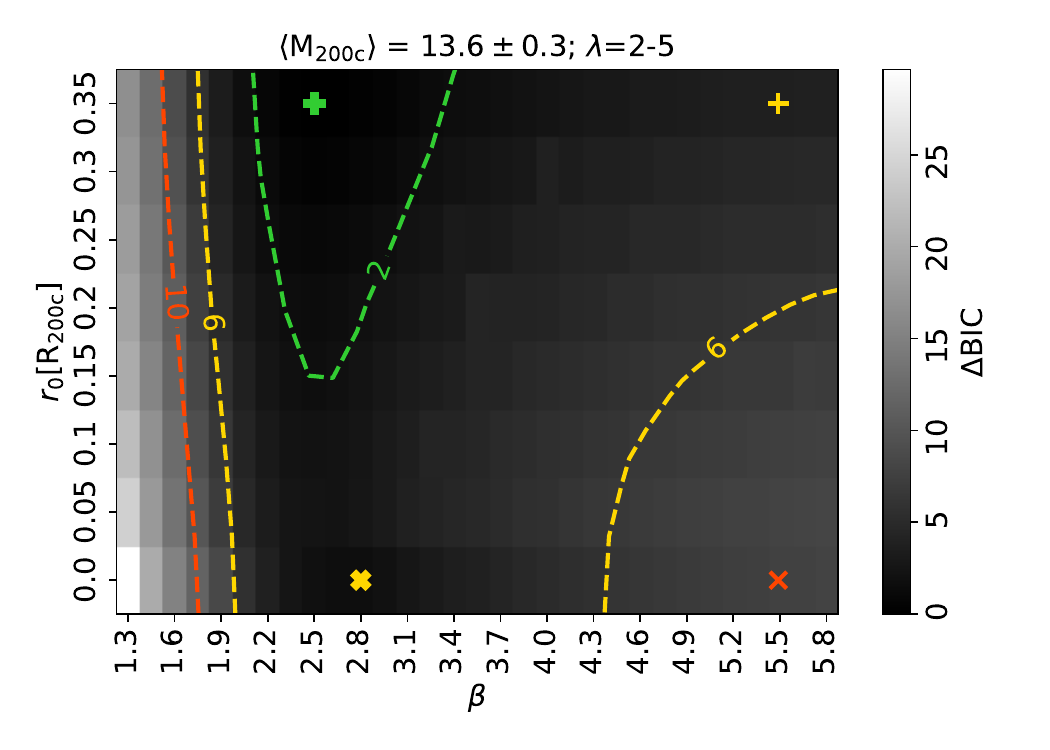}{0.25\linewidth}{(e)}
          \fig{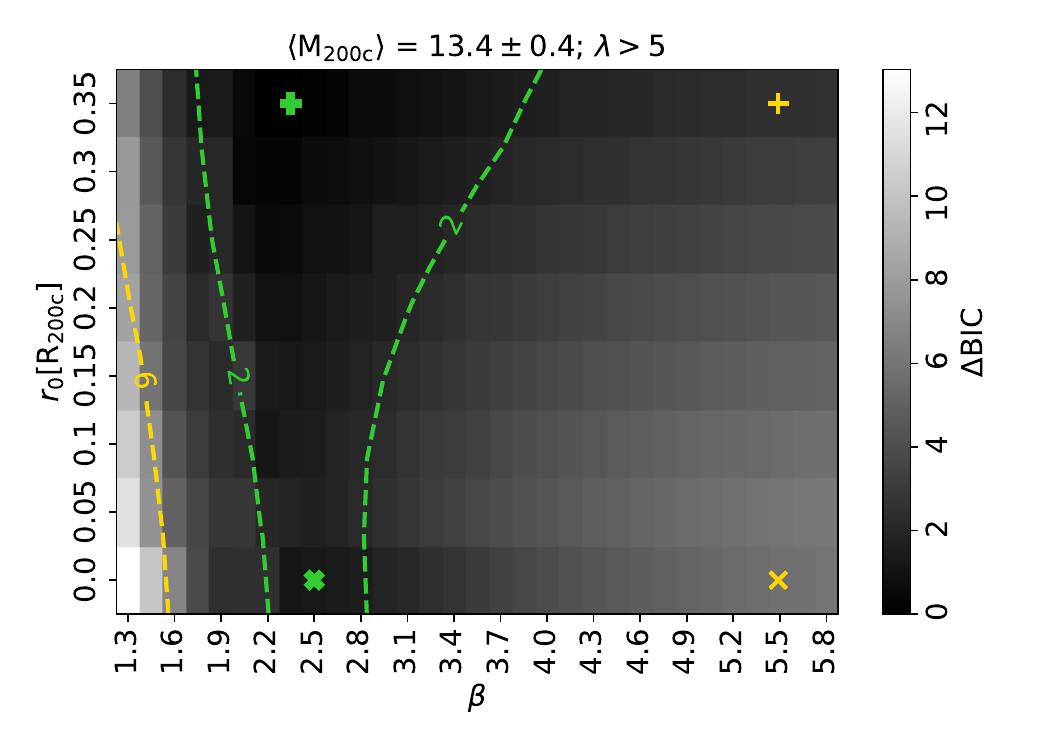}{0.25\linewidth}{(f)}
          \fig{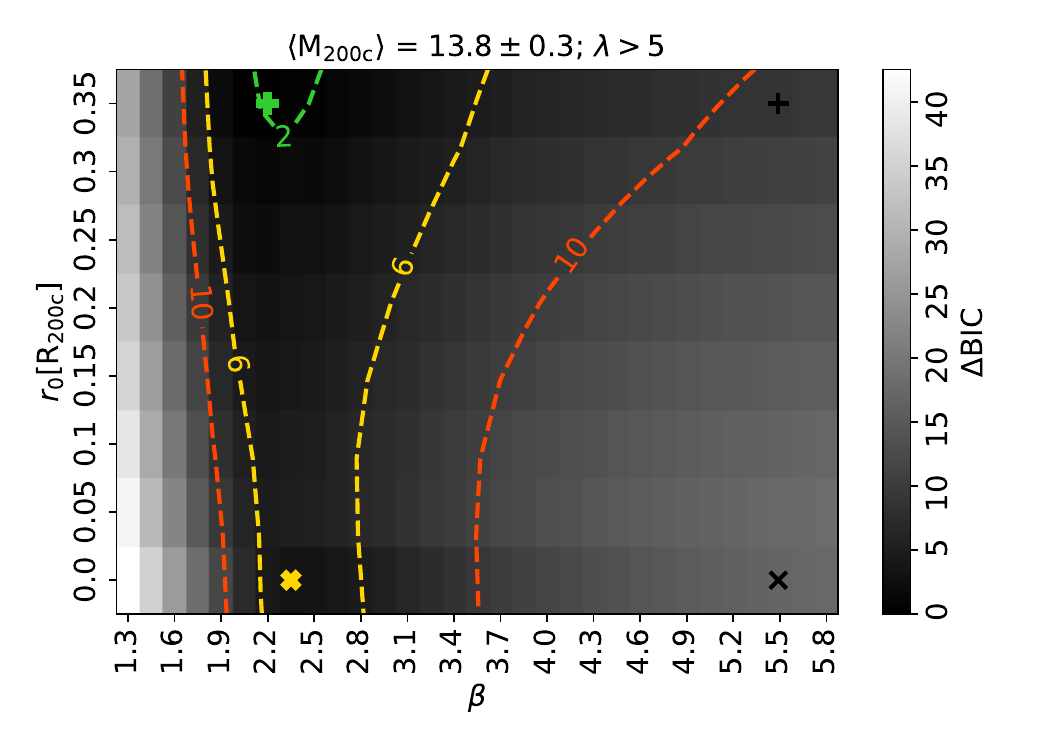}{0.25\linewidth}{(g)}
          \fig{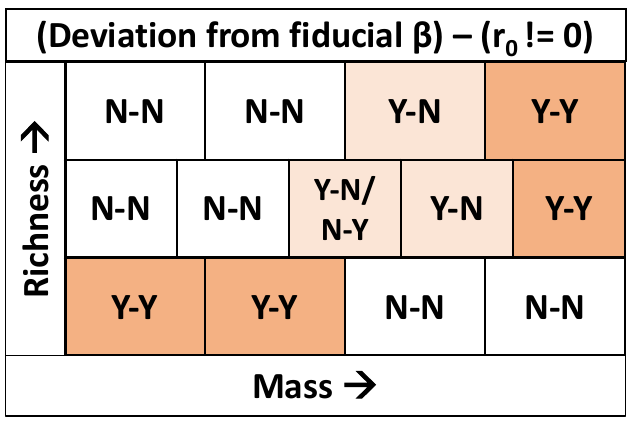}{0.25\linewidth}{(h)}}
\caption{\textbf{Panels a to g:} Fit statistics of the variable centroid ($r_o$) and slope ($\beta$) for the mass and richness bins where a flatter-than-GNFW $\beta$ \textit{is required} with a fixed centroid ($r_o$=0). Regions within the green contours are \textit{statistically indistinguishable} from the best-fit. Regions between the green-and-yellow/yellow-and-red contours have \textit{moderate/strong evidence} of deviation from the best-fit. Regions outside the red contours are \textit{ruled out}. The best-fit and the fiducial $\beta$ at the best-fitted $r_o$ are shown with the thick and thin `+' signs, respectively; the corresponding parameters at $r_o$=0 are shown with `$\times$' signs.  \textbf{Bottom right:} The summary of whether a deviation from the GNFW slope and/or miscentering \textit{is required} by the data for different mass and richness bins. Y/N implies yes/no.}
\label{fig:betax0}
\end{figure*}

The baryon fraction, $f_b$, of halos cannot exceed the cosmological fraction. Thus, for the maximum possible baryon content, we can obtain the \textit{temperature floor}, \Tfloorn, (equation~\ref{eq:TCGM}), i.e., the minimum volume-average temperature allowed by the measured thermal pressure. Without the measurement of ionized baryon density, this provides us with a zeroth-order idea of the thermodynamic state. 

\begin{subequations}
\begin{equation}\label{eq:TCGM}
   \frac{\rm \langle T_{gas} \rangle  _{min}}{\rm  T_{200c}} = \frac{\rm M_{gas}}{f_{b,cosmo}\rm M_{200c}} \equiv \frac{f_b}{f_{b,cosmo}}
\end{equation}
\begin{equation}\label{eq:mass}
   {\rm where},\;\; {\rm M_{gas}} = \int_{0}^{\rm R_{200c}} \mu_e m_p n_e(r) 4\pi r^2 dr E(z)^{-1}
\end{equation}
\begin{equation}\label{eq:nTvir}
    n_e(r)  = {\rm P_e}(r) \rm k_B^{-1} T_{gas}^{-1} \;
   {\rm  and, T_{200c}} = \frac{\mu m_p \rm G M_{200c}}{\rm 2k_B R_{200c}} 
\end{equation}
\end{subequations}

Here, $m_p$, $n_e$, and $\rm k_B$ are the rest mass of the proton, the number density of electrons, and Boltzmann's constant. 

\section{Results and discussion}\label{sec:result}

\subsection{Slope of centered profiles}\label{sec:beta}

First, we discuss the results of the best-fitted pressure profiles when the center, $r_o$, is frozen at zero, shown in Figure~\ref{fig:main}. Unlike the three main free parameters, the probability distribution of $\beta$ is far from a Gaussian in all mass and richness bins. It is asymmetric, skewed to larger $\beta$. This is reasonable because above a certain value of $\beta$, the pressure profile beyond the core radius, $r_c$, barely contributes anything; it is primarily the flatter part within $r_c$ that constrains the data, reducing the sensitivity of the model to $\beta$. Thus, getting the constraint on the upper end of $\beta$ is more challenging when the true best-fit is close to the fiducial value. Also, as suspected before (\S\ref{sec:model}), we often notice non-monotonic variation of likelihood with the value of $\beta$, and in some cases, the likelihood is almost uniform across the entire range of $\beta$. Naturally, freeing $\beta$ in these cases could lead to spurious results with weaker constraints on the other free parameters. 

Out of thirteen mass and richness bins, the data \textit{statistically require} a deviation from the fiducial $\beta$ in seven cases ({shown in Figure~\ref{fig:others}}, {but the one in Figure~\ref{fig:main} does not require a deviation}). However, the \textit{requirement} is not monotonic with mass. The CGM of isolated galaxies prefers a flatter pressure profile at lower mass, but is consistent with the fiducial GNFW profile at higher masses. In the IGrM across richness, this \textit{requirement} flips: low-mass halos follow the fiducial GNFW profile, but the higher-mass halos reveal a flatter pressure profile. Thus, the shape of the pressure profile is mass- as well as richness-dependent.  

\subsection{Miscentering}\label{sec:miscentering}

In the subset of cases where a deviation from the fiducial $\beta$ is \textit{statistically required} by the data, we test if the deviation is physical or an artifact of miscentering, i.e., whether $r_o$ differs from zero, and show it in Figure~\ref{fig:betax0}. The likelihood is not entirely monotonic in both $\beta$ and $r_o$, emphasizing why a fit-by-part of three+two parameters in this case worked more efficiently than a simultaneous five-parameter fit. The results vary widely across mass and richness bins, so we discuss each of them separately. 

\paragraph{Isolated, low mass (panel a) and high richness, high mass (panel g)} These two are the most well-constrained mass and richness bins with the least degeneracy. Here, the fiducial $\beta$ is \textit{ruled out} across $r_o$. The data show \textit{moderate evidence} even against the best-fit $\beta$ at $r_o=0$ considered in the previous section. It emphasizes why we must explore the miscentering in conjunction with a variable $\beta$. Overall, the thermal pressure of the ionized CGM/IGrM of these galaxies/groups is miscentered at $r_o>0.25$\rvirial and the profile shape is flatter than the default expectation. The mass of high-richness groups is well explored in the literature in tSZ (and kSZ), but we are not aware of any investigation of miscentering, let alone the simultaneous exploration of $\beta$ and $r_o$; this is a new result.    

\paragraph{Isolated, high mass (panel b) and Low richness, high mass (panel e)} Here, the profile with fiducial $\beta$ at $r_o = 0$ has \textit{strong evidence} to be disfavored. The miscentered profile at fiducial $\beta$ is \textit{statistically indistinguishable} from the best-fitted flatter profile at $r_o=0$ considered in \S\ref{sec:beta}, and both are disfavored at \textit{moderate evidence}. It establishes that simultaneous variation of $\beta$ and $r_o$ is indeed essential; otherwise, a miscentered profile could be misinterpreted as a flatter profile, leading to incorrect physical inference. Overall, the best-fitted thermal pressure profiles of the ionized CGM/IGrM of these galaxies/groups are flatter-than-fiducial and miscentered at $r_o>0.15$\rvirialn. 

\paragraph{Low richness, low mass (panel c)} Here, the profile with fiducial $\beta$ at $r_o = 0$ has \textit{moderate evidence} to be disfavored. The miscentered best-fit is \textit{statistically indistinguishable} from the best-fit at $r_o=0$ considered in \S\ref{sec:beta}, as well as the miscentered profile at fiducial $\beta$. In fact, the upper end of $\beta$ cannot be constrained when the pressure profile is allowed to be miscentered. Thus, we cannot conclude whether the pressure profile is flatter or miscentered or both, but we can conclude that it is NOT centered at fiducial $\beta$. 

\paragraph{Low richness, intermediate mass (panel d)} Here, the profile with fiducial $\beta$ at $r_o = 0$ has \textit{strong evidence} to be disfavored. In fact, the miscentered profile with fiducial $\beta$ has \textit{moderate evidence} to be disfavored, implying the profile is truly flat and not an artifact of miscentering. The best-fit in \S\ref{sec:beta} is \textit{statistically indistinguishable} from the miscentered best-fit here; therefore, we retain the previous best-fit. Thus, this mass bin does not require a miscentered pressure profile.  

\paragraph{High richness, low mass (panel f)} Here, the fiducial $\beta$ across $r_o$ is disfavored with \textit{moderate evidence}. We retain the best-fit at $r_o=0$ obtained in \S\ref{sec:beta} as it is \textit{statistically indistinguishable} from the miscentered best-fit. Thus, this mass bin does not require a miscentered pressure profile but does require a flatter-than-fiducial shape.   

\begin{figure*}
\centering
\includegraphics[trim=55 267 90 50,clip,width=0.99\linewidth]{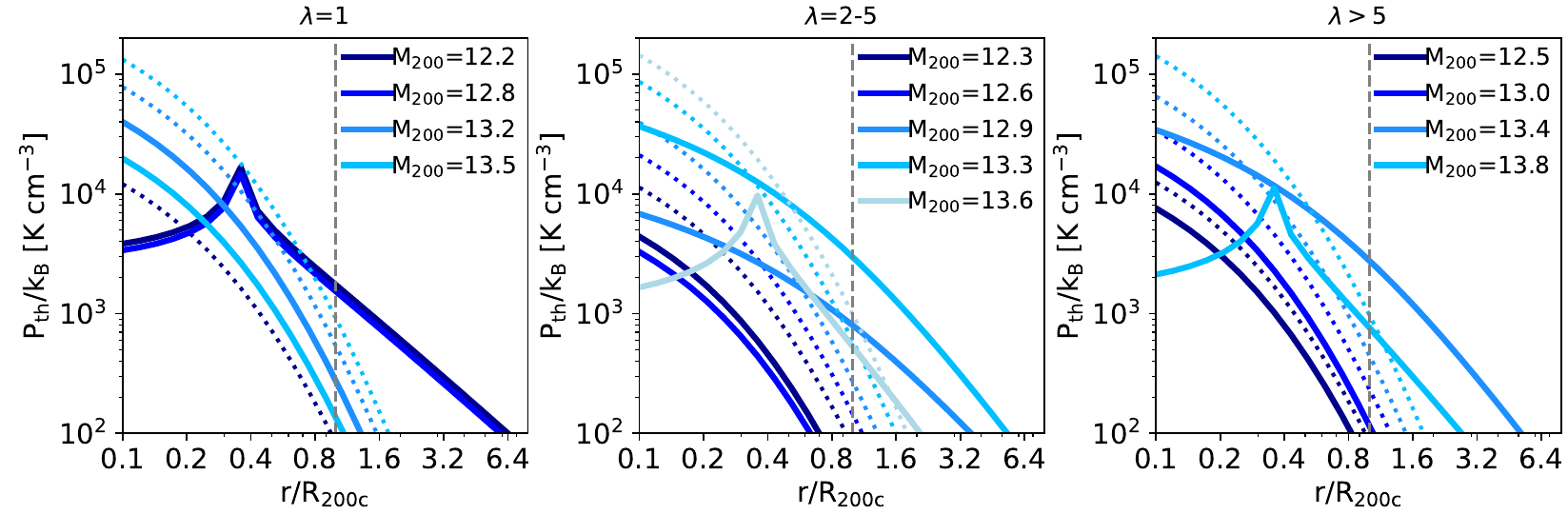}
\includegraphics[trim = 0 15 0 0,clip,width=0.975\linewidth]{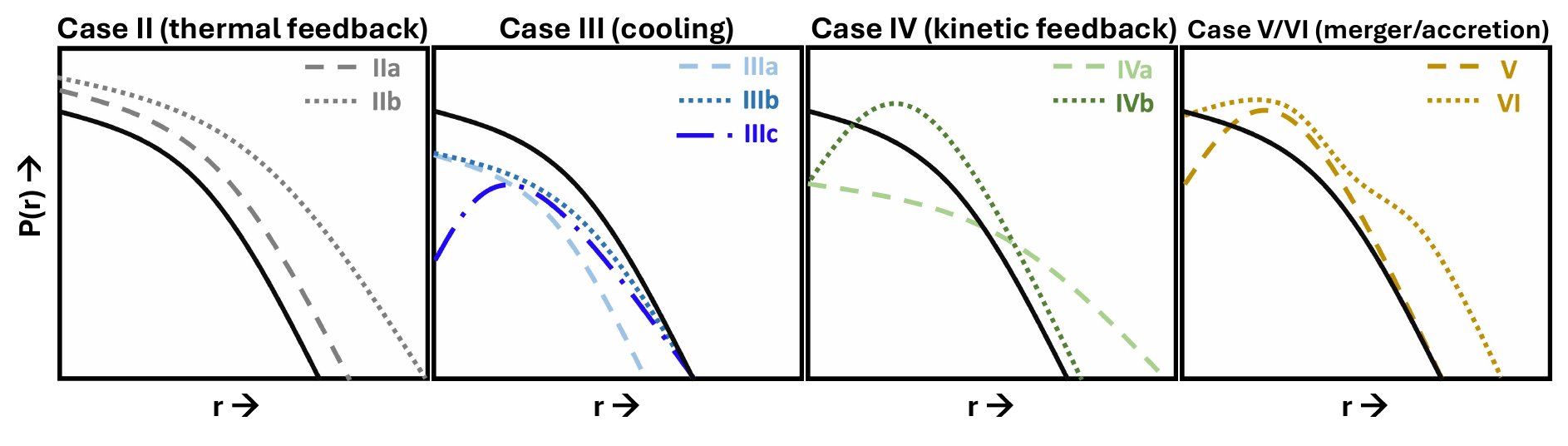}
\caption{\textbf{Top:} Radial profile of the best-fitted thermal pressure for all the mass and richness ($\lambda$) bins (solid curves). The dotted curves of similar colors are drawn for the self-similar solution. The vertical gray dashed line is drawn at \rvirialn. The pressure profiles show a range of amplitude, shape, and centroid across richness at similar masses, and across mass at similar richness, justifying the necessity to explore the mass dependence of the tSZ effect as a function of richness. \textbf{Bottom:} A cartoon of possible pressure profiles for different physical processes affecting their amplitude, shape, and centroid. Case I (gravity-driven regime) is shown with the black solid line in all plots.}
\label{fig:P}
\end{figure*}

With the revised best-fits, we summarize the results for the shape and centroid of the pressure profile in terms of $\beta$ and $r_o$ in the bottom right panel of Figure~\ref{fig:betax0}. The first letter denotes whether a deviation from the fiducial $\beta$ is \textit{required}. The second letter denotes whether the data \textit{require} a miscentered pressure profile. Clearly, at similar mass, the shape and centroid of the pressure profile of all richnesses are not the same, and how it differs across richness also changes across mass. Thus, the pressure profiles depend on mass as well as richness in a complex way; these are plotted in Figure~\ref{fig:P} (top).  
 
\subsection{Pressure profiles}\label{sec:pressure}

Before moving on to the integrated properties (thermal energy and the \textit{temperature floor}), we interpret the best-fit radial profile of thermal pressure in all mass and richness bins (Figure~\ref{fig:P}, top). In principle, some of the miscentering could be systematic artifacts in optical identification, e.g., stray light from bright stars in the close vicinity of our targets, insufficient deblending of individual galaxy members, and misidentification of the central galaxy \citep[e.g.,][]{Ding2025}. However, these issues are more prevalent in galaxy clusters with high richness that are individually detected in tSZ; these are already excluded from our analysis. Assessing whether the systematics in individual halos of lower mass and richness cancel each other or pile up in stacking is nontrivial and beyond the scope of this paper. In the following, we discuss our results assuming the effect is entirely physical. 

We start by considering possible processes that can generally affect the default amplitude, shape, and centroid, i.e., $P_o, \beta, r_o$, of the pressure profile. We should be cognizant of several aspects here. First, tSZ probes thermal pressure, i.e., the product of density and temperature of ionized baryons. Physical mechanisms influencing the density and temperature are not necessarily the same. Secondly, we have ignored any angular information in the 2-D distribution of density or temperature in the sky plane throughout the analysis. As the azimuthal asymmetry, if any, is expected to be randomly distributed across targets, the effect on the stacked radial profile could be manifested as a disturbed shape and/or centroid. The true centers of the 2-D distribution of temperature and density may not be aligned in the first place. Thus, the derived center of the thermal pressure may not represent the true center of either temperature or density. Thirdly, most of the following discussions are on one global effect by one component. In reality, most physical processes have multiple effects, and there are multiple physical processes at play. We make an effort to disentangle these effects as much as possible through six cases (Figure~\ref{fig:P}, bottom). 
\\

\noindent
\textbf{Case-I: Gravity-dominated regime} This is the default or ideal scenario. Here, heating and cooling might balance each other, or be too weak to dominate over gravity in the first place. The observable effect of merger or baryon stripping might not be significant either. 
\\
\textbf{Case-II: Thermal feedback.} \textbf{(a)} Thermal feedback of low mass-loading factor increases the amplitude without observationally significant changes in the shape or centroid. \textbf{(b)} Thermal feedback with a high mass-loading factor can increase the amplitude as well as flatten the shape. In both cases, the thermal pressure will have an overall excess.  
\\
\textbf{Case-III: Cooling.} \textbf{(a)} In the absence of dominant feedback, halos can globally cool for the right combination of metallicity, density, and temperature. Here, the amplitude reduces without noticeably affecting the shape. \textbf{(b)} Cooling within a critical radius can be more efficient where the baryons are accreted/recycled from past outflows to the disk, fueling star formation. This reduces the amplitude, leading to an apparently flattened shape. \textbf{(c)} In a substantial cooling scenario in the inner halo, there might be a turnover in the thermal pressure profile, leading to a shifted centroid in addition to reduced amplitude and a flattened shape. Here, the inner halo might have non-negligible pressure support from non-thermal sources. In all of these cases, there will be an overall deficit in the thermal pressure. 
\\
\textbf{Case-IV: Kinetic feedback.} \textbf{(a)} Strong kinetic feedback can shovel baryons from small to large radii. Here, the thermal pressure will be reorganized by flattening the shape and reducing the amplitude. \textbf{(b)} Kinetic feedback of moderate-strength snowplows the preexisting CGM/IGrM to create a cavity surrounded by a shock-compressed shell. This shifts the centroid of the pressure profile and may affect the shape.  
\\
\textbf{Case-V: Mergers.} If the halos are going through mergers or are not back to a dynamically relaxed state post-merger, the center of ionized baryons can be shifted \citep[e.g.,][]{George2012}. Here, the thermal pressure will be repositioned without an obvious change in the amplitude or shape. 
\\
\textbf{Case-VI: Accretion.} A non-negligible amount of baryons could be stripped from member galaxies through ram-pressure/tidal interaction \citep[e.g.,][]{Rhee2026}. This would cause an apparent excess of ionized baryons that the fiducial \twohalo cannot account for. This will affect the shape and, in extreme situations, the centroid of the pressure profile.      

\subsubsection{Identifying the physical cases} 

Below, we identify the dominant case as a function of mass for three richness bins. 

\paragraph{CGM of isolated galaxies (Figure~\ref{fig:P}, top left)} Here, as the mass decreases, the amplitude of the pressure profile, $P_o$, increases, the shape gets flatter, i.e., $\beta$ decreases, and miscentering becomes more prominent, i.e., $r_o$ increases. Thus, lower-mass galaxies are a combination of Case-IIb and Case-IVb, while higher-mass galaxies are consistent with Case-IIIa, implying dominant heating and cooling at lower and higher masses, respectively. 

\paragraph{IGrM of low richness groups (Figure~\ref{fig:P}, top middle)} Lower two mass bins are consistent with case-IIIa. In the intermediate mass bin, two different pressure profiles representing case-IIIb+VI and case IVb+V are \textit{statistically indistinguishable} (see \S\ref{sec:miscentering}). The second highest mass bin is consistent with case-IIb+VI. The thermal pressure profile of the highest mass bin has reduced amplitude, a flatter shape, and is miscentered. Case-IIIc+V is consistent with this. Thus, there is evidence of cooling, merger/accretion, and thermal feedback in different mass bins, but without any trend with mass in the driving astrophysical process in this richness bin. 

\paragraph{IGrM of high richness groups (Figure~\ref{fig:P}, top right)} Lower two mass bins are consistent with case-IIIa. Same as the low richness groups, the second-highest and the highest mass bins are consistent with case-IIb+VI and case-IIIc+V, respectively. Note that this situation is different from the weak lensing measurements and tSZ/kSZ stacking at the center of galaxies in the literature, where the miscentering is not explored, but the flattening is interpreted as an effect of dominant kinetic feedback. In our targets, the stacking is performed at the optically identified barycenter of the groups, which does not necessarily coincide with a galaxy, and the flattening is \textit{required} even after accounting for miscentering. Also, we explore multiple physical scenarios that could cause flattening of the thermal pressure profile instead of assuming it is an impact of ejective feedback. Thus, our physical interpretation is less ambiguous and more comprehensive. 

Next, we reorganize the above discussion for low vs. high masses and identify the dominant case for different richness.

\paragraph{Low mass (Figure~\ref{fig:P}, top, dark blue curves)} Here, the effect of heating (case-IIb) is prominent at lower richness. At higher richness, the pressure profiles indicate suppressed feedback and undisturbed cooling (case-IIIa), possibly triggered by merger/accretion \citep[see, e.g.,][]{Yang2026}. 

\paragraph{High mass (Figure~\ref{fig:P}, top, light blue curves)} Here, the effect of heating (case-IIb) is sporadically present in some bins. Other bins prefer mild (case-IIIa) or even substantial cooling (case-IIIc). We do not see a clean trend of a dominant physical process with richness, and we provide a plausible explanation. In the isolated and low-richness halos, the richness may be underestimated due to the faintness of stellar light from satellite galaxies, and/or the current galaxy sample could form from past completed merger(s) that coalesced the merging galaxies into one or a few. If true, the prevalence of cooling and the lack of any (thermal or kinetic) feedback in these massive halos could be driven by merger/accretion.   

\begin{figure*}
\centering
\includegraphics[trim=0 0 0 0,clip,width=0.85\linewidth]{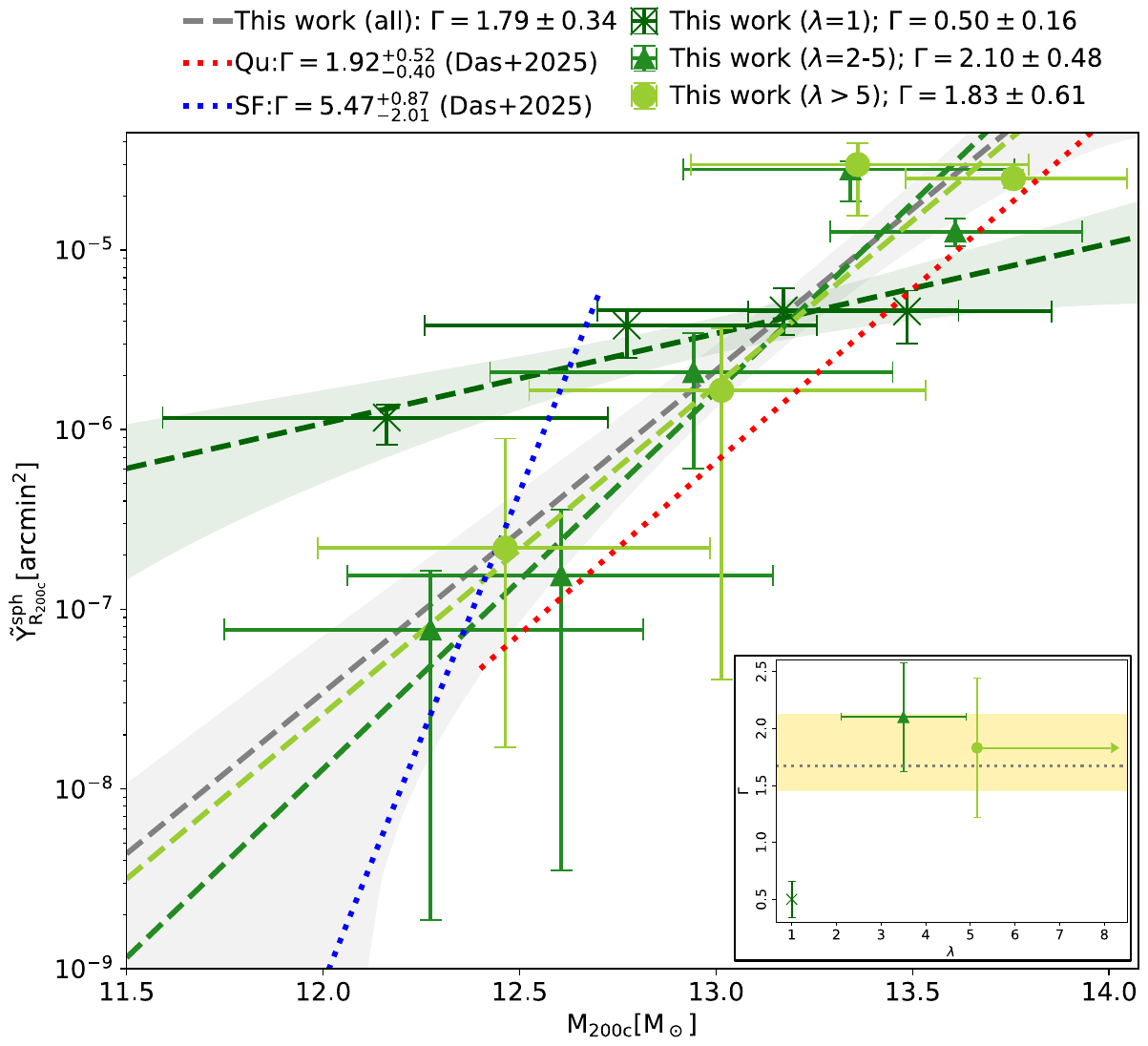}
\caption{Thermal energy within \rvirial as a function of mass. The CGM ($\lambda$=1) and IGrM of low vs high richness are shown in triangles, squares, and circles, with the shade of green becoming lighter with increasing richness. The best-fitted global scaling relation (gray dashed line) and the same for different richnesses (in green) are plotted. The results for quiescent and star-forming galaxies from \citetalias{Das2025} are shown for comparison. In the inset, the best-fitted slope for different richnesses is plotted with the same symbols, with the horizontal yellow patch showing the slope for all samples together. The horizontal dotted line denoting self-similarity is plotted for reference.}
\label{fig:Y}
\end{figure*}

The reason behind the discrepancies between our findings and (most of) the literature is a combination of differences in sample selection, difference/incompleteness in pressure profile modeling, and differences
in physical considerations. In the literature, the true slope of the pressure profile may not be flat but is misinterpreted due to miscentering, or the true slope is flat but not necessarily due to dominant feedback, or the true slope might be flat and due to strong feedback but is representative of different galaxy populations. Overall, the feedback-vs-gravity scenario is not universal; miscentering should be investigated in conjunction with flattening, and halos must be separated based on mass as well as richness to correctly identify the dominant physical situation. 

To assess the robustness of our results and their physical interpretations, we explore the possible effect of residual dust contamination (see Appendix~\ref{sec:cib}). {Under the assumptions that the residual dust contamination and the systematics in the methods of component separation do not depend on sample mass, richness, or scale,} the physical interpretation does not change qualitatively. The quantitative difference in thermal energy and temperature floor {due to different values of dust emissivity} is at most by a factor of $\approx$2. The main source of error in our measurements remains halo-to-halo scatter, followed by the uncertainties in halo mass and photometric redshift; the systematics due to dust are a tertiary effect.

\subsection{Thermal energy}

We plot the thermal energy integrated within \rvirial as a function of mass and richness in Figure~\ref{fig:Y}. At $>$10$^{13}$\msunn, the thermal energy of the IGrM of higher richness groups (circles) is larger than that of the IGrM of lower richness groups and the CGM of isolated galaxies (crosses and triangles), and the overall spread in thermal energy is within an order of magnitude. However, at $<$10$^{13}$\msunn, the scatter in the thermal energy is more than two orders of magnitude, with the CGM of isolated galaxies (crosses) containing more thermal energy than the IGrM of similar-mass halos (triangles and circles). Thus, the thermal energy dominance of halos of different richness flips with mass. This affects the value of $\Gamma$, the slope of the thermal energy-mass relation, discussed below. 

The best-fit global scaling relation (gray dashed line) is steeper than the self-similar relation of $\Gamma$=1.67, but is consistent with the latter within $1\sigma$. Halos of different richness exhibit a distinct thermal energy-mass relation. The CGM of isolated galaxies follows a relation that is $7.3\sigma$ flatter than the self-similar relation, which has never been found or predicted before. The best-fitted $\Gamma$ of the IGrM is flatter than the self-similarity across richness, but it is consistent with the latter within $1\sigma$ error. More precise halo mass and mm data would be essential to precisely constrain the value of $\Gamma$ in the IGrM. 

We compare the results for quiescent and star-forming galaxies (red and blue dotted lines) from \citetalias{Das2025} with this work. The best-fitted slope for the IGrM is qualitatively similar to that of quiescent galaxies, although with a larger normalization (or amplitude). But if we consider the actual measurements instead of the scaling relation, some high-richness groups in low masses, low-richness groups across all masses, and the CGM of isolated galaxies at the highest mass bin are consistent with the quiescent galaxies. Star-forming galaxies, on the other hand, are similar to the groups and host less thermal energy than the CGM of isolated galaxies at similar mass. This can be inferred in two ways. The most massive galaxies (or the central galaxies) in halos of similar richness do not necessarily have similar star-forming properties and hence similar feedback history. In the halos of low richness, there may not always be a prominent central galaxy, but the collective feedback history of all members would still affect the thermal pressure profile. On the other hand, galaxies of similar star-forming activities may not reside in halos of similar richness and may have different accretion/merger histories. This illustrates why a simplified sample selection of luminous infrared galaxies, often considered in the literature, contains multiple astrophysical inhomogeneities that may bias inferences about the pressure profile and thermal energy of the diverse galaxy populations. 

Stacking provides average estimates of the thermal energy. Multiplying it by the sample size, we get the total thermal energy. On average, IGrM contains $\approx7\times$ more thermal energy than the CGM, but the thermal energy content of high-richness groups is similar to the low richness groups and isolated galaxies. But as galaxies outnumber groups, the total thermal energy of galaxies is $\sim$2(18) times larger than that of the low(high) richness groups. Also, galaxies are less massive than groups on average (Figure~\ref{fig:Mz}). This implies that the halos of lower mass and richness are the dominant contributors to the thermal energy budget of the universe at $z\leqslant1$. 

\subsection{Temperature floor}
\begin{figure}
\centering
\includegraphics[trim=30 25 50 50,clip,width=\linewidth]{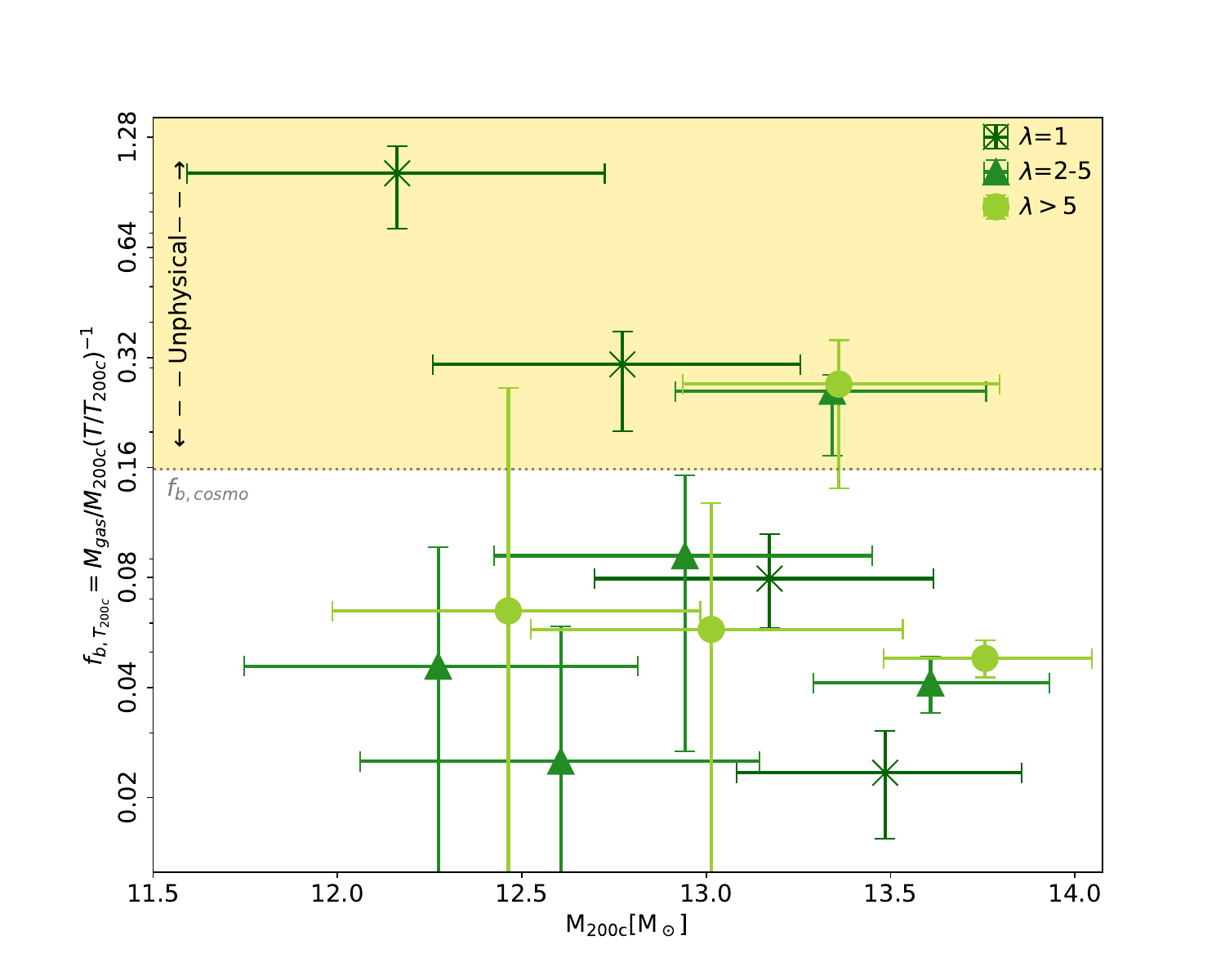}
\includegraphics[trim=30 25 50 50,clip,width=\linewidth]{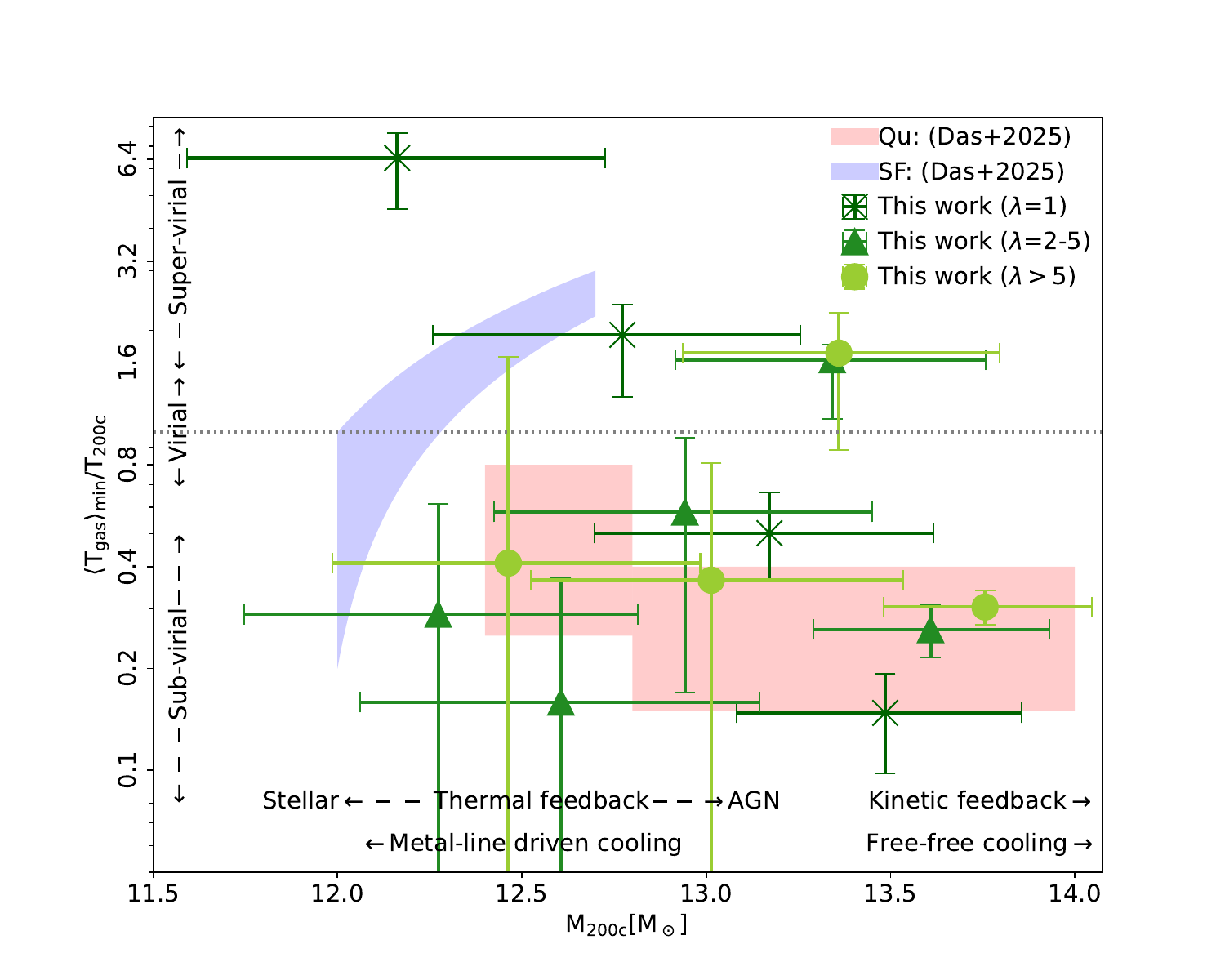}
\caption{\textbf{Top:} Baryon fraction assuming the volume average temperature to be \Tvirialn. $f_{b,cosmo}$ is shown with the gray horizontal dotted line. The presence of three mass bins in the unphysical zone (yellow shaded region) shows that the \Tfloor must be above \Tvirial in these halos. \textbf{Bottom:} \Tfloor as a function of mass for different richness. The results for quiescent and star-forming galaxies from \citetalias{Das2025} are shown for comparison.}
\label{fig:T}
\end{figure}

We plot \Tfloorn, the lowest possible volume-average temperature of the CGM/IGrM for different mass and richness allowed by the data in Figure~\ref{fig:T} (bottom panel). If this is the true temperature of the halos, they would be baryon-sufficient. Below this temperature, the baryon fraction of the halos would exceed the cosmological baryon fraction, which is unphysical. If the halos are hotter, that would imply a smaller baryon fraction, indicating both thermal and kinetic feedback at play, where the former heats the halo, and the latter ejects baryons beyond the virial radius. Thus, even before constraining the actual baryon distribution, this exercise gives us an idea of possible volume-average temperatures and densities, which will inform other observational probes of the ionized CGM/IGrM. 

To further justify the importance of \Tfloorn, we show the $f_b$ (calculated for the volume-average temperature, \Tavgn, to be at the virial temperature, \Tvirialn, in the top panel of Figure~\ref{fig:T}. $f_b$ is calculated using the ionized halo gas we probe using tSZ; the stellar mass of the galaxies in these halos is not included; the true $f_b$ will be larger once that is included. As a note of caution, $f_b$ values that are smaller than $f_{b,cosmo}$ could deceivingly indicate strong ejective feedback, but as we have discussed in \S\ref{sec:pressure}, the best-fitted pressure profiles in these halos indicate cooling rather than ejective feedback. Thus, the actual \Tavg of these halos is likely smaller than \Tvirialn, and the actual $f_b$ is larger than the shown values. Most strikingly, in three mass bins, $f_b$ exceeds $f_{b,cosmo}$, which is unphysical. Thus, in these halos, \Tavg must be above \Tvirialn, and attain a critical value where $f_b$ is at least comparable to $f_{b,cosmo}$. That critical value is the \textit{temperature floor}. This justifies why we should consider \Tfloor in conjunction with $f_b$ to gain a holistic insight into the nature of the ionized CGM/IGrM.   

Here, we discuss \Tfloor in high vs. low masses. Above $10^{13.5}$\msunn, we find that the halos are ``cooled" at sub-virial temperatures, but due to their large \Tvirialn, the \Tfloor is still $\gtrapprox10^6$~K, making the CGM/IGrM hot on an absolute level. Below $10^{13.5}$\msunn, the halos are scattered evenly around their \Tvirialn. Thus, here, shock-heating during gravitational collapse of these halos remains the primary determinant of the thermodynamic state, while feedback-induced heating to prevent runaway cooling plays a secondary role. In three cases with unphysically large $f_b$ for \Tavg = \Tvirialn, \Tfloor is super-virial ($\approx$2--7~\Tvirialn). Here, the thermal feedback heats the halo while maintaining baryon sufficiency. 

Comparing with and extrapolating from the trends for star-forming and quiescent galaxies in the same mass range \citepalias{Das2025}, we deduce the following. Halos with their \Tfloor around or above their respective \Tvirialn, if isolated, contain a star-forming galaxy, and if a galaxy group, are influenced by their star-forming galaxy member(s). On the other hand, halos with sub-virial \Tfloor are likely quiescent. 

We find that the mass dependence of \Tfloor differs with richness. The IGrM does not exhibit any correlation of \Tfloor with mass. However, a strong anti-correlation is visible in the CGM of isolated galaxies, where \Tfloor decreases from $\approx7$\Tvirial to $\approx 0.2$\Tvirial as the mass increases. We infer this in two ways. One, the CGM is truly baryon-sufficient, and the halos at lower/higher masses are overheated/cooled. Or, \Tavg of the CGM of higher-mass galaxies is larger than their \Tfloorn, and their $f_b$ is lower than $f_{b,cosmo}$, resulting in an increasing baryon fraction with decreasing mass. Thus, the ionized CGM of our galaxy sample exhibits an inversion in the mass dependence of either temperature or density, a trend opposite to that of their higher-mass counterparts.     

The implications for super-virial temperature \Tfloor in some halos are huge. It means the entire halos are out of thermal equilibrium. The possibility of non-equilibrium is investigated in the literature mostly in light of cooling to explain the sub-virial temperature CGM observed with absorption lines of metal ions in X-ray/UV/optical, but that phase is not volume-filling. Our observation on 7 million halos shows the reverse, i.e., a deviation from thermal equilibrium due to heating is equally likely. We do not find any evidence of \textit{missing thermal energy} in these halos: either the energy is recycled back into the galactic disks, or retained in the halo gas within \rvirialn, even including a surplus of thermal energy in halos with super-virial \Tfloorn.

\subsection{Truncation} 

The best-fit values of $\beta$ for the six mass and richness bins that \textit{require} a flatter-than-fiducial profile are less than 3, and three of them disfavor $\beta=3$ at \textit{moderate evidence}. $\beta<3$ has been found before for the central galaxy in high-mass galaxy groups \citep[e.g.,][]{Amodeo2021}, although its physical implications at large radii were not explicitly discussed. The data measure the surface density of projected energy, not the total energy, so the pressure modeling cannot decide whether the best-fit solution is physically viable. Thermal energy is proportional to $\int P r^2 dr \propto \frac{r^{3-\beta}}{3-\beta}$ at large radius. Thus, while the best-fit $\beta$ can be smaller than 3 in principle, the thermal energy will not converge to a finite value at large impact parameter, $r_\perp,$ in that case. One of the plausible solutions to this anomaly is that the density and/or temperature have to truncate at a finite radius, $r_t$, which will imply the existence of a shock. Whether it is feedback- or merger/accretion-driven, and whether the ionized baryons are shock-heated and/or shock-compressed, will unveil valuable physical insights. 

If the true $r_t$ is smaller than the observed region in the sky plane (in our case 10\rvirialn), a stronger \twohalo would be required to fit the data at larger $r_\perp$, making the \twohalo stronger at all $r_\perp$ as its shape remains unchanged. That would reduce the fractional contribution of \onehalo at smaller $r_\perp$.  If $r_t$  becomes smaller, the line-of-sight integration of thermal pressure also becomes smaller at the same values of $P_o$ (the amplitude of \onehalon) and $\beta$ (the shape of \onehalon). Thus, there would be a runaway effect where $r_t$ keeps on becoming smaller, \twohalo larger, and some combination of $P_o$ and $\beta$ still fits the data. There has to be a lower limit of $r_t$, beyond which the excess of the data over the \twohalon, if any, will be fitted with \onehalo featuring a large $\beta$, where the truncation radius $r_t$ asymptotically converges to the core radius $r_c$. Illustrating this is beyond the scope of the existing instruments, as most halos are unresolved and \onehalo is dominated by \twohalo beyond 3$\rm R_{200c}$. In future, for an instrument that will resolve the halos down to $\lessapprox r_c$ and also achieve extraordinary S/N, the pressure model should constrain eight free parameters: $P_o, r_o, \alpha, r_c, \beta, r_t, A_{2h}, y_{zp}$, although extracting physically meaningful information would be nontrivial as most of the free parameters would be strongly correlated with each other. 

\subsection{Back to Galactic home}\label{sec:MW}

The Milky~Way is the only galaxy where the multiphase CGM has been observed in detail. It serves as an empirical reference for the CGM of external galaxies and also provides the omnipresent foreground to extragalactic observations in SZ and X-ray. Below, we discuss the implications of super-virial \Tfloor and miscentering in our stacked targets for existing observations of our Galaxy and future observations of individual external galaxies. 

In the Milky~Way, the super-virial phase of the CGM has been widely detected with high column density H-like metal ions, e.g., \oviiin, \nex, \mgxii, \sixiv, \sxvi, etc., in $z=0$ X-ray absorption toward individual as well as stacked quasar sightlines \citep{Das2019a,Das2021b,Lara2023a,Lara2024,McClain2024,Singh2025}. Empirically, this reveals a significant reservoir of \textit{missing metals}, but whether it also harbors a significant amount of \textit{missing Galactic baryons} and \textit{missing energy} heavily relies on ionization modeling and the assumption of absolute metallicity. 

So far, there has been no observational evidence of \Tavg of the CGM being super-virial in \textit{any} galaxy. Therefore, it was assumed that the super-virial phase traced by H-like metal ions only represents a subdominant component tracing past Galactic outflows with a high metal-loading factor, while the phase around \Tvirial traced by He-like metal ions like \ovii is volume-filling. The tSZ effect traces all the ionized baryons, not just metals. Thus, our observation of the excess thermal energy measured from the tSZ effect shows, \textit{for the first time}, that even the volume-filling CGM can be super-virial, especially around Milky Way-like L$^\star$ galaxies. Whether our tSZ observation on $z=0.1-1$ stacked galaxies is directly applicable to an individual galaxy like the Milky~Way at $z=0$ is a nontrivial question to answer, but it suggests that we cannot rule out the possibility of a super-virial temperature volume-filling Milky~Way CGM. 

X-ray absorption by the hot CGM of the Milky~Way can be used to estimate the Galactic contribution to the dispersion measure (DM) toward extragalactic fast radio bursts (FRB), especially at higher Galactic latitude ($|b|>20^\circ$). The Galactic DM spans more than an order-of-magnitude range and is anisotropic across the sky without any trend with Galactic coordinates \citep{Das2021a}. Generally, it is interpreted as an intrinsically amorphous hot CGM centered at the Galactic center. Similar to our stacked tSZ targets in this paper, if the Milky~Way CGM is also miscentered by $>$0.2\rvirialn$\equiv 50$~kpc, even for a spherically symmetric hot CGM, the observed DM would be anisotropic due to our `miscentered' vantage point in the Galactic disk. Thus, ignoring this effect could lead to an incorrect estimate of extragalactic DM toward FRBs, as well as a misinterpretation of the Milky~Way CGM.     

\subsection{What is the principal component?}

The scaling relation between the thermal energy and mass has been the conventional marker of tSZ studies. This works well in the gravity-dominated regime of massive halos, as mass controls the thermal pressure of the ionized baryons. However, when the effect of non-gravitational factors in lower-mass halos is no longer just perturbative, it raises the question: what drives the physical properties of halos? More quantitatively, should we still expect the thermal energy and mass to follow a power-law relation? Should we even consider thermal energy to be scaled with only mass, or should we normalize it with other astrophysical traits, e.g., star-forming and nuclear activities (an indirect tracker of feedback history), and richness (a tracker of merger history and main halo-subhalo interaction)? Immediately answering this is not in the scope of this paper, but we are initiating an effort to build evidence on which physical aspects contribute to the thermal energy of the ionized CGM/IGrM, and which subset of them is the presiding factor in different galaxies. 

{Our sample spans a broad redshift range, and the thermal energies reported in this paper are cosmology-corrected values at the median redshift of the sample. The intrinsic thermal energy of the halos may depend on both mass and richness. In practice, however, halos in higher-redshift bins subtend smaller angular sizes and are therefore less well resolved, which makes the slope and centroid of their pressure profiles less reliably constrained. Consequently, it would be difficult to determine whether a trend with redshift, if detected, reflects true cosmic evolution or an artifact of poorly constrained profile fits at high redshift. This problem is exacerbated at lower masses, where halos are physically smaller, and the one-halo term is weaker. The redshift evolution should therefore be constrained only above a critical mass, at which the amplitude, slope, and centroid of the pressure profile can be measured meaningfully. Future mm instruments with better angular resolution will allow this threshold to be pushed to lower masses. Because such a subsample would not span the full range of mass and richness presented here, we defer this analysis to a follow-up study.} 

Inspired by the literature on higher mass halos, we have so far assumed that the thermal energy and mass of our sample follow a power-law relation. However, in \citetalias{Das2023}, we noticed that the $\Gamma$ could be steeper/flatter than the global value below/above a critical mass, indicating a broken power-law might be a better description. The larger/smaller slopes of star-forming/quiescent galaxies at smaller/larger masses in \citetalias{Das2025} validated that a galaxy sample with mixed star-forming properties would indeed feature a broken power-law-like relation. In this work, again, we see that halos of different richness have different slopes if described with a power-law relation. This reinstates that the assumption of a global power-law dependence is oversimplifying; the sample must be split based on star-forming properties and richness (and yet unidentified aspects), and the scaling relation in each subset must be explored separately.  

\section{Summary and future directions}\label{sec:end}
In this paper, we have explored the dependence of the thermal pressure of the ionized gas on the richness of 7 million $10^{12-14}$ \msun halos at $z\leqslant 1$ by cross-correlating the DESI Legacy Survey catalog of galaxies and galaxy groups and ACT+\planck data. We improve upon the pressure modeling compared to previous studies by simultaneously constraining its slope and centroid. This leads us to a series of new and interesting scientific results listed below:
\begin{enumerate}
    \item Most of the mass and richness bins do not \textit{require} miscentering and/or deviation from the fiducial shape. These halos are prevalent in higher-mass isolated galaxies and lower-mass galaxy groups. However, lower-mass isolated galaxies and higher-mass groups \textit{require} both miscentering and a flatter-than-fiducial shape. Without separating the sample by richness, we could not identify these patterns. 
    \item Based on the pressure profiles, we find broadly four different situations in the thirteen mass and richness bins. a) Six prefer halo-wide mild cooling and subsidiary thermal feedback without dominant kinetic feedback, indicating the formation of a multiphase medium through an active disk-halo baryon cycle; b) One is driven by moderate cooling and accretion; c) Two have evidence of substantial cooling/merger in the inner/outer halo; d) Four are suggestive of dominant thermal feedback that keeps the CGM/IGrM hot but does not shovel it out of the virial radius. These results rule out the universality of the simplified (kinetic) feedback vs gravity scenario and establish that multiple non-gravitational processes of varied strength are at play in different masses and different environments.  
    \item The slope of the thermal energy-mass relation, $\Gamma$, increases with increasing richness. It implies that the relative strength of non-gravitational to gravitational forces is not environment-agnostic. The CGM reveals a slope $7.3\sigma$ flatter than self-similarity, without a significant deviation from the default normalization. This suggests that the ionized CGM in higher/lower mass galaxies is efficiently cooled/overheated without their baryon content being significantly affected. The IGrM is consistent with self-similarity. 
    \item The minimum volume-average \Tvirialn-normalized temperature allowed by the data decreases with increasing mass for the CGM of isolated galaxies, with the lower mass halos requiring a super-virial phase. This has strong implications for the metal, mass, and energy-loading efficiency of galactic outflows.  
\end{enumerate} 

{Following common practice, we have assumed that the correction for dust contamination is uniform across halo mass, richness, and circumgalactic scale. Testing this assumption is beyond the scope of this paper and current mm instruments. However, as discussed in Appendix~\ref{sec:cib}, residual dust contamination is not the dominant source of systematic uncertainty in our measurements, so any deviation from the assumed uniformity of the dust is unlikely to change our conclusions.}

tSZ effect has a promising potential to be effectively combined with other complementary probes of ionized CGM/IGrM across the electromagnetic spectrum: from the DM of FRBs in radio and kSZ in mm constraining the density of total ionized baryons, to optical/UV/X-ray quasar absorption line spectroscopy of metals across all ionization states constraining the phase distribution of the multiphase ionized baryons, and X-ray emission imaging/spectroscopy constraining the baryons, metals, and energetics in the hot phase. However, these observations are often performed in discrete populations of galaxy/group samples driven by distinct science goals, technical limitations, and observational design (e.g., metal absorption lines and DM rely on the presence of background quasars and FRBs, while other observing techniques are direct). This limits our ability to holistically understand CGM/IGrM properties and their causal connection to host galaxies through multi-wavelength observations. In our cross-correlation studies, we have so far used catalogs not necessarily built for these observations, so the best catalog for each project varies. This makes the comparison even among our own results nontrivial. Catalogs of robustly estimated core properties (redshift, stellar and halo mass, star formation rate, metallicity, nuclear activity, environmental information, etc.) of galaxies/groups would be crucial for homogenized target selection across the aforementioned observational probes. This will be enabled by wide-field surveys with Euclid, Rubin, Roman (and in the future UVEX), thanks to their extraordinary accuracy and precision. As far as the SZ effect is concerned, next-generation telescopes, e.g., CCAT-Prime (FYST), Simons Observatory, and AtLAST, with their increased sensitivity, footprint, angular resolution, and frequency coverage, will enable us to improve dust correction, reduce systematics due to halo-to-halo scatter, and, most importantly, resolve the thermal pressure and density of the inner ionized CGM/IGrM from their outer counterparts. This will lay the path forward for a comprehensive multi-wavelength picture of the interplay between galaxies and the cosmic ecosystem surrounding them.


\section*{acknowledgments}
{We thank the anonymous reviewer for constructive suggestions that helped improve the paper.} S.D. acknowledges support provided by NASA through Hubble Fellowship grant HST-HF2-51551.001-A, awarded by the Space Telescope Science Institute, which is operated by the Association of Universities for Research in Astronomy, Inc., for NASA, under the contract NAS 5-26555. S.M. is grateful for the NASA ADAP grant 80NSSC22K1121. 



\facilities{\textit{DESI}, \planckn, \actn}
\software{\texttt{AstroPy} \citep{Astropy2013,Astropy2018,Astropy2022}, \texttt{Colossus} \citep{Diemer2018}, \texttt{corner} \citep{Foreman-Mackey2016}, \texttt{emcee} \citep{Foreman-Mackey2013},  \texttt{Halotools} \citep{Hearin2016}, \texttt{Jupyter} \citep{jupyter2016}, \texttt{Matplotlib} \citep{Hunter2007}, \texttt{NumPy} \citep{numpy2020}, \texttt{pandas} \citep{pandas2020},  \texttt{Pixell} \citep{Naess2021}, \texttt{Python math} \citep{mathpy2020},  \texttt{SciPy} \citep{scipy2022}}

\appendix 
\counterwithin{figure}{section}

\section{Other halos}\label{sec:others}
\begin{figure}
\centering
    \includegraphics[trim=15 0 45 40, clip,width=0.16\linewidth]{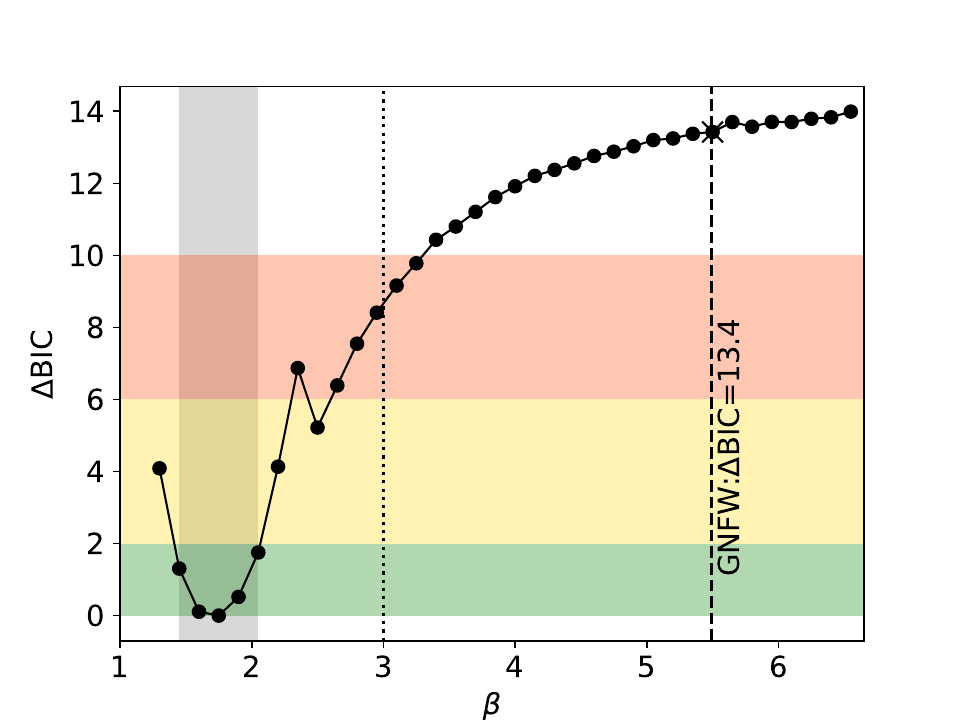}
    \includegraphics[width=0.16\linewidth]{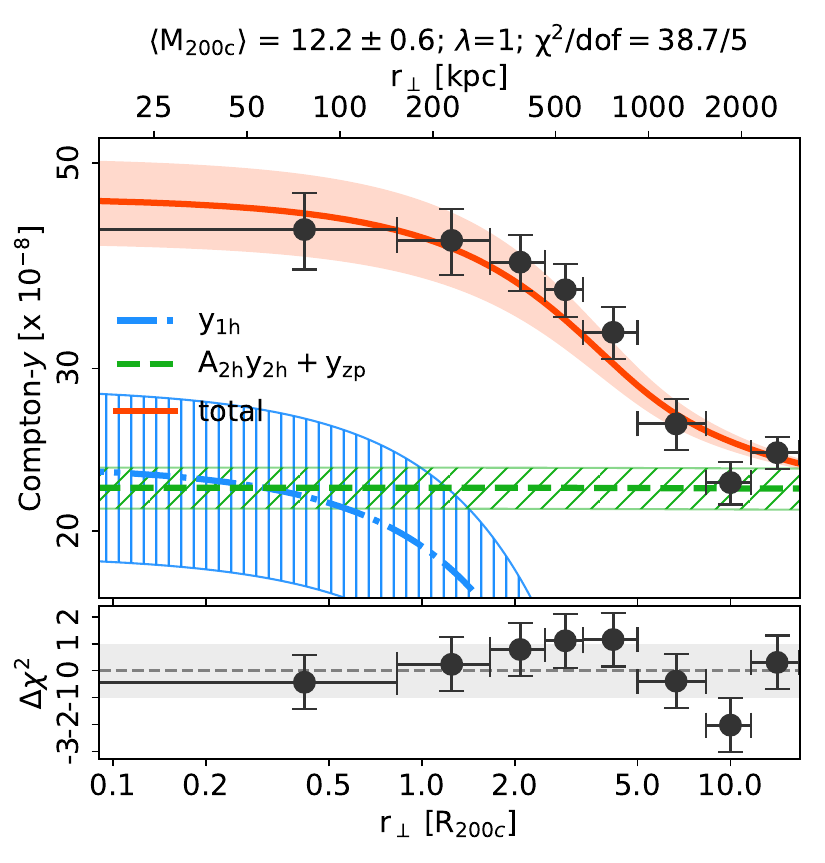}
    \includegraphics[width=0.16\linewidth]{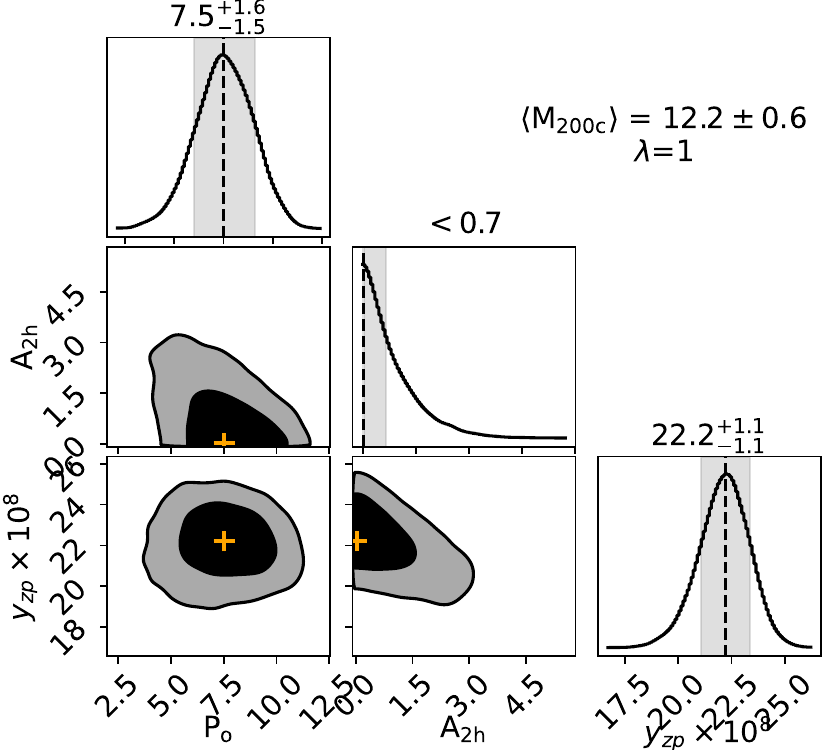}
    \includegraphics[trim=15 0 45 40, clip,width=0.16\linewidth]{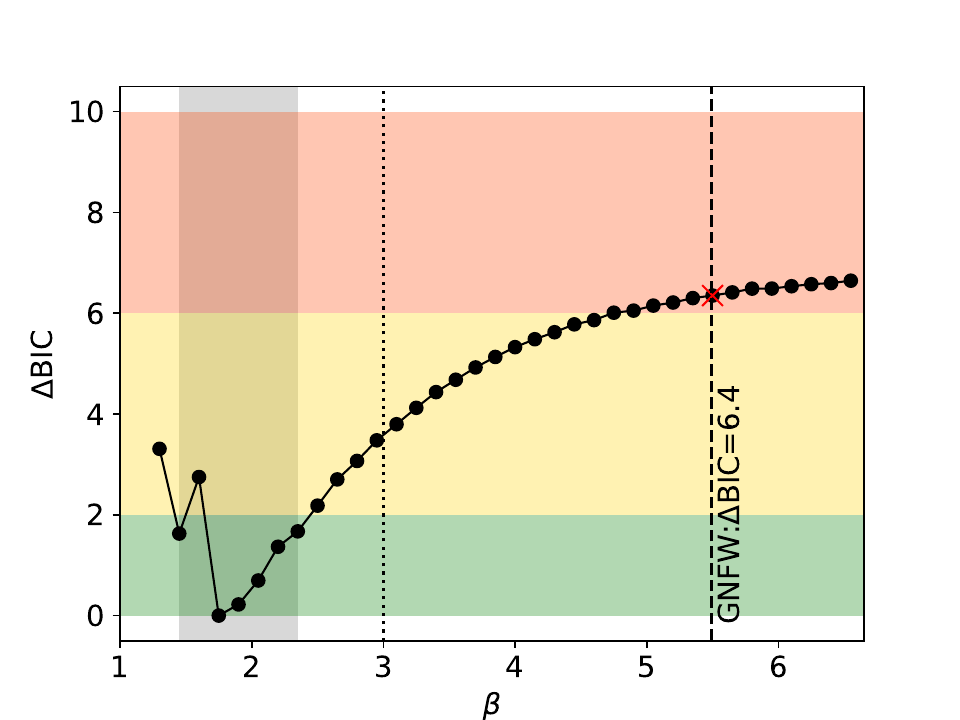}
    \includegraphics[width=0.16\linewidth]{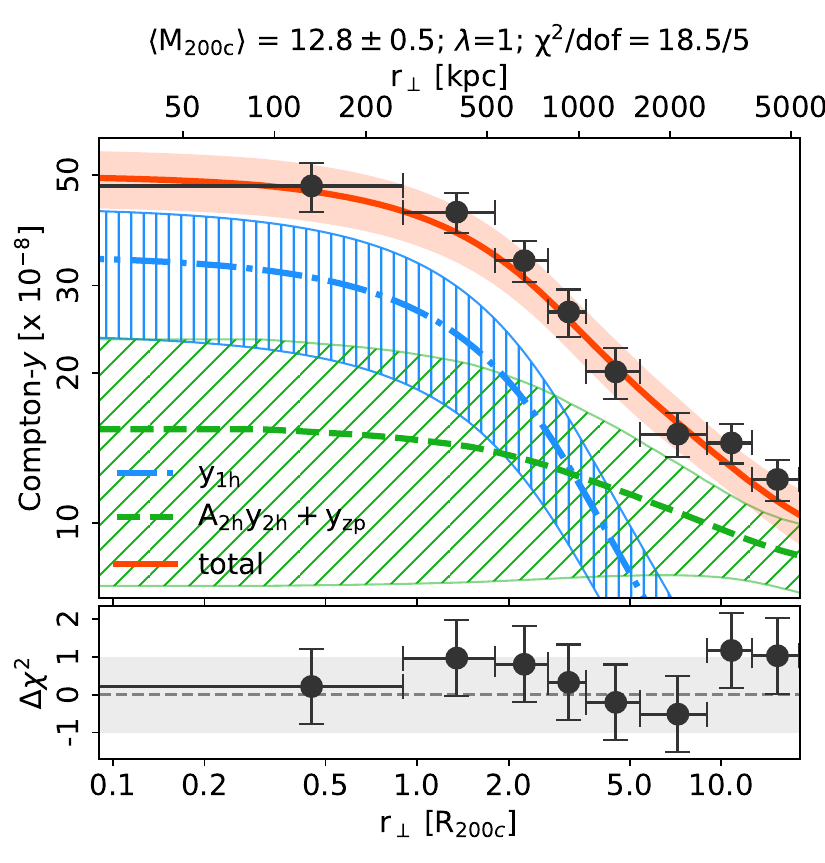}
    \includegraphics[width=0.16\linewidth]{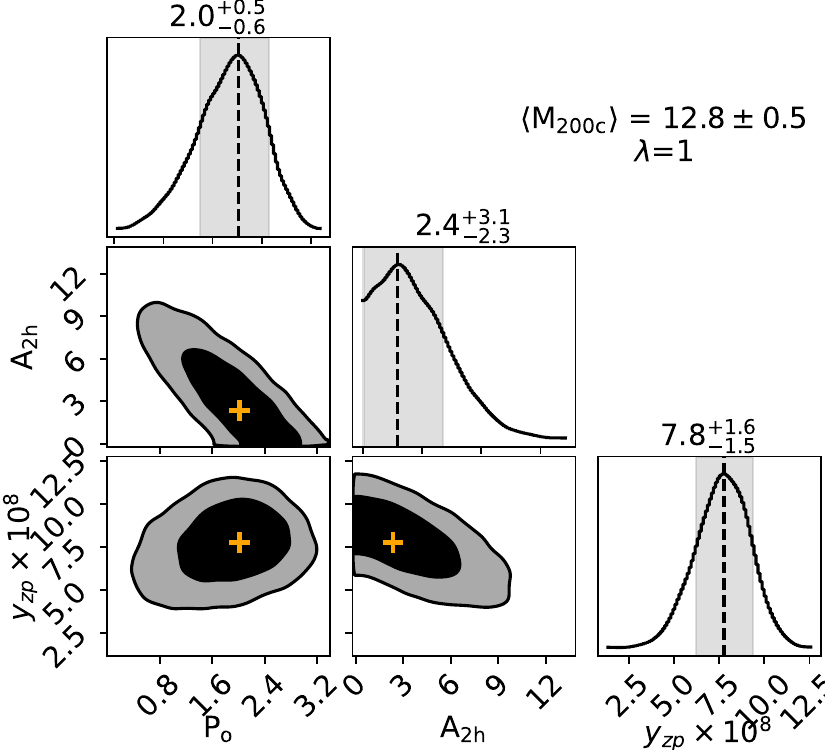}
    \includegraphics[trim=15 0 45 40, clip,width=0.16\linewidth]{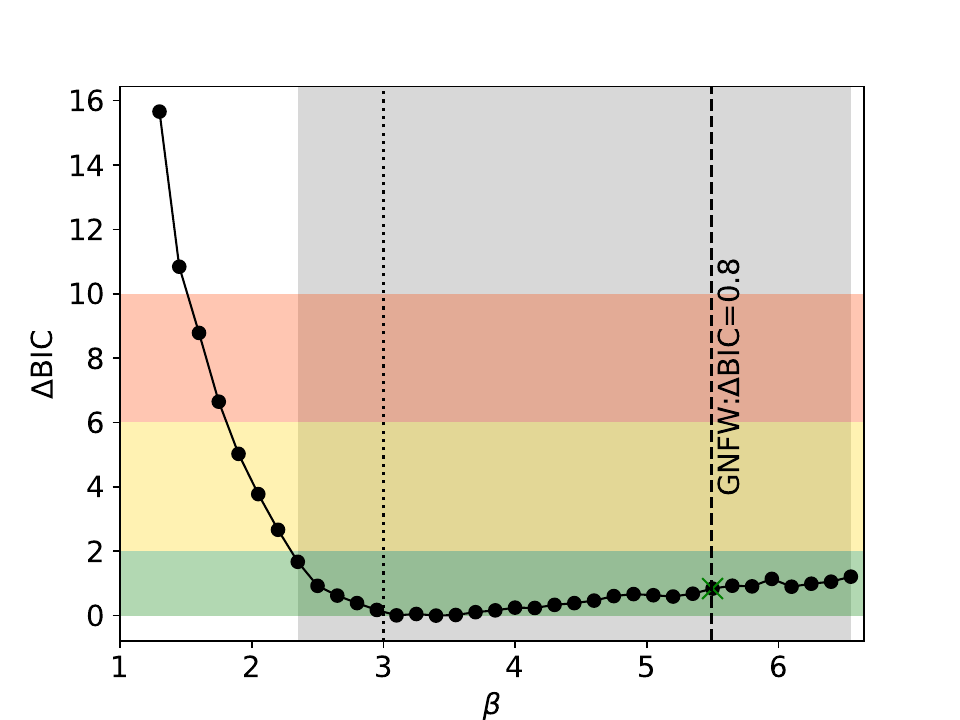}
    \includegraphics[width=0.16\linewidth]{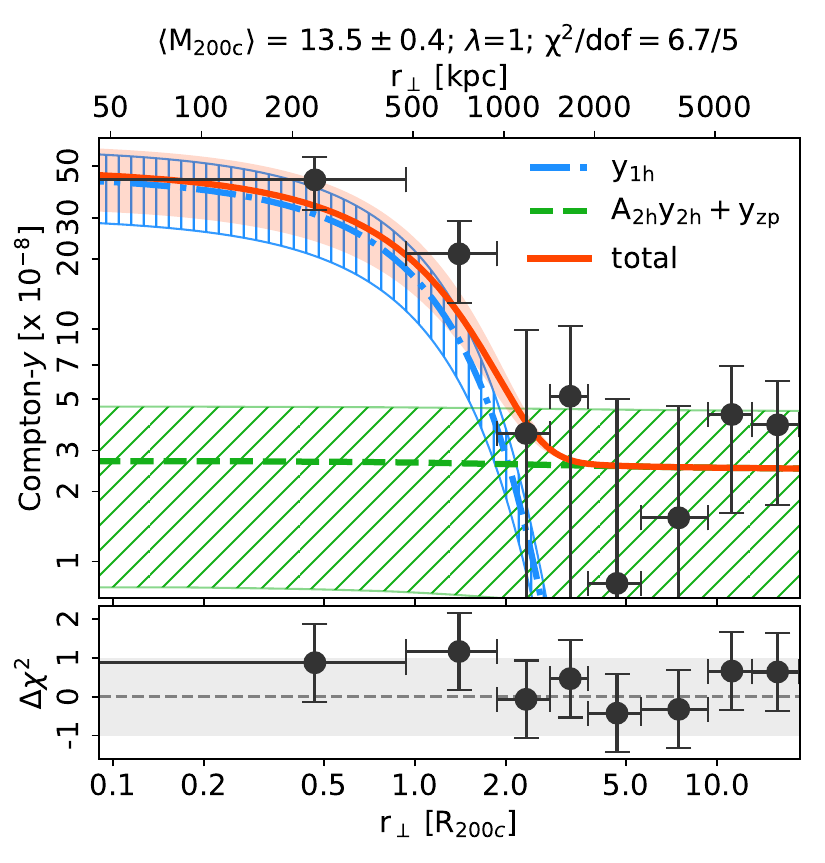}
    \includegraphics[width=0.16\linewidth]{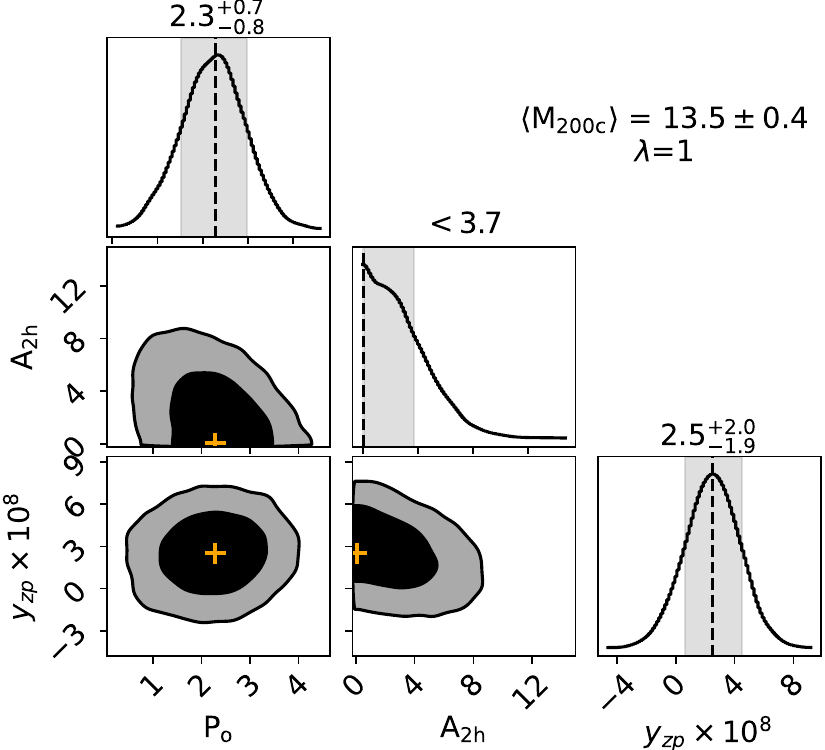}
    \includegraphics[trim=15 0 45 40, clip,width=0.16\linewidth]{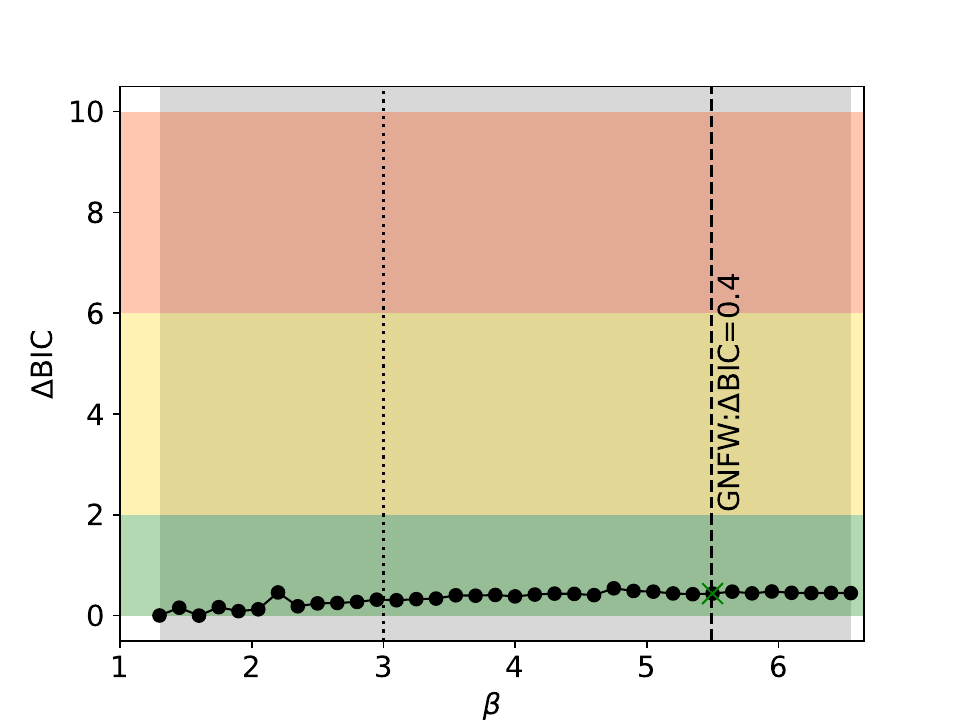}
    \includegraphics[width=0.16\linewidth]{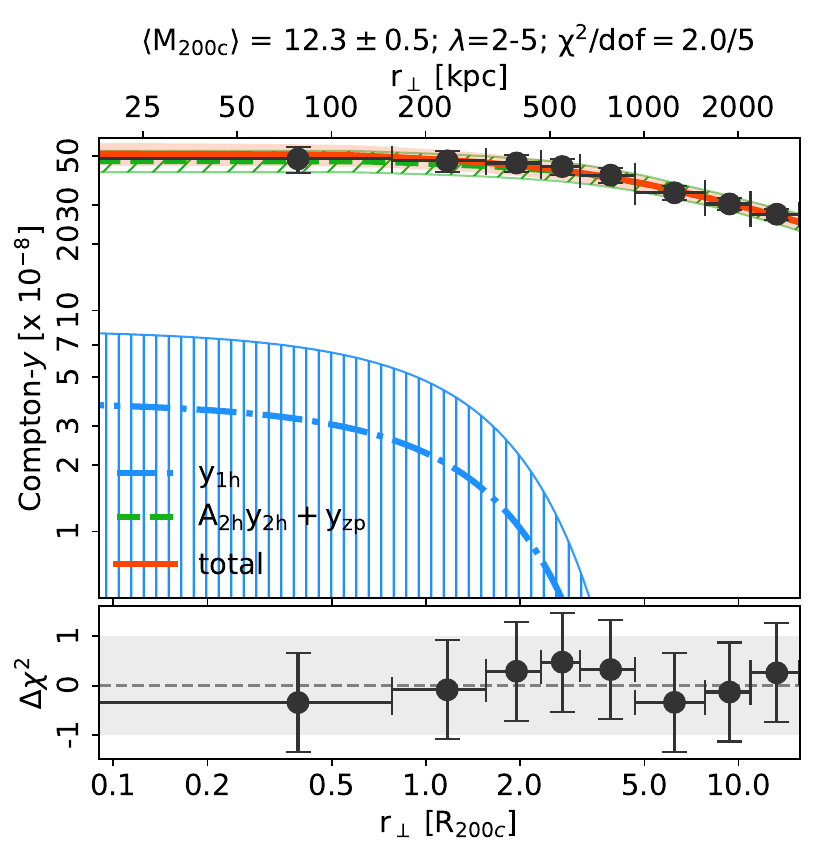}
    \includegraphics[width=0.16\linewidth]{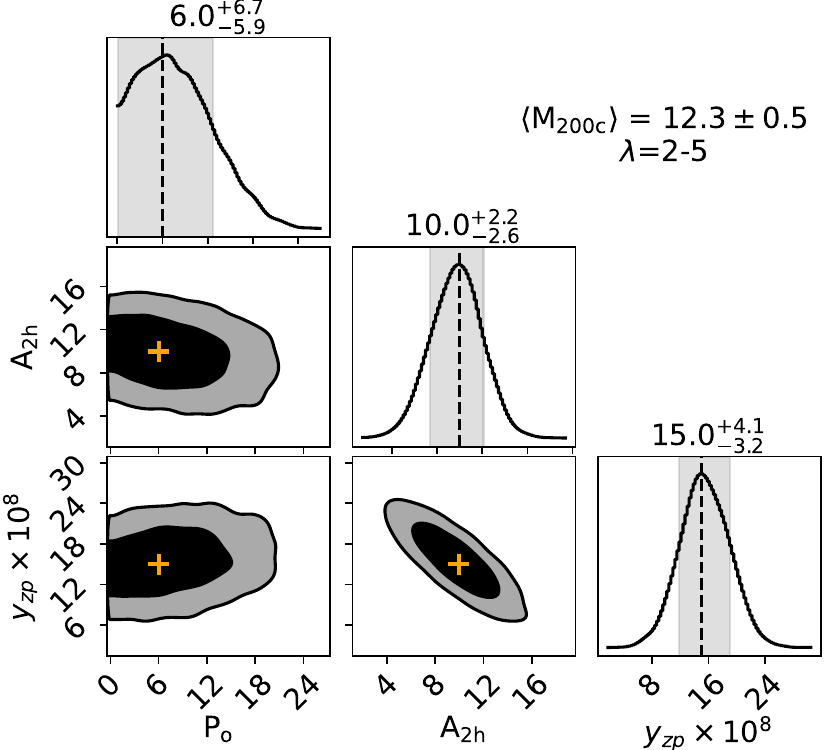} 
    \includegraphics[trim=15 0 45 40, clip,width=0.16\linewidth]{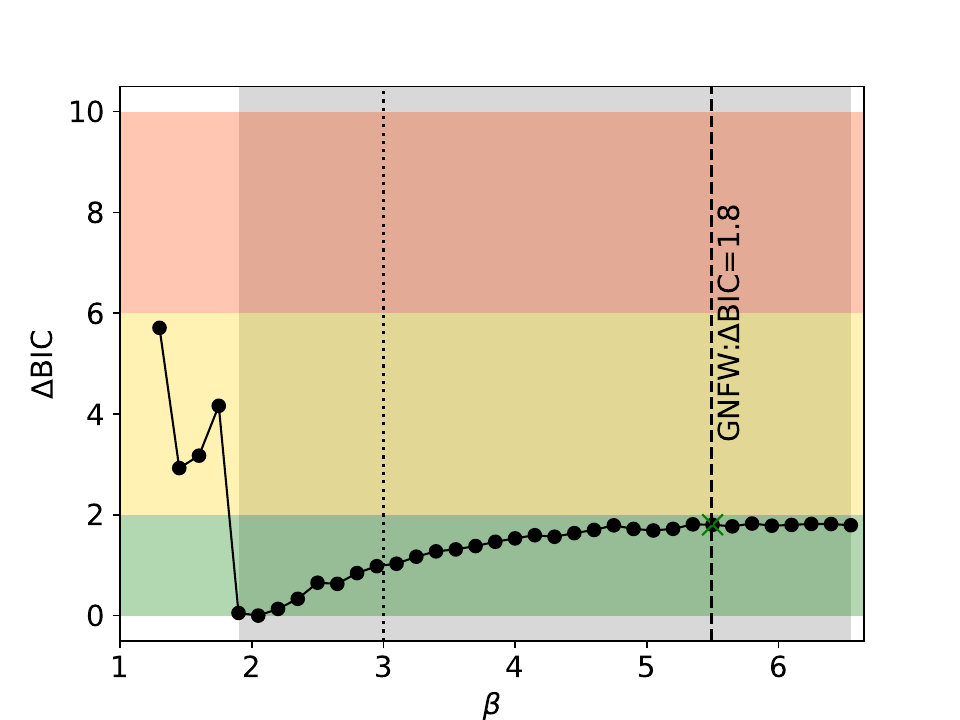}
    \includegraphics[width=0.16\linewidth]{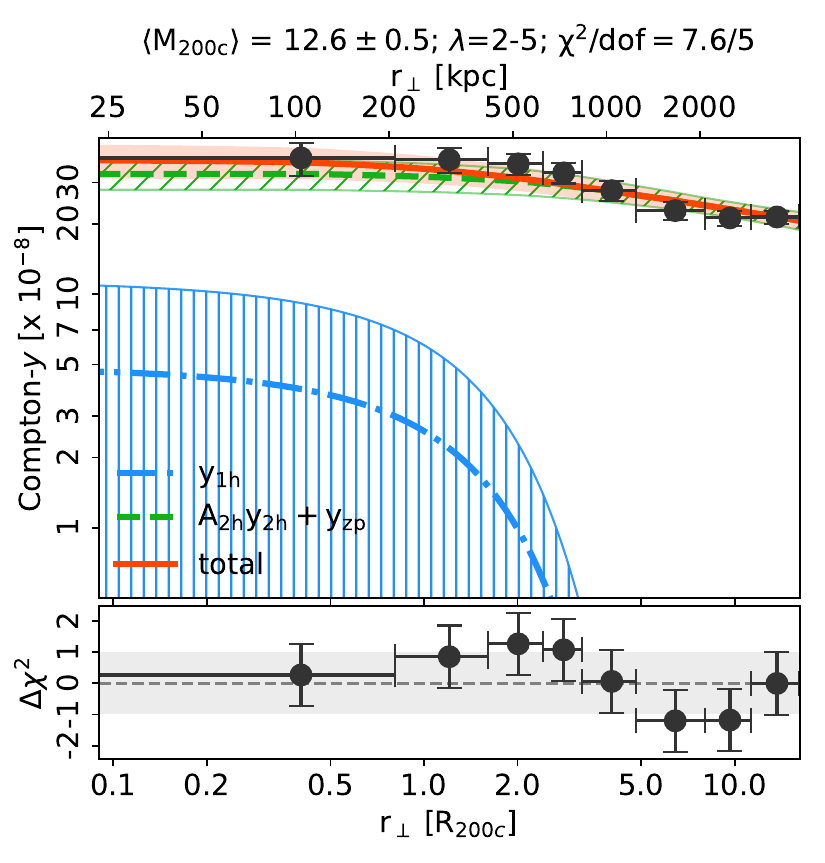}
    \includegraphics[width=0.16\linewidth]{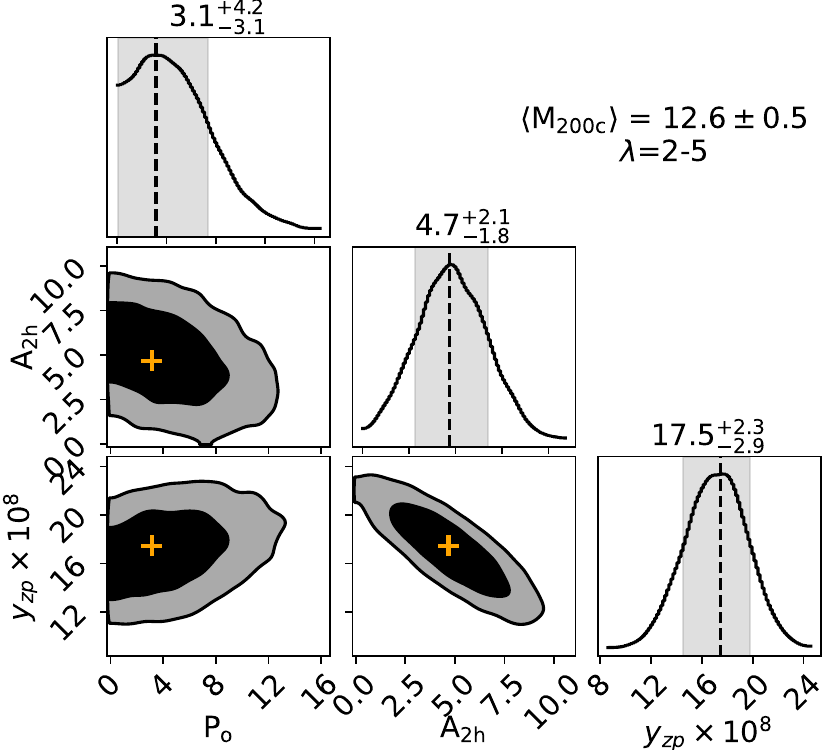}
    \includegraphics[trim=15 0 45 40, clip,width=0.16\linewidth]{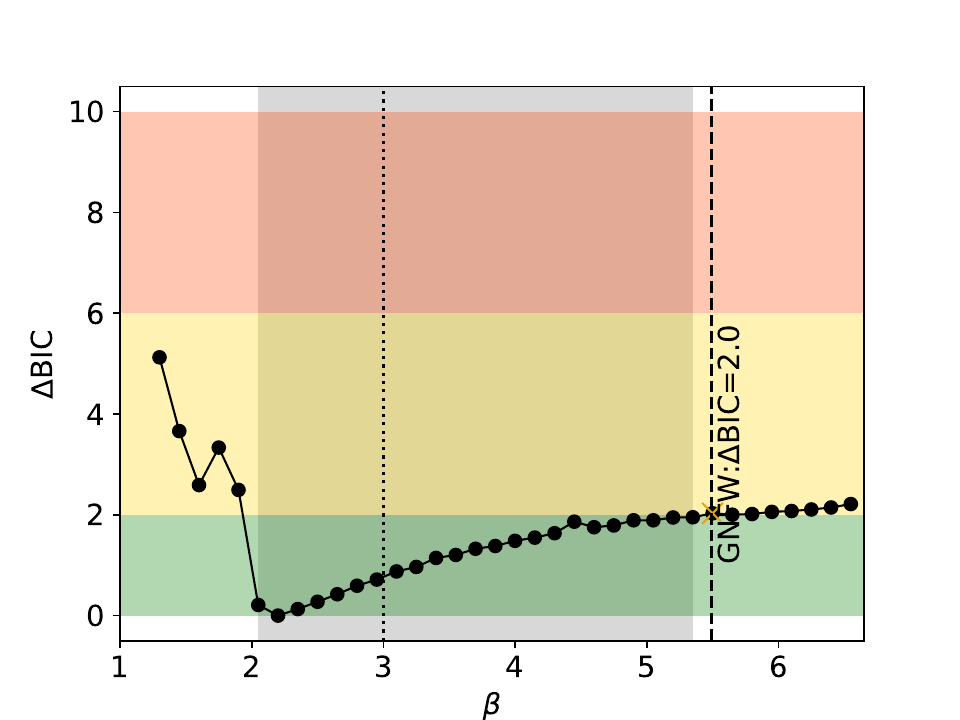}
    \includegraphics[width=0.16\linewidth]{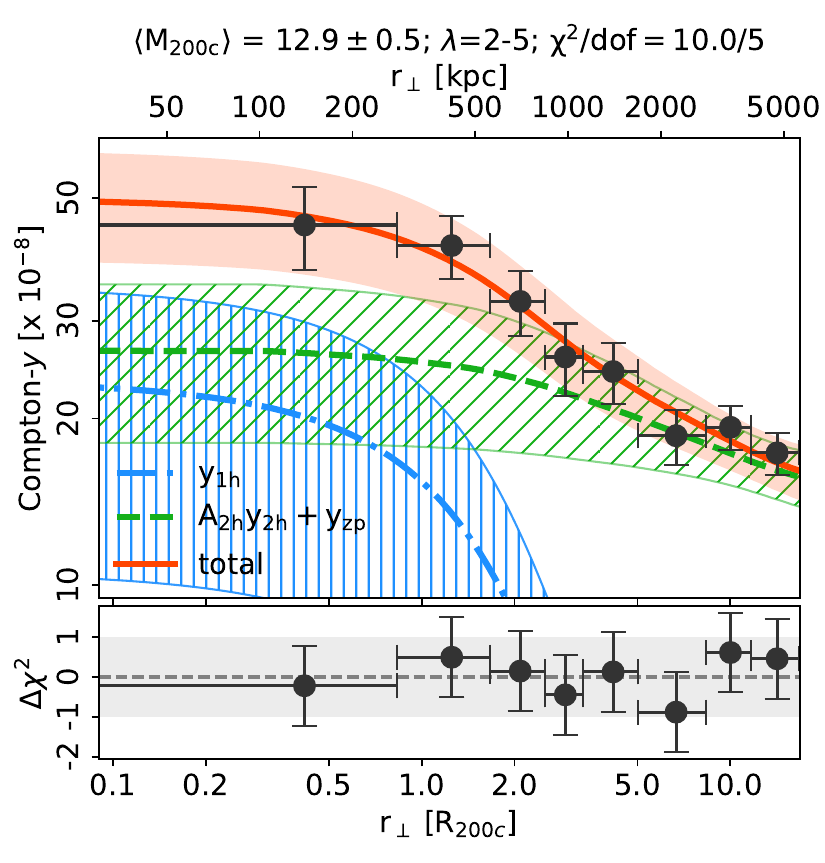}
    \includegraphics[width=0.16\linewidth]{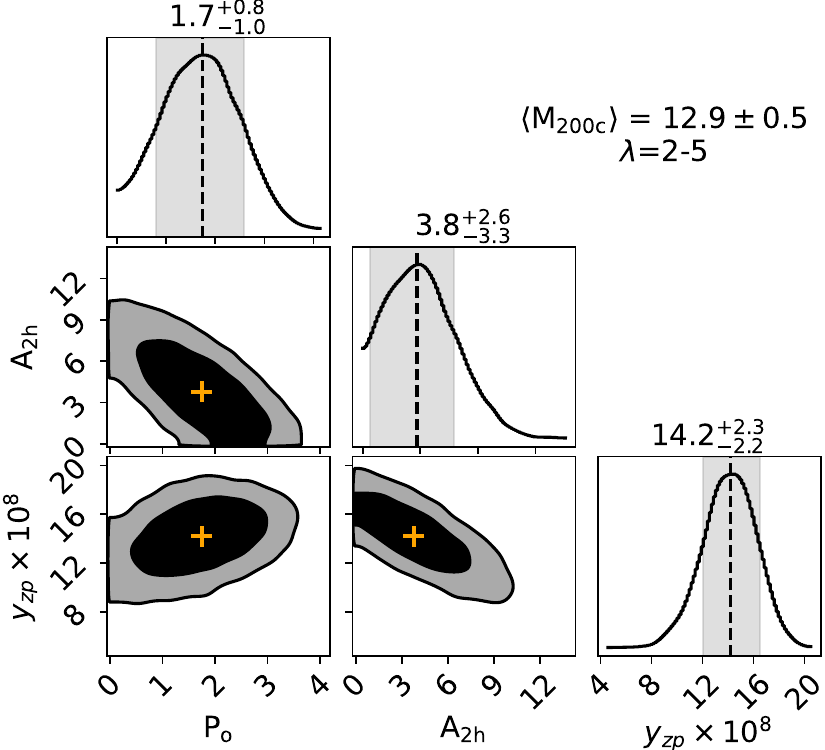}
    \includegraphics[trim=15 0 45 40, clip,width=0.16\linewidth]{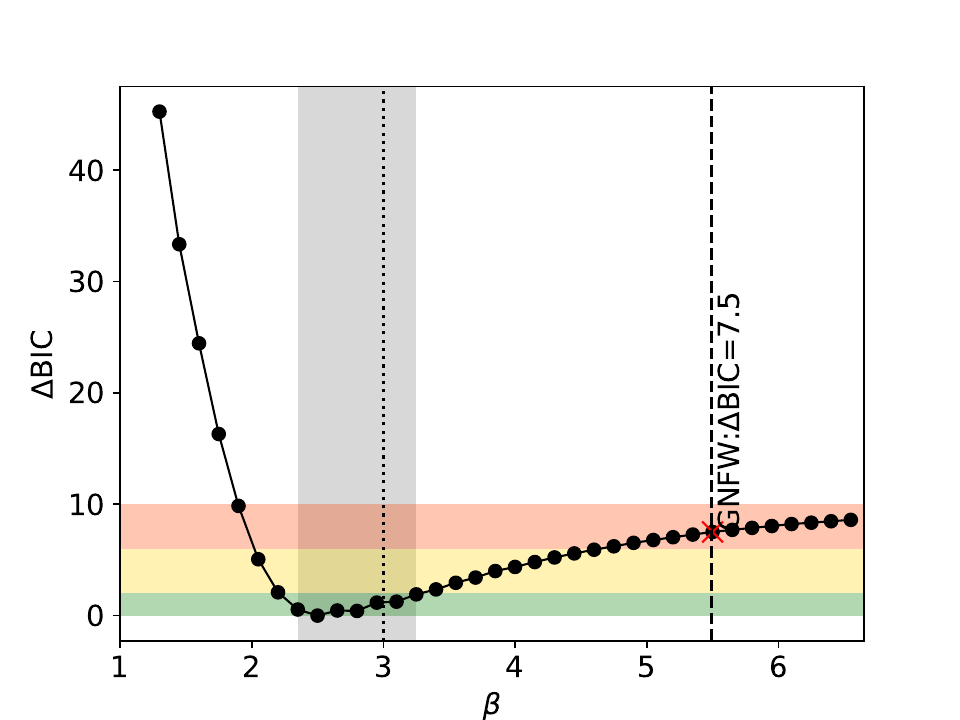}
    \includegraphics[width=0.16\linewidth]{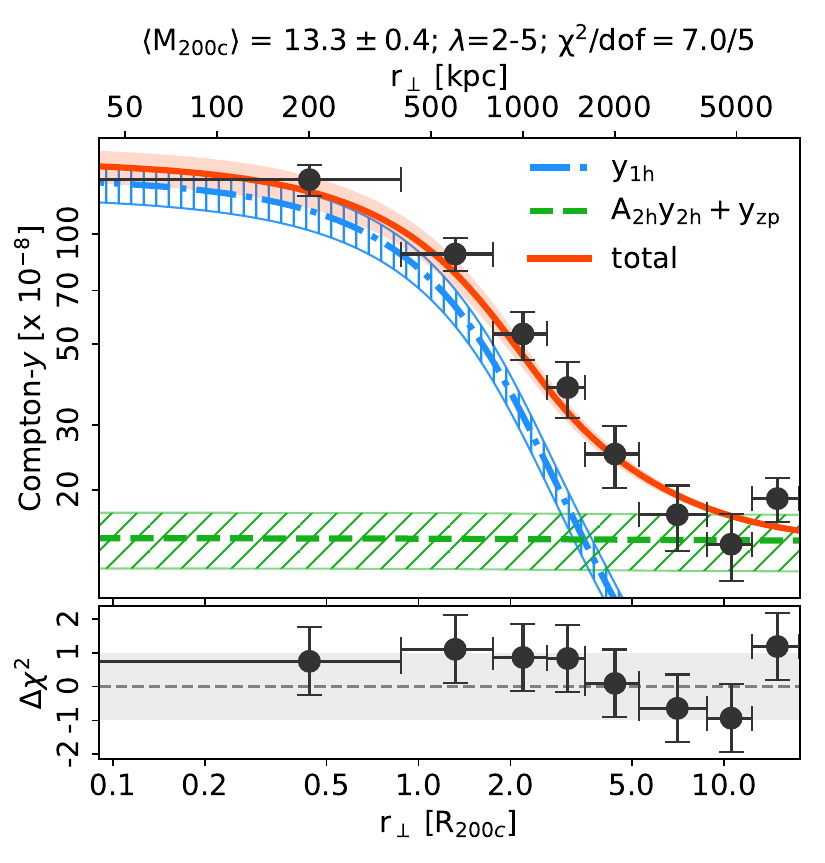}
    \includegraphics[width=0.16\linewidth]{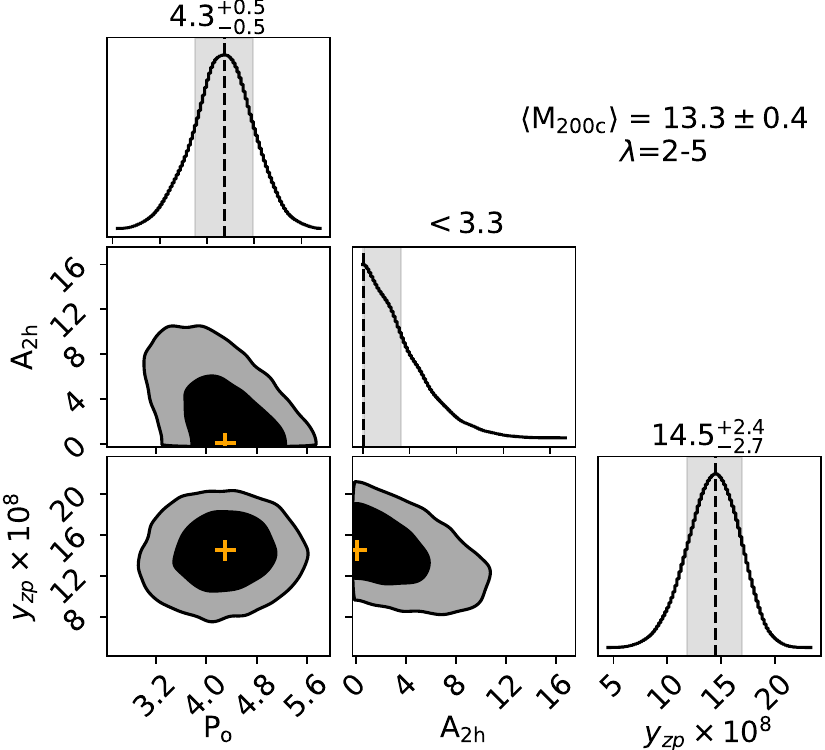}
    \includegraphics[trim=15 0 45 40, clip,width=0.16\linewidth]{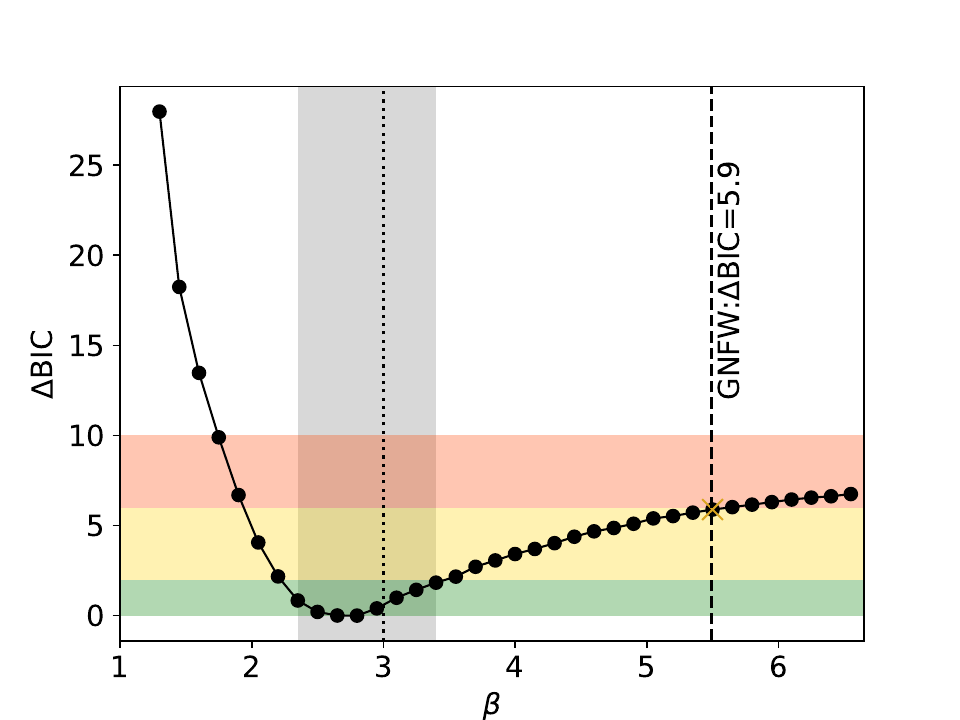}
    \includegraphics[width=0.16\linewidth]{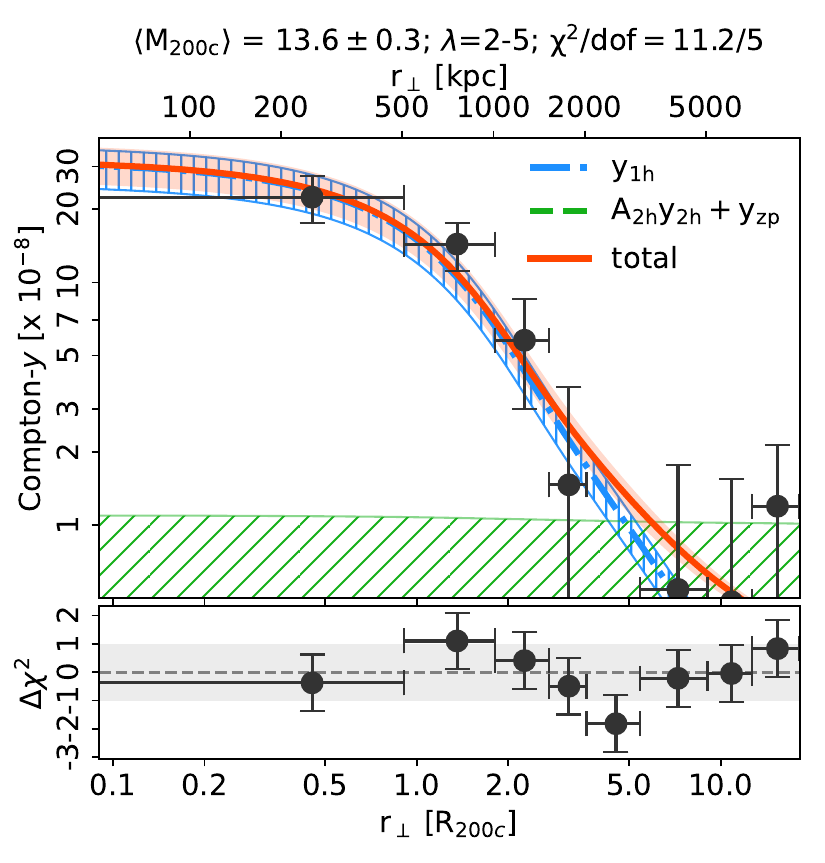}
    \includegraphics[width=0.16\linewidth]{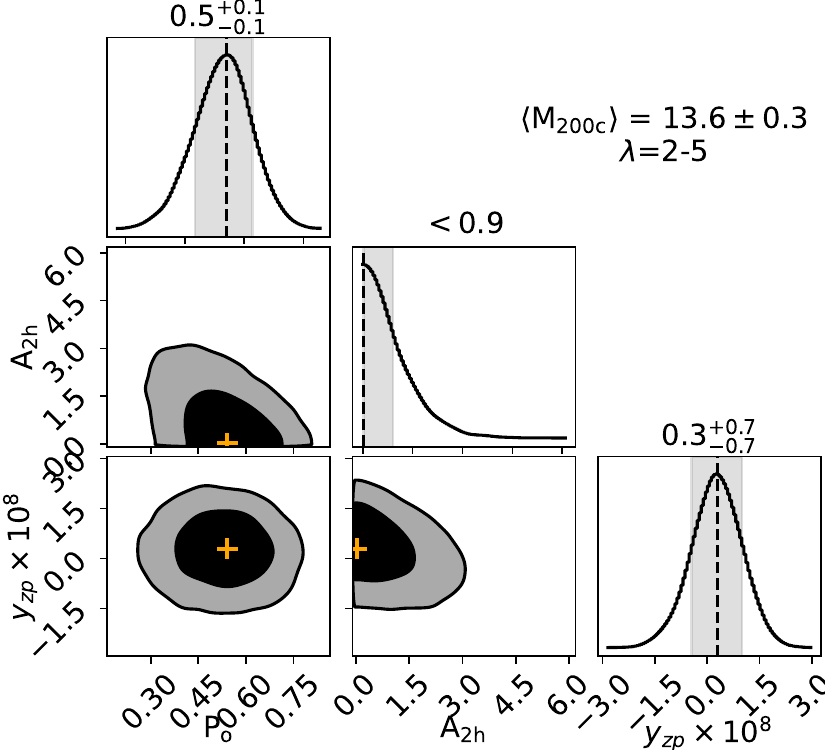}
    \includegraphics[trim=15 0 45 40, clip,width=0.16\linewidth]{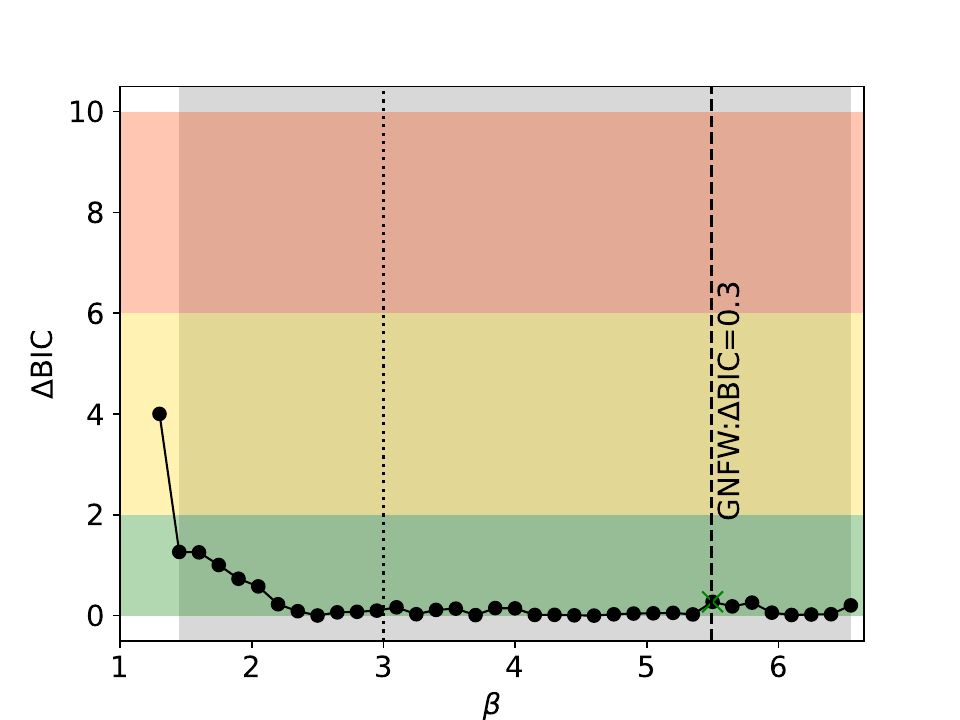}
    \includegraphics[width=0.16\linewidth]{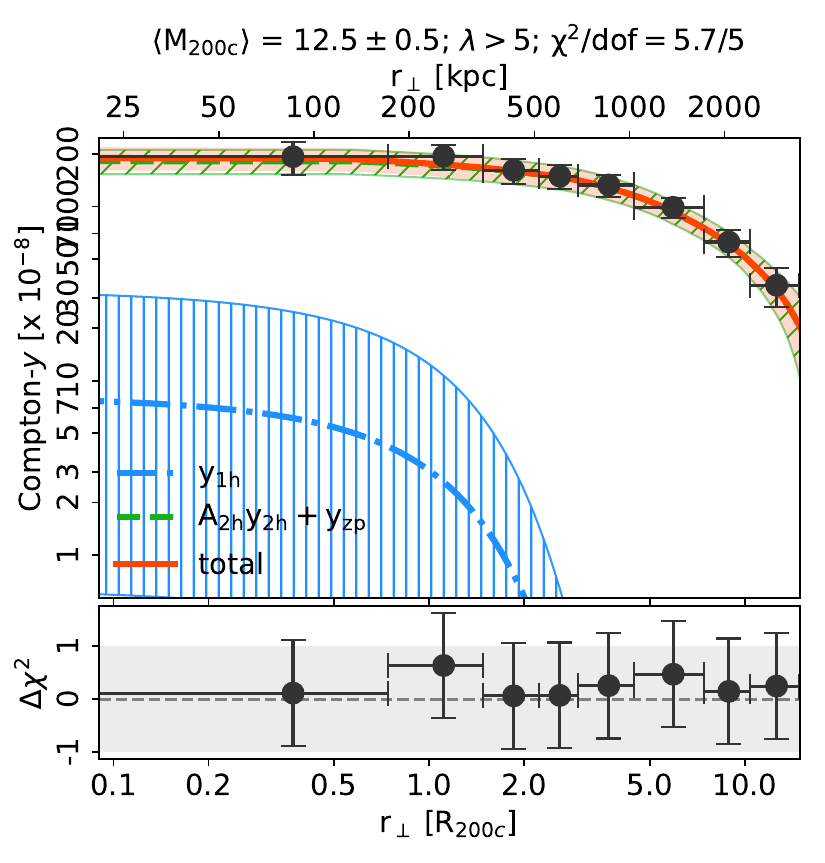}
    \includegraphics[width=0.16\linewidth]{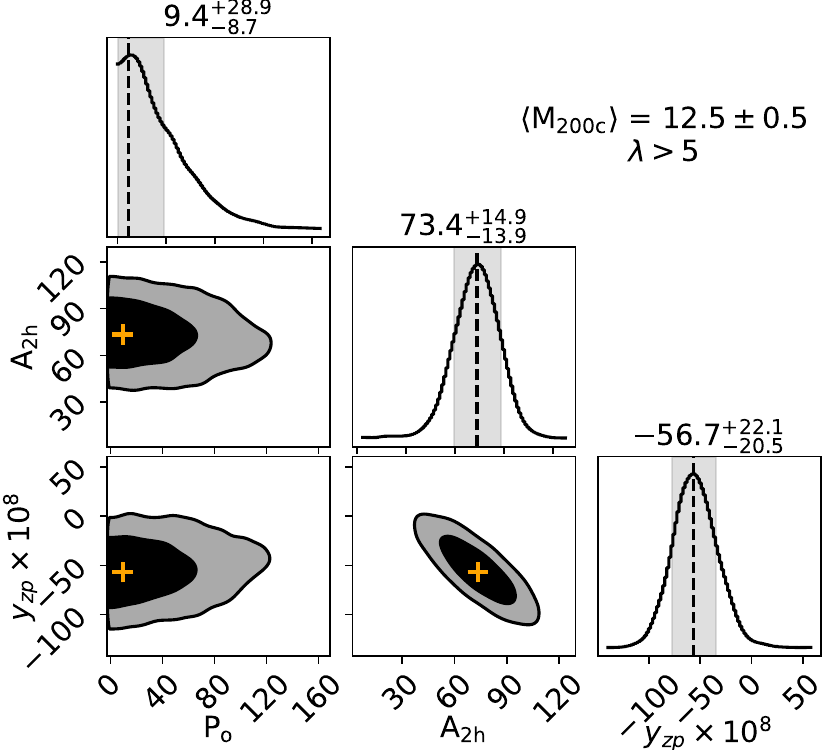}
    \includegraphics[trim=15 0 45 40, clip,width=0.16\linewidth]{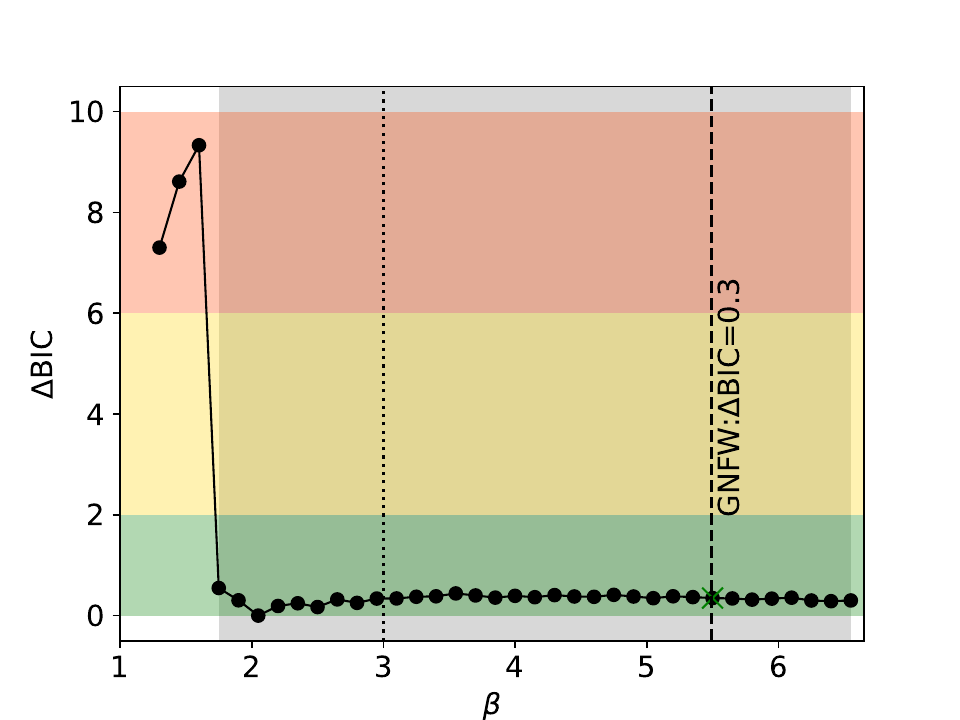}
    \includegraphics[width=0.16\linewidth]{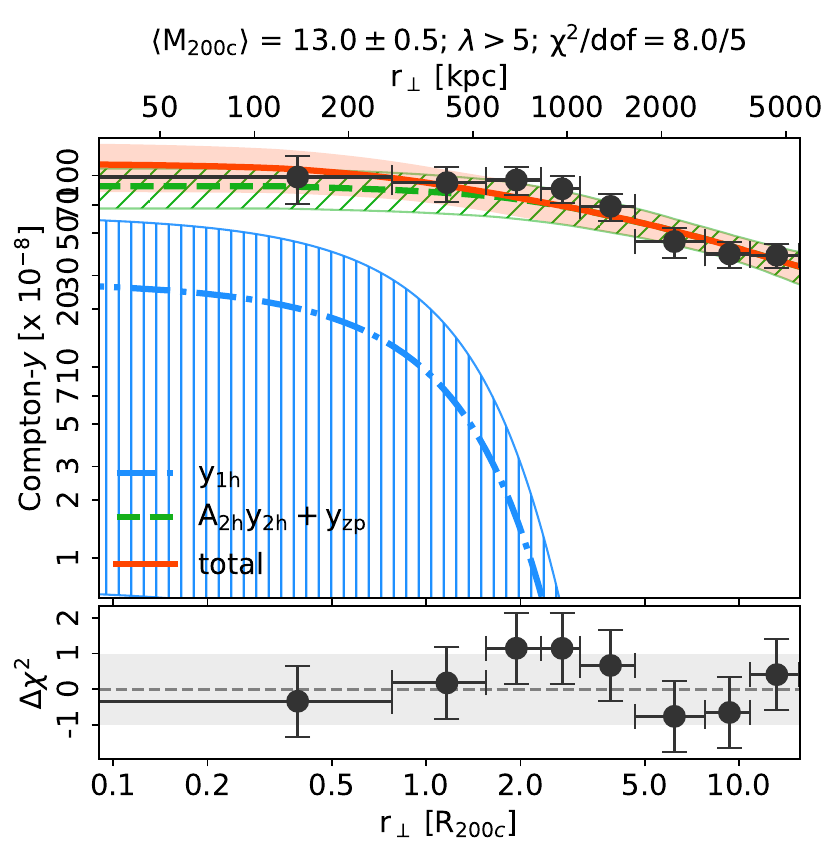}
    \includegraphics[width=0.16\linewidth]{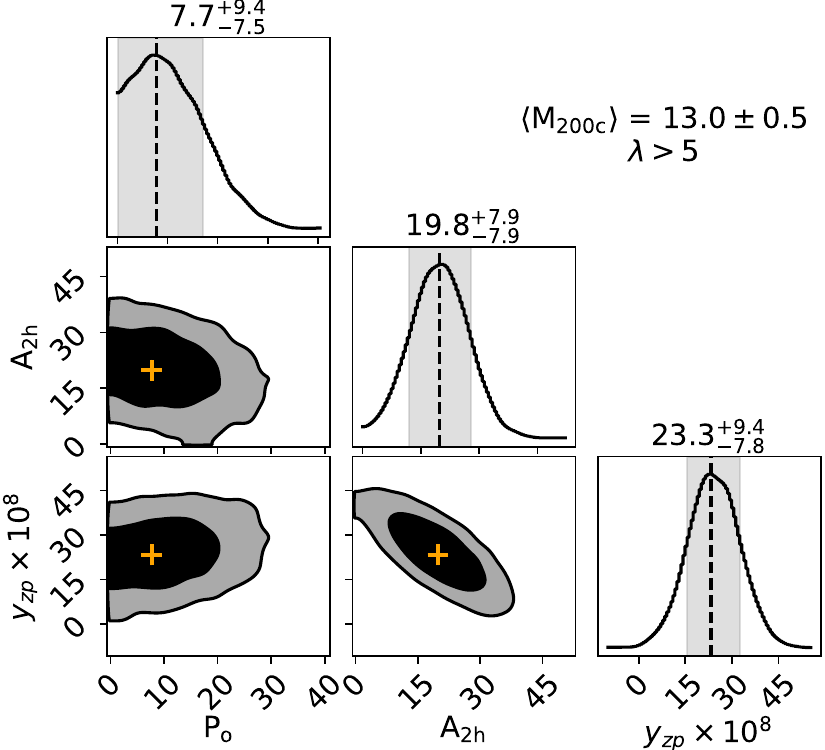}
    \includegraphics[trim=0 0 45 40, clip,width=0.16\linewidth]{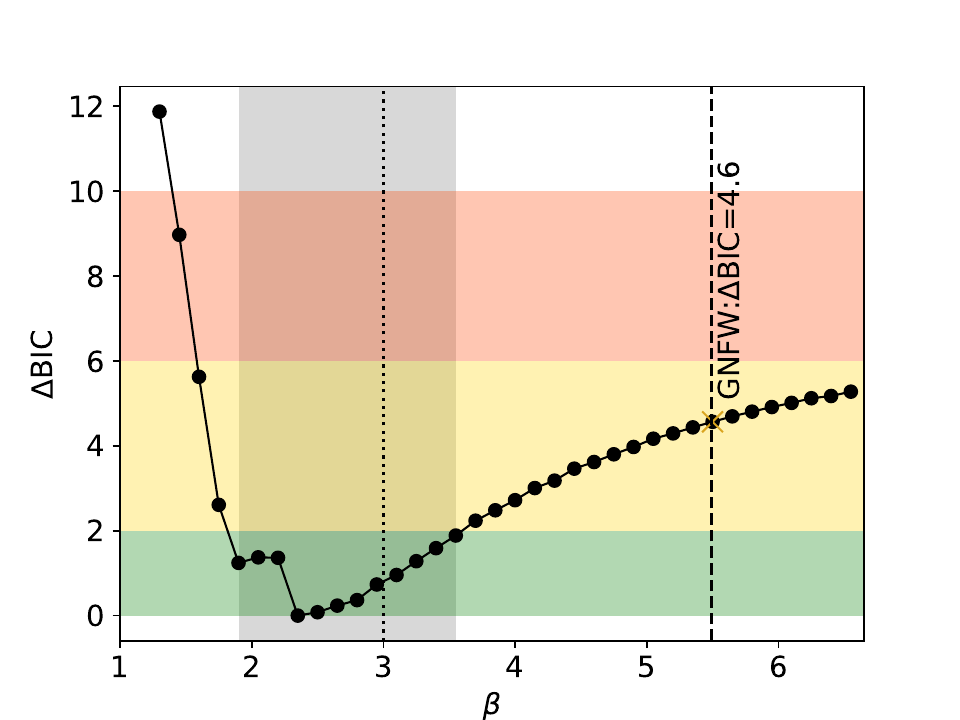}
    \includegraphics[width=0.16\linewidth]{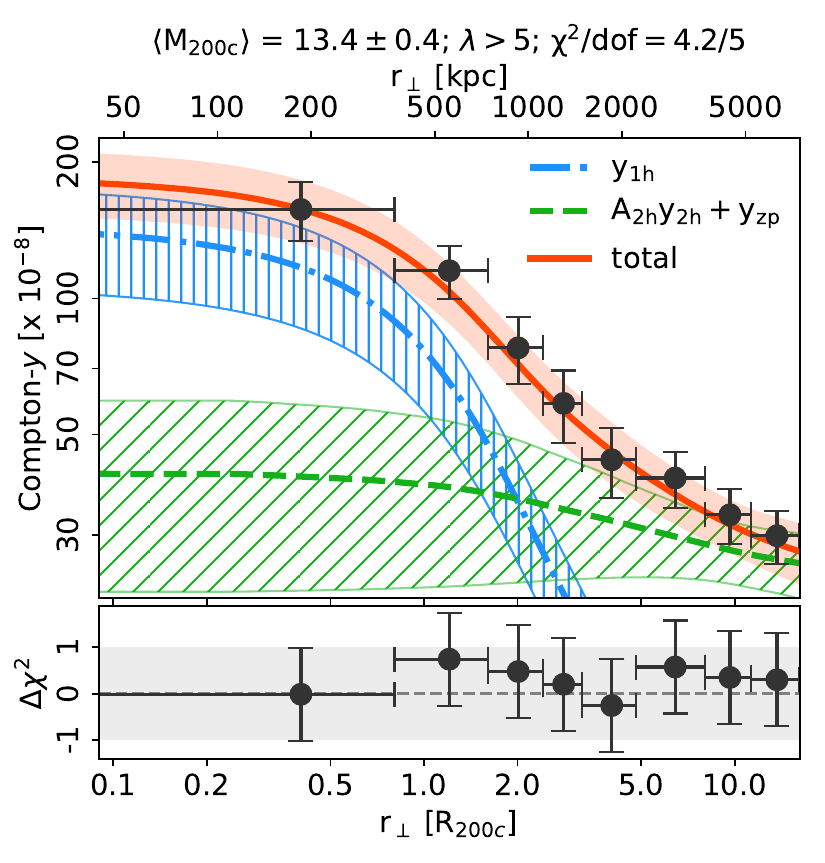}
    \includegraphics[width=0.16\linewidth]{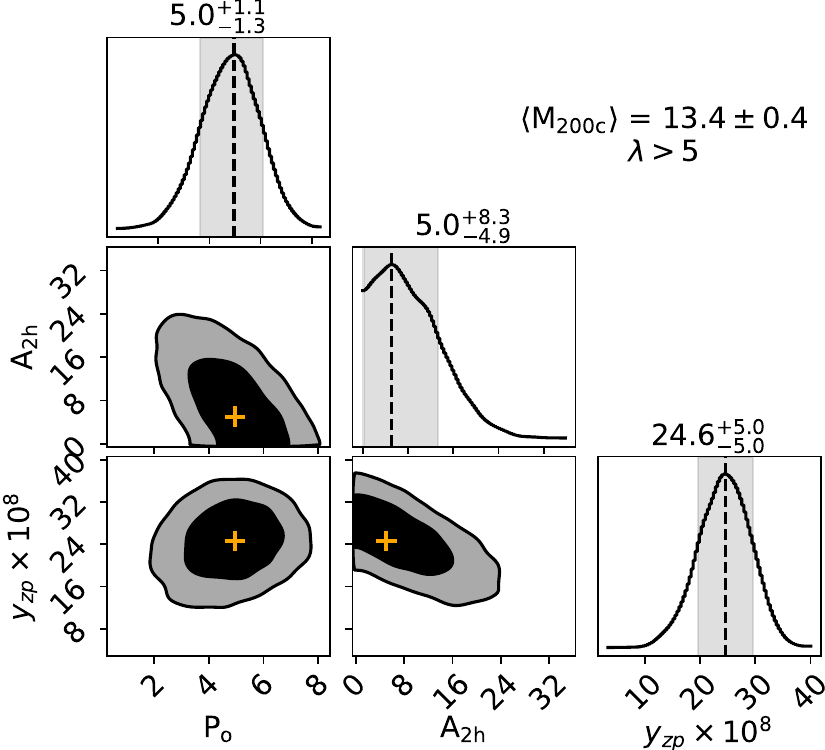}
    \includegraphics[trim=15 0 45 40, clip,width=0.16\linewidth]{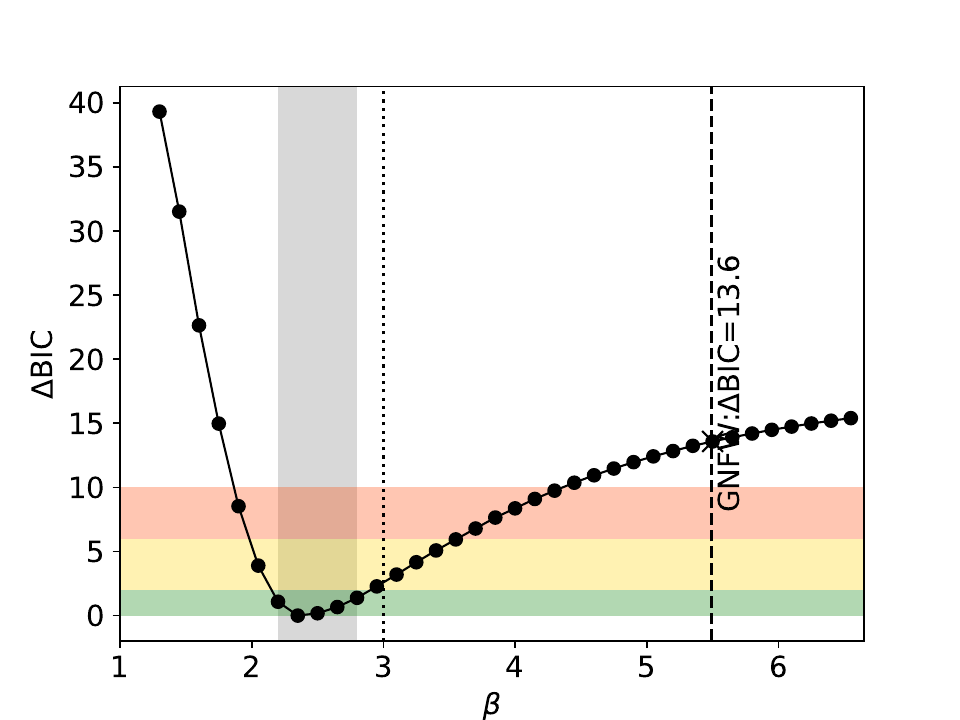}
    \includegraphics[width=0.16\linewidth]{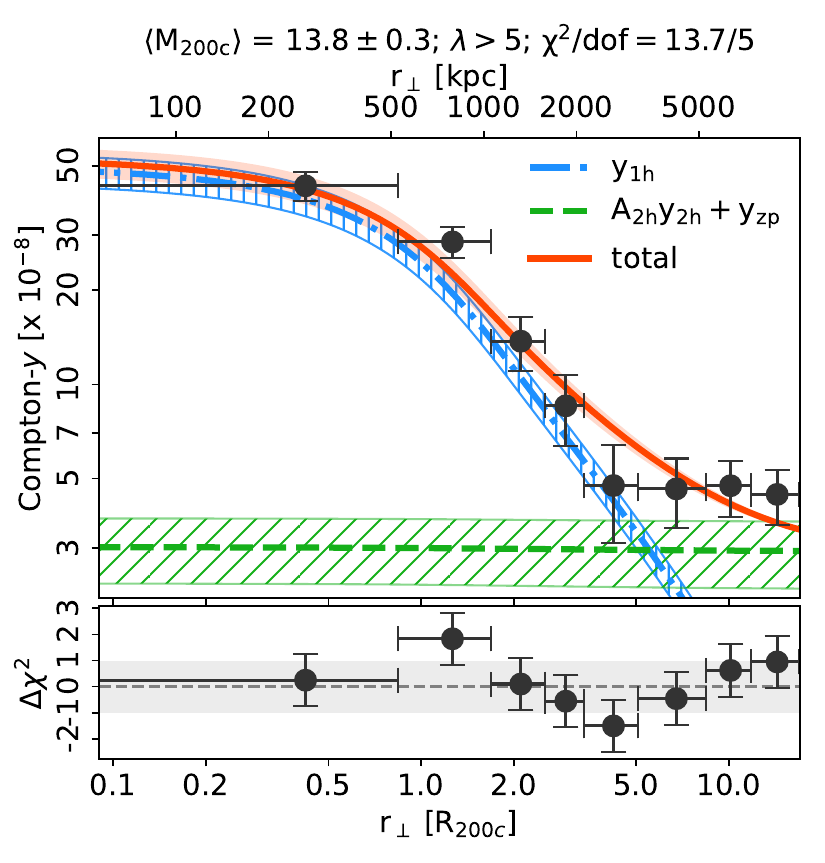}
    \includegraphics[width=0.16\linewidth]{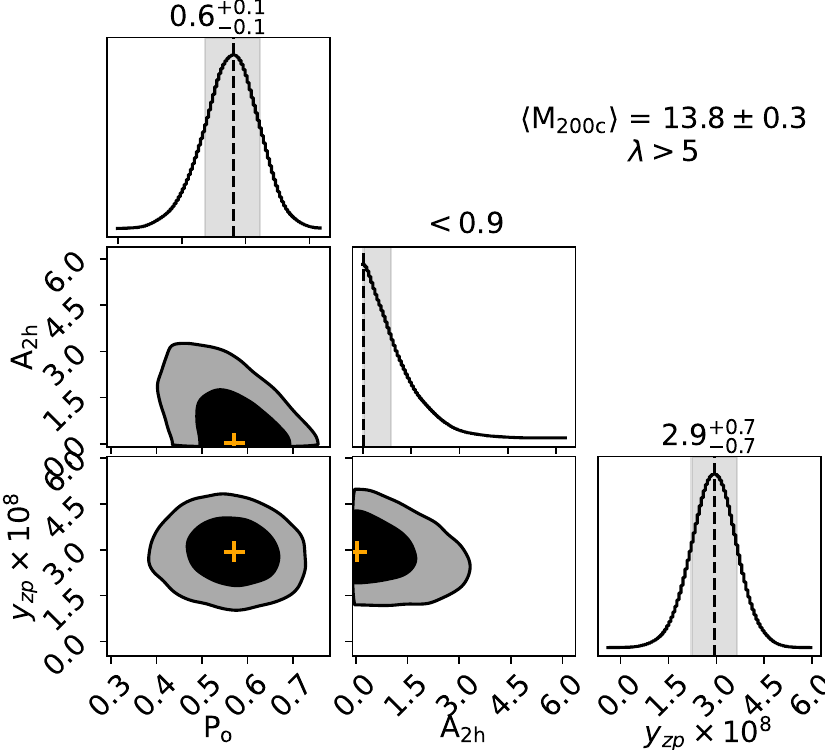}
    \caption{Fit statistic, radial profiles, and posterior parameter distribution for all other mass bins at different richness, $\lambda$. The description of the figures is similar to that in Figure~\ref{fig:main}. Top row and left panel of the 2nd row: isolated ($\lambda$=1) halos. Right panel of the 2nd row, 3rd and 4th rows (from top): low-richness ($\lambda$=2--5) halos. Bottom two rows: high-richness ($\lambda>$5) halos.}\label{fig:others}
\end{figure}

We show the results for all other mass and richness bins except the one shown in Figure~\ref{fig:main}, in Figure~\ref{fig:others}. Out of the 13 cases (including the one shown in the main text), there are three \textit{moderate evidence} ($\rm 2<\Delta BIC\leqslant6$) and two \textit{strong evidence} ($\rm 6<\Delta BIC\leqslant10$) of a flatter-than-GNFW thermal pressure profile. There are two cases of the fiducial GNFW being \textit{ruled out} ($\rm \Delta BIC>10$). In all other six cases, a deviation from GNFW is \textit{not required} ($\rm \Delta BIC<2$). The flat profile of $\beta=2.6$ often claimed in the literature \citep[e.g.,][]{Amodeo2021,Liu2025} is moderately and strongly disfavored in two cases. Although not intuitively obvious, the S/N and the relative strength of \onehalo to the tSZ background (\twohalo and $y_{zp}$) do not play a key role in determining this \textit{requirement}. The S/N of $y_{stacked}$ ranges from $4.7\sigma$ to $16.8\sigma$ in different mass and richness bins, with an average around $10.2\sigma$, but the constraint on \onehalo never exceeds $9\sigma$, indicating the S/N of the total data is barely correlated with the constraint on \onehalo (correlation coefficient 0.18, p-value=0.53). Some flatter profiles are dominated by the tSZ background, while in some cases the dominant one-halo term is not necessarily flatter than the fiducial GNFW profile. The \textit{requirement} of a flatter pressure profile is both mass- and richness-dependent, underscoring the importance of studying halos of different richness at different masses separately. 


\section{Assessing any residual effect of CIB}\label{sec:cib}

\begin{figure*}
\includegraphics[trim = 15 0 50 45, clip,width=0.485\linewidth]{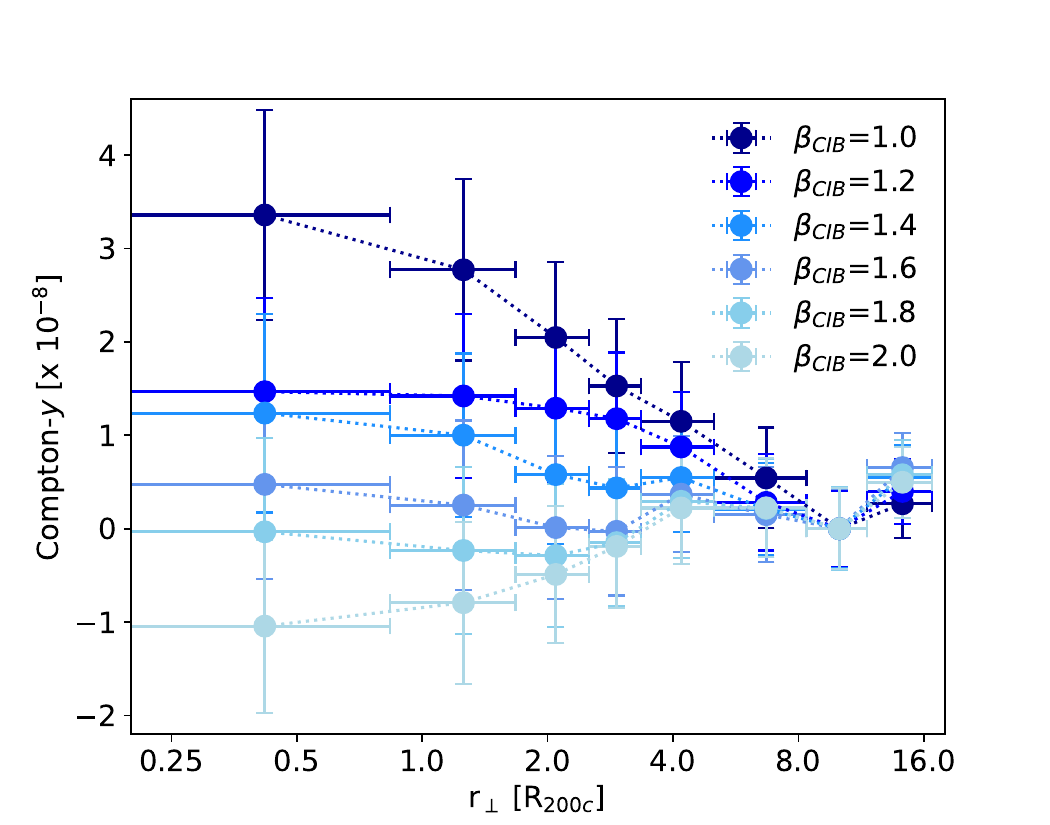}
    \includegraphics[trim = 15 0 40 40, clip, width=0.51\linewidth]{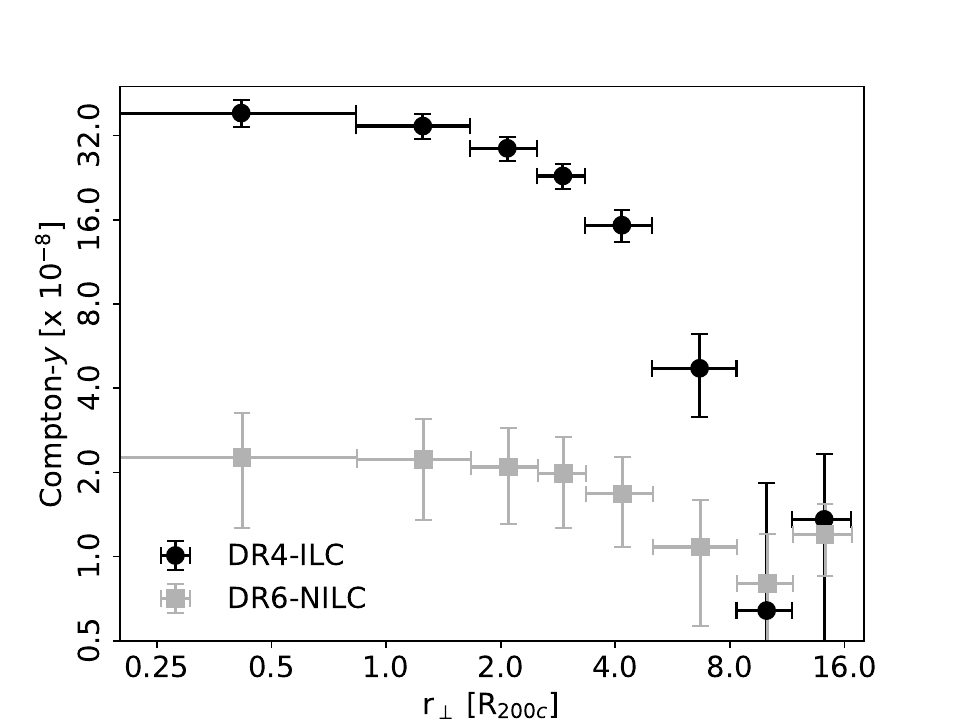}
    \includegraphics[trim = 60 0 80 25, clip,width=0.61\linewidth]{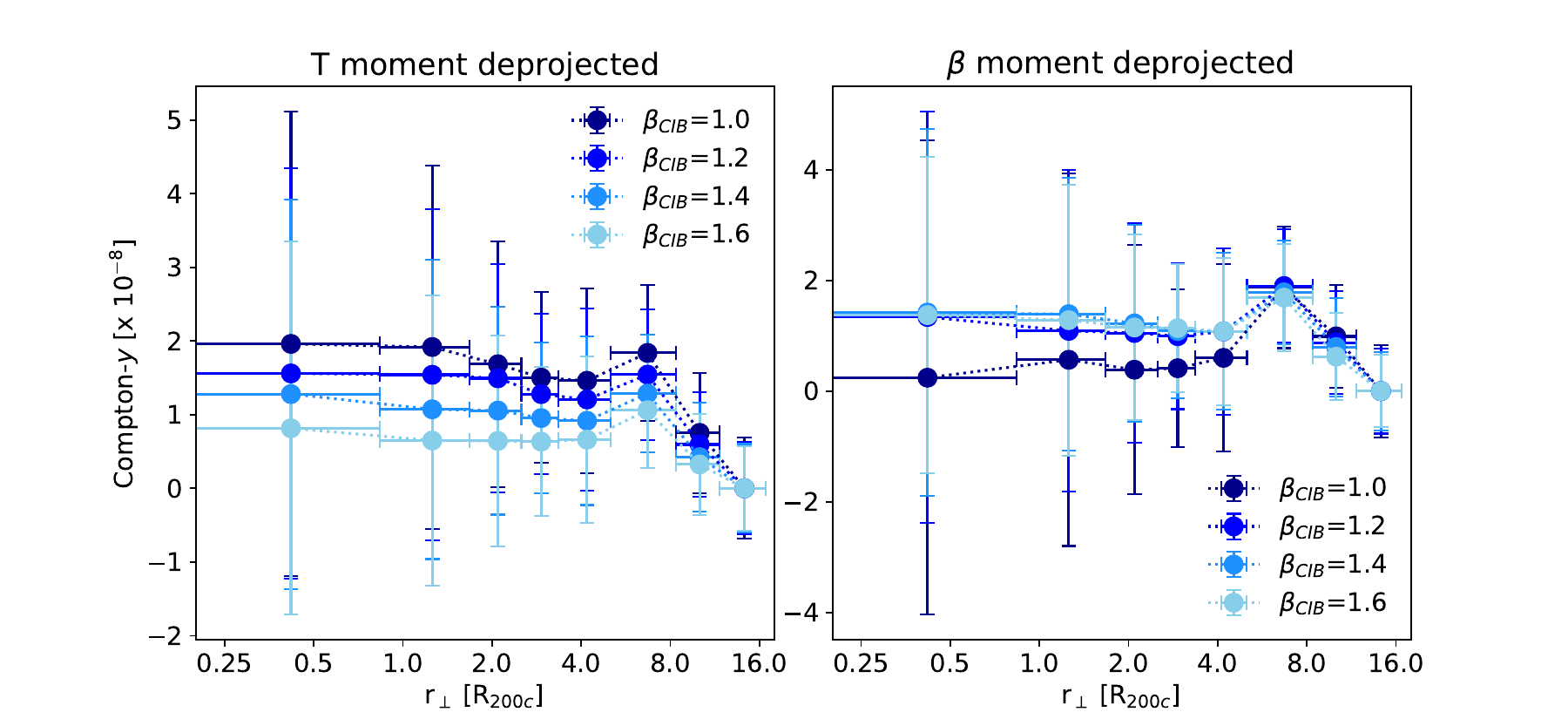}
    \includegraphics[trim = 15 0 40 40, clip,width=0.38\linewidth]{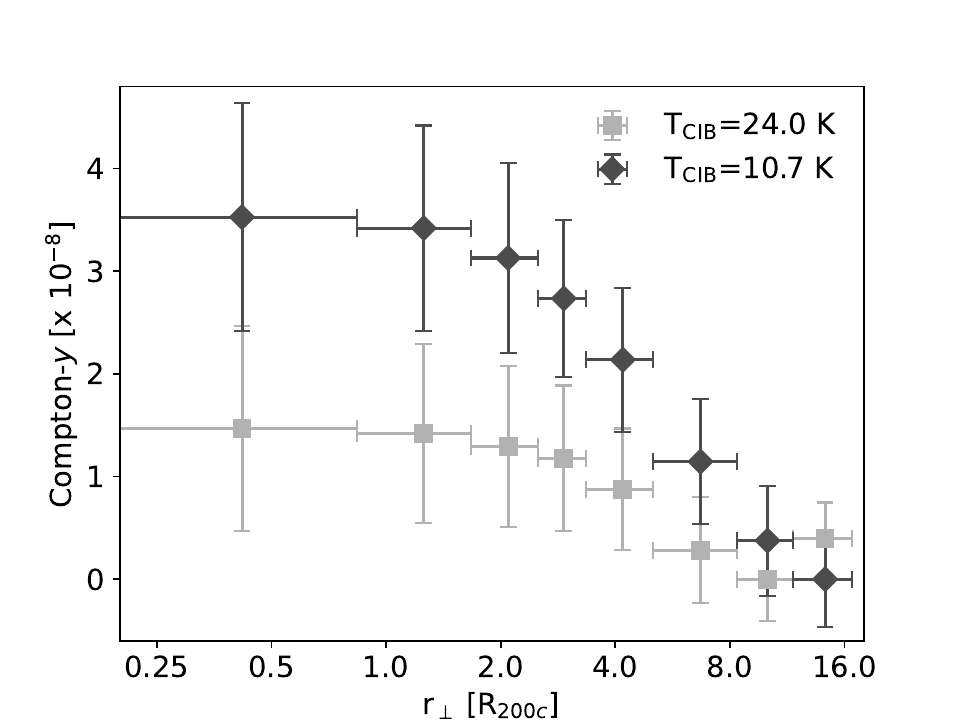}
    \caption{{Comparison of the projected radial profile of Compton-$y$ of our entire sample cross-correlated with different $y$-maps. \textbf{Top left:} DR6 $y$-map CIB-deprojected at different values of $\beta_{CIB}$ (mentioned in the legend) and at $\rm T_{CIB}$ = 24~K. \textbf{Bottom left and middle:} DR6 $y$-map deprojected with CIB and the moment of $\rm T_{CIB}$ (left panel) or the moment of $\beta_{CIB}$ (middle panel), at different values of $\beta_{CIB}$ (mentioned in the legend) and at $\rm T_{CIB}$ = 24~K. \textbf{Bottom right:} DR6 $y$-map CIB-deprojected at different values of $\rm T_{CIB}$ (mentioned in the legend) and at $\beta_{CIB}$ = 1.2. \textbf{Top right:} ACT DR4 maps (black circles) vs DR6 maps (gray squares), where the tSZ signal was extracted employing internal linear combination (ILC) vs a needlet-based internal linear combination (NILC) method, respectively, but both were deprojected using the same CIB parameters of $\beta_{CIB}=1.2$ and $\rm T_{CIB}$ = 24~K. In this panel, all $y$-values have been shifted up by 1e-8 to enable plotting the y-axis on a log scale and help visualize the large difference between the two profiles.}}\label{fig:cib}
\end{figure*}

{The cosmic infrared background (CIB) is one of the major sources of systematics in the thermal SZ signal extracted from the CMB data. The CIB is usually modeled with a modified blackbody spectrum, with its two main variables being temperature ($\rm T_{CIB}$) and the spectral index of dust emissivity ($\beta_{CIB}$). The $y$-map from ACT DR4 that we have used in the main text \citepalias{Madhavacheril2020} was CIB-deprojected at $\rm T_{CIB}$ = 24~K and $\beta_{CIB}$ = 1.2. However, the accurate value of these quantities and their spatial as well as cosmic evolution is debatable \citep[e.g.,][and references therein]{Planck2014}. Thus, the effective true CIB for our sample may not be the same as the one we have considered in the main text. To assess any residual effect of CIB on our results, we use the section of the $y$-maps from ACT DR6 overlapping with the footprint of DR4, CIB-deprojected with a range of $\rm T_{CIB}$ and $\beta_{CIB}$ \citep{Coulton2024}, and cross-correlate those $y$-maps with our entire sample. First, we compare the results for different values of $\beta_{CIB}$ at $\rm T_{CIB}$ = 24~K (Figure~\ref{fig:cib}, top left). At $\beta_{CIB}>1.6$, the residual effect of dust leading to unphysically negative Compton-$y$ values is clearly visible; therefore, we discard those higher values of $\beta_{CIB}$. We further test whether we need to deproject the $y$-maps with the moments of $\rm T_{CIB}$ or $\beta_{CIB}$. The resultant profiles are convergent across $\beta_{CIB}$ (Figure~\ref{fig:cib}, bottom left and middle), implying that further deprojection with the moments is not necessary. We explore the difference between $\rm T_{CIB}$ = 24~K vs 10.7~K for $\beta_{CIB}$ = 1.2 as well (Figure~\ref{fig:cib}, bottom right). It shows that the thermal energy within \rvirial would be $2.4\pm0.8\times$ larger at the lower temperature. However, the average $\rm T_{CIB}$ is usually found to be $\approx15$~K at $z=0$, and increases with redshift \citep[e.g.,][]{Chiang2025}; indicating a choice of $\rm T_{CIB}$ = 10.7~K might not be physically wise. Also, the profiles for $(\beta_{CIB},\rm T_{CIB}$) = (1.2, 10.7~K) and (1.0, 24~K) are similar (Figure~\ref{fig:cib}, black diamonds in the bottom-left and dark-blue circles in the top-left panel), implying that the effect of different $\beta_{CIB}$ and $\rm T_{CIB}$ is highly degenerate. To avoid over-accounting for the systematic effect, we thus discard $\rm T_{CIB}$ = 10.7~K, and consider the difference due to $\beta_{CIB}$ between 1.0 and 1.6 at $\rm T_{CIB}$ = 24~K. Thermal energies within \rvirial for $\beta_{CIB}$ = 1.2--1.6 are consistent with each other within 1$\sigma$. Thus, at the lower end, residual dust is unlikely to affect the thermal energy. On the higher end, the thermal energy at $\beta_{CIB}$ = 1.0 is $2.1\pm0.8\times$ higher than that at $\beta_{CIB}$ = 1.2.} 

{Finally, we compare the results between DR4 and DR6 for the same $\rm T_{CIB}$ and $\beta_{CIB}$ of 24~K and 1.2 (Figure~\ref{fig:cib}, top right). Ideally, these profiles should be identical, as it is stacked upon the same targets and the $y$-maps are deprojected with the same CIB parameters. However, there is more than an order-of-magnitude difference between the profiles. The remaining difference is in the techniques of preparing the $y$-maps. $y$-values in DR4 were extracted using the Internal Linear Combination (ILC) method in the Fourier domain, taking into account anisotropy in atmospheric noise, and incorporating explicit deprojection to mitigate the CIB contamination. On the other hand, $y$-values in DR6 were extracted using the needlet (localized wavelet) based ILC method in spherical harmonic space, assuming isotropic noise in each needlet, and suppressing total foreground power using a blind variance minimization approach, which could leave residual CIB contamination, especially at smaller length scales. This explains the stark deficit in the Compton-$y$ values from DR6 compared to those from DR4. This difference at the same CIB parameters makes translating the systematic assessment from DR6 maps to DR4 maps nontrivial. Nonetheless, if we assume that the fractional differences among stacked profiles due to different CIB parameters remain the same in DR4 vs DR6, that the same systematics is valid across all mass and richness bins, and that the correction is spatially uniform across our considered length scales, then we can claim that all thermal energy estimates in the main text could be underestimated by a factor of $2.1\pm0.8$ but are unlikely to be overestimated. Testing whether these simplified assumptions are entirely valid is beyond the scope of this paper and current mm facilities; this will be explored in a follow-up investigation. The inclusion of this additional systematics won't affect the slope ($\Gamma$) of the thermal energy-mass relation, but will enhance the normalization, leading to a larger baryon fraction and temperature floor. Thus, some halos might not be as cooled as they were suspected to be, and the overheating by thermal feedback in some halos might be more intense (see \S\ref{sec:physical}). The relative comparison across richness at similar masses will also remain unchanged if all halos are affected by the same factor from the residual dust, if any.}


\bibliographystyle{aasjournal}


\end{document}